# LASER DRIVEN NUCLEAR PHYSICS AT ELI–NP

F. NEGOITA[1,*], M. ROTH[2], P.G. THIROLF[3], S. TUDISCO[4], F. HANNACHI[5], S. MOUSTAIZIS[6], I. POMERANTZ[7] , P. MCKENNA[8], J. FUCHS[9], K. SPHOR[10], G. ACBAS[1], A. ANZALONE[4], P. AUDEBERT[9], S. BALASCUTA[1], F. CAPPUZZELLO[4,11], M.O. CERNAIANU[1], S. CHEN[9], I. DANCUS[1], R. FREEMAN[12], H. GEISSEL[13], P. GHENUCHE[1], L. GIZZI[14], F. GOBET[5], G. GOSSELIN[15], M. GUGIU[1], D. HIGGINSON[9], E. D'HUMIÈRES[16], C. IVAN[1], D. JAROSZYNSKI[8], S. KAR[17], L. LAMIA[4,11], V. LECA[1], L. NEAGU[1], G. LANZALONE[4,18], V. MÉOT[15], S.R. MIRFAYZI[17], I.O. MITU[1], P. MOREL[15], C. MURPHY[19], C. PETCU[1], H. PETRASCU[21], C. PETRONE[21], P. RACZKA[20], M. RISCA[1], F. ROTARU[1], J.J. SANTOS[16], D. SCHUMACHER[12], D. STUTMAN[1], M. TARISIEN[5], M. TATARU[1], B. TATULEA[1], I.C.E. TURCU[1], M. VERSTEEGEN[5], D. URSESCU[1], S. GALES[1], N.V. ZAMFIR[1]

[1] ELI-NP, "Horia Hulubei" Institute for Physics and Nuclear Engineering, 30 Reactorului Street, RO-077125 Bucharest-Magurele, Romania
[2] Institut fur Kernphysik, Technische Universitat Darmstadt, Schloßgartenstrasse 9, D-64289 Darmstadt, Germany
[3] Ludwig-Maximilians-Universität Munich, D-85748 Garching, Germany
[4] INFN - Laboratori Nazionali del Sud – Via S. Sofia 62, 95123 Catania, Italy
[5] Centre d'Etudes Nucleaires de Bordeaux Gradignan, Universite Bordeaux1, CNRS-IN2P3, Route du solarium, 33175 Gradignan, France
[6] Technical University of Crete, Chania, Crete, Greece
[7] Tel Aviv University, P.O. Box 39040, Tel Aviv 6997801, Israel
[8] Department of Physics, University of Strathclyde, Glasgow, G4 0NG, UK
[9] Laboratoire pour l'Utilisation des Lasers Intenses, UMR 7605 CNRS-CEA-École Polytechnique-Université Paris VI, 91128 Palaiseau, France
[10] University of the West of Scotland, Paisley, PA1 2BE Scotland, UK
[11] Dip. di Fisica e Astronomia, Univ. degli Studi di Catania – Via S. Sofia 64, 95123 Catania, Italy
[12] Department of Physics, The Ohio State University, 191 West Woodruff Avenue, Columbus, Ohio 43210, USA
[13] Justus-Liebig-University Giessen, Ludwigstrasse 23, 35390 Giessen, Germany
[14] Istituto Nazionale di Ottica - UOS, Area della Ricerca del CNR, Via G. Moruzzi 1 - 56124 Pisa, Italy
[15] Commissariat à l'énergie atomique, Service de Physique Nucléaire Boite Postale 12, F-91680 Bruyères-le-Châtel, France
[16] CELIA, Université Bordeaux1, 351 Cours de la Libération, F-33405 Talence cedex, France
[17] School of Mathematics and Physics, The Queen's University of Belfast, Belfast BT7 1NN, UK
[18] Università degli Studi di Enna "Kore" – Via delle Olimpiadi, 94100 Enna, Italy
[19] Department of Physics, University of York, York YO10 5D, UK
[20] Institute of Plasma Physics and Laser Microfusion, Hery Street 23, 01-497 Warsaw, Poland
[21] "Horia Hulubei" Institute for Physics and Nuclear Engineering, 30 Reactorului Street, RO-077125 Bucharest-Magurele, Romania
[*] Corresponding author *E-mail:* florin.negoita@eli-np.ro

*Abstract.* High power lasers have proven being capable to produce high energy γ-rays, charged particles and neutrons, and to induce all kinds of nuclear reactions. At ELI, the studies with high power lasers will enter for the first time into new domains of power and intensities: 10 PW and $10^{23}$ W/cm$^2$. While the development of laser based radiation sources is the main focus at the ELI-Beamlines pillar of ELI, at ELI-NP the studies that will benefit from High Power Laser System pulses will focus on *Laser Driven Nuclear Physics* (this TDR, acronym LDNP, associated to the E1 experimental area), *High Field Physics and QED* (associated to the E6 area) and fundamental research opened by the unique combination of the two 10 PW laser pulses with a gamma beam provided by the Gamma Beam System (associated to E7 area). The scientific case of the LDNP TDR encompasses studies of laser induced nuclear reactions, aiming for a better understanding of nuclear properties, of nuclear reaction rates in laser-plasmas, as well as on the development of radiation source characterization methods based on nuclear techniques. As an example of proposed studies: the promise of achieving solid-state density bunches of (very) heavy ions accelerated to about 10 MeV/nucleon through the RPA mechanism will be exploited to produce highly astrophysical relevant neutron rich nuclei around the N~126 waiting point, using the sequential fission-fusion scheme, complementary to any other existing or planned method of producing radioactive nuclei.

The studies will be implemented predominantly in the E1 area of ELI-NP. However, many of them can be, in a first stage, performed in the E5 and/or E4 areas, where higher repetition laser pulses are available, while the harsh X-ray and electromagnetic pulse (EMP) environments are less damaging compared to E1.

A number of options are discussed through the document, having an important impact on the budget and needed resources. Depending on the TDR review and subsequent project decisions, they may be taken into account for space reservation, while their detailed design and implementation will be postponed.

The present TDR is the result of contributions from several institutions engaged in nuclear physics and high power laser research. A significant part of the proposed equipment can be designed, and afterwards can be built, only in close collaboration with (or subcontracting to) some of these institutions. A Memorandum of Understanding (MOU) is currently under preparation with each of these key partners as well as with others that are interested to participate in the design or in the future experimental program.

## 1. INTRODUCTION

The present Technical Design Report (TDR) is rather meant as an advanced conceptual design report, similar to all the other TDRs for experiments prepared at this stage within the Extreme Light Infrastructure – Nuclear Physics (ELI-NP) project, for the purpose of evaluation of feasibility and overall project coherence before going into detailed design of experimental devices.

The ELI-NP High Power Laser System (HPLS) is composed of two amplification chains working in parallel. Each arm has three outputs (to be used only one at once):

- 10 PW with a repetition rate of 1 pulse per minute or higher
- 1 PW with a repetition rate of 1 Hz
- 100 TW with a repetition rate of 10 Hz

All outputs are expected to have their central wavelength at approximately 800 nm, a pulse duration of about 25 fs (if larger, the energy per pulse will be increased to reach the specified power), a pre-pulse contrast of $1:10^{12}$ and a Strehl ratio of 0.7.

At present there are two major laser systems operational or under construction that deliver intense pulses of laser light, the National Ignition Facility (NIF) at the Lawrence Livermore National Laboratory (LLNL) in the US and the Laser Megajoule (LMJ) in France. Both laser systems are dedicated to the compression and heating of matter using energies up to the megajoule range to explore exotic states of matter, perform classified research for defense applications and ignite a burning fusion capsule for energy research. In contrast, ELI-NP not only is a pure civilian facility for fundamental research only, but also exceeds the capabilities of both laser systems in terms of beam intensity by orders of magnitude. Currently NIF is augmented with the addition of a short pulse laser system (ARC, Advanced Radiography Capability). The design goal of this system is kJ in energy delivered in picosecond pulse duration, resulting in a PW class laser power. However, as this system is attached to NIF and suffers from limitations in focusing, the maximum intensity achievable will be of the order of $10^{19}$-$10^{20}$ W/cm$^2$, about three orders of magnitude below the design goal of ELI-NP. ELI-NP therefore complements the large systems, as it exchanges energy for intensity to explore novel aspects of nuclear phenomena not accessible by the other systems.

## 2. PHYSICS CASES

### 2.1 NUCLEAR FUSION REACTIONS FROM LASER-ACCELERATED FISSILE ION BEAMS

Elements like platinum, gold, thorium and uranium are produced via the rapid neutron capture process (r-process) at astrophysical sites like merging neutron star binaries or (core collapse) supernova type II explosions. We aim at improving our understanding of these nuclear processes by measuring the properties of heavy nuclei on (or near) the r-process path. While the lower-mass path of the r-process for the production of heavy elements is well explored, the nuclei around the N = 126 waiting point critically determine this element production mechanism. At present, basically nothing is known about these nuclei. Fig. 1 shows the nuclides chart marked with different nucleosynthesis pathways for the production of heavy elements in the Universe: the thermonuclear fusion processes in stars producing elements up to iron (orange arrow), the slow neutron capture process (s-process) along the valley of stability leading to about half of the heavier nuclei (red arrow) and the rapid neutron capture process (r-process). The astrophysical site of the r-process nucleosynthesis is still under debate: it may be cataclysmic core collapse supernovae (II) explosions with neutrino winds [1-4] or mergers of neutron-star binaries [5-7]. For the heavier elements beyond barium, the isotopic abundances are always very similar (called universality) and the process seems to be very robust. Perhaps also the recycling of fission fragments from the end of the r-process strengthens this stability. Presently, it seems more likely that a merger of neutron star binaries is the source for the heavier r-process branch, while core collapsing supernova explosions contribute to the lighter elements below barium.

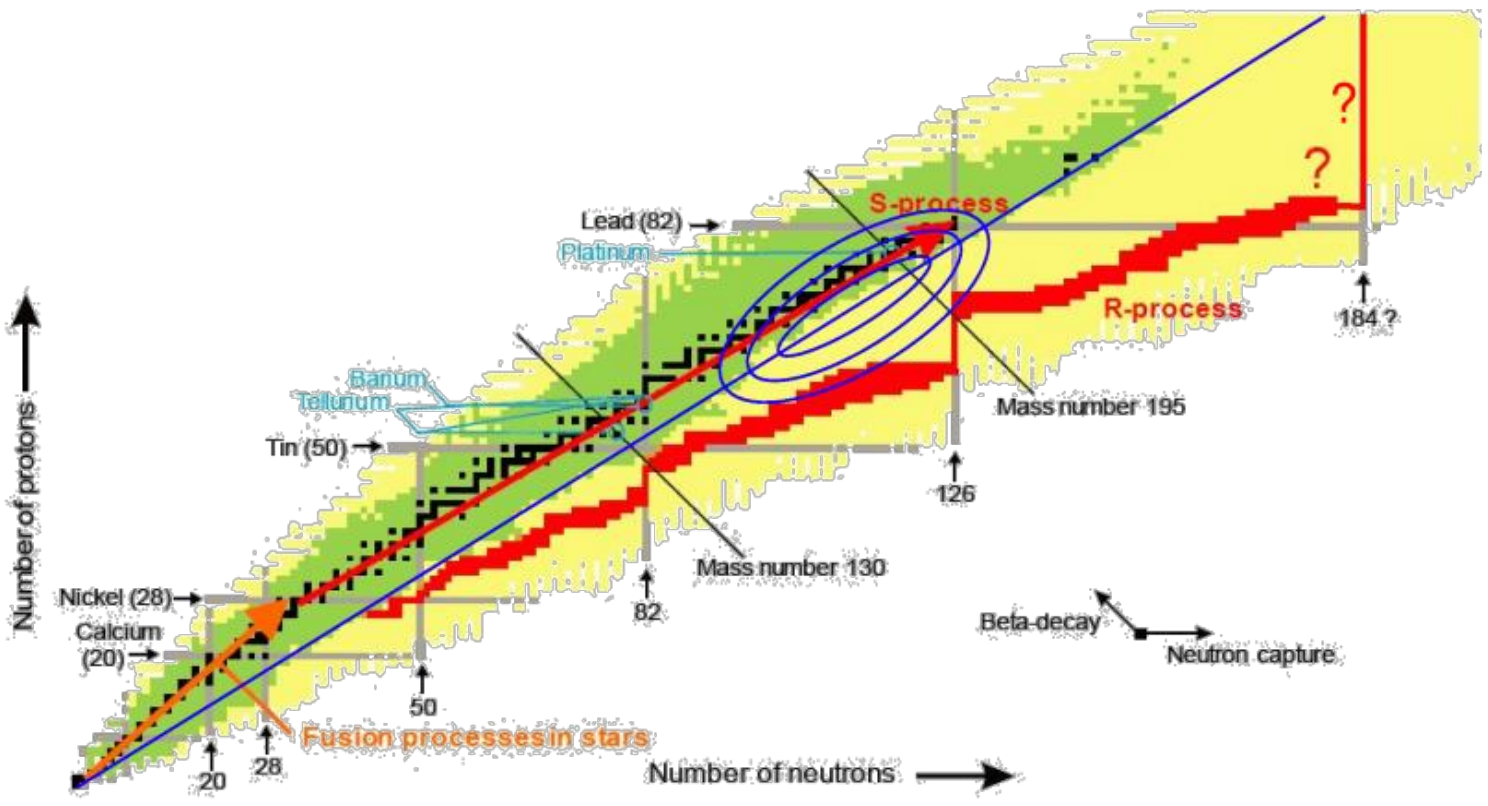

Figure 1 – Chart of the nuclides indicating various pathways for astrophysical nucleosynthesis: thermonuclear fusion reactions in stars (orange vector), s-process path (red vector) and the r-process generating heavy nuclei in the Universe (red pathway). The nuclei marked in black indicate stable nuclei. For the green nuclei some nuclear properties are known, while the yellow, yet unexplored regions extend to the neutron and proton drip lines. The blue line connects nuclei with the same neutron/proton ratio as for (almost) stable actinide nuclei. On this line the maximum yield of nuclei produced via fission-fusion (without neutron evaporation) will be located. The elliptical contour lines correspond to the expected maximum fission-fusion cross sections decreased to 50%, 10% and 0.1%, respectively, for primary $^{232}$Th beams.

The modern nuclear equations of state, neutrino interactions and recent supernova explosion simulations [2] lead to detailed discussions of the waiting point N=126. Here measured nuclear properties along the N=126 waiting point may help to clarify the sites of the r-process.

Fig. 2 shows the measured solar elemental abundances of the r-process nuclei together with a theoretical calculation, where masses from the Extended Thomas-Fermi plus Strutinski Integral (ETFSI) mass model [8] have been used together with several neutron flux components, characterized by a temperature $T_9$, neutron densities $n_n$ and expansion time scales. A quenching of shell effects [9] was assumed in the nuclear mass calculations to achieve a better agreement between observed and calculated abundances. The three pronounced peaks visible in the abundance distribution seem to be of different origin, which is also reflected in the theoretical calculations shown in Fig. 2, where contributions from different temperatures and neutron densities are superimposed to the observed data. We note the pronounced third peak in the abundance distribution around A = 180−200, corresponding to the group of elements around gold, platinum and osmium, where until now no experimental nuclear properties have been measured for r-process nuclei. Several astrophysical scenarios try to explain this third abundance peak.

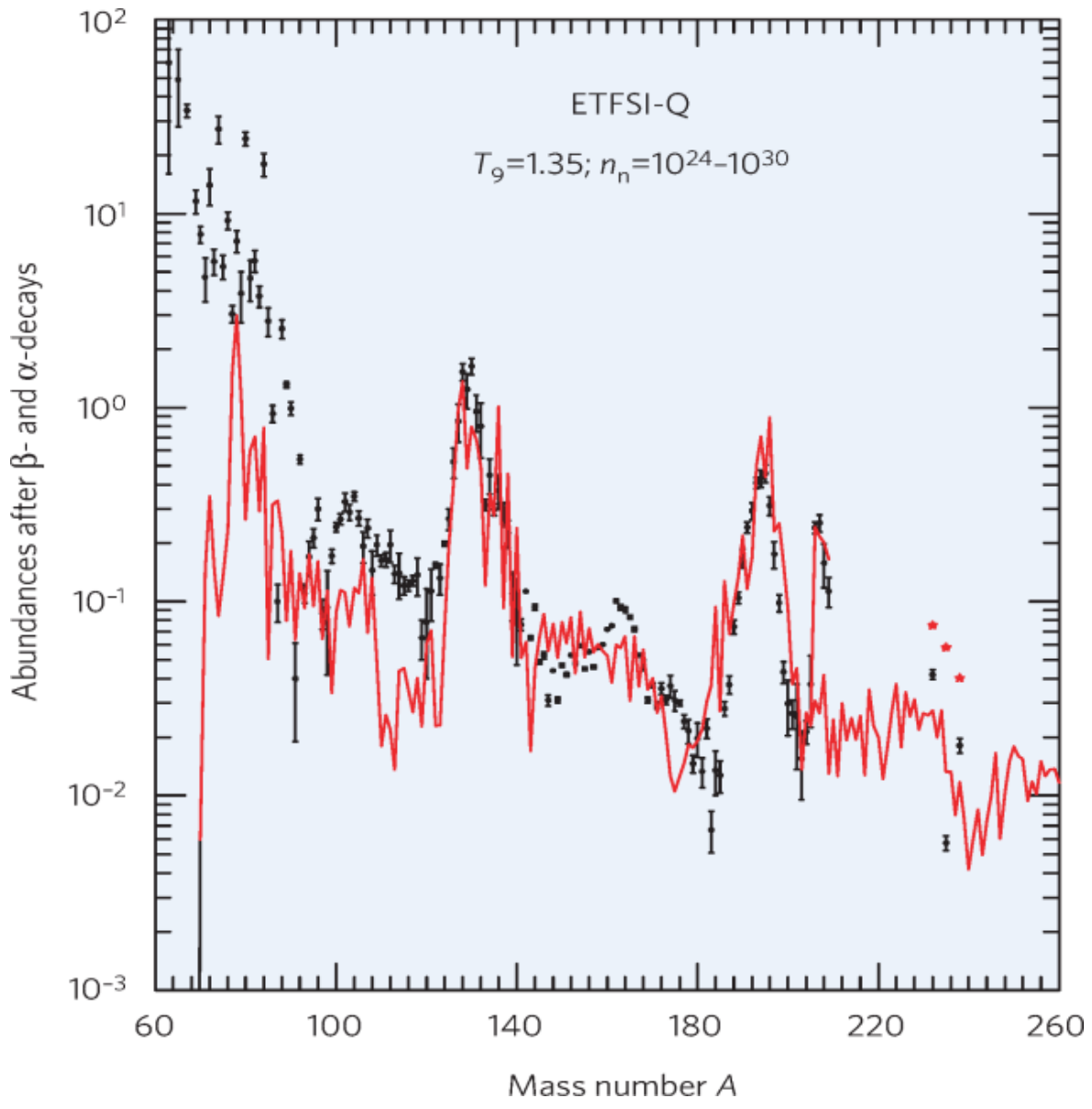

Figure 2 – Observed elemental solar abundances in the r-process mass range (black symbols) in comparison with calculated abundances (red line and symbols), normalized to silicon=$10^6$. The theoretical predictions show the elemental abundances for stable isotopes after α and β decay as obtained in the ETFSI-Q mass model [8,10] for a wide range of neutron densities $n_n$ (in 1/cm$^3$) and temperatures $T_9$ (in units of $10^9$K) and including shell quenching effects. Included with permission from [11].

A detailed knowledge of nuclear lifetimes and binding energies in the region of the N=126 waiting point will narrow down the possible astrophysical sites. If, e.g., no shell quenching could be found in this mass range, the large dip existing for this case in front of the third abundance peak would have to be filled up by other processes like neutrino wind interactions. Considering the still rather large difficulties to identify convincing astrophysical sites for the third peak of the r-process with sufficiently occurrence rates, measurements of the nuclear properties around the N=126 waiting point will represent an important step forward in solving the difficult and yet confusing site selection of the third abundance peak of the r-process.

The key bottleneck nuclei of the N=126 waiting point around Z~70 are about 15 neutrons away from presently known nuclei (see Fig. 1), with a typical drop of the production cross section for classical radioactive beam production schemes of about a factor of 10-20 for each additional neutron towards more neutron-rich isotopes. Thus presently nothing is known about these nuclei and even next-generation large-scale 'conventional' radioactive beam facilities like FAIR [12], SPIRAL II [13] or FRIB [14] will hardly be able to grant experimental access to the most important isotopes on the r-process path. The third peak in the abundance curve of r-process nuclei is due to the N = 126 waiting point as visible in Fig. 1. These nuclei are expected to have rather long half-lives of a few 100 ms. This waiting point represents the bottleneck for the nucleosynthesis of heavy elements up to the actinides. From the view point of astrophysics, it is the last region, where the r-process path gets close to the valley of stability and thus can be studied with the new isotopic production scheme discussed below. While the waiting point nuclei at N = 50 and N = 82 have been studied rather extensively [15, 16, 17, 18], nothing is known experimentally about the nuclear properties of waiting point nuclei at the N=126 magic number. Nuclear properties to be studied here are nuclear masses, lifetimes, beta-delayed neutron emission probabilities $P_n$ and the underlying nuclear structure. If we improve our experimental understanding of this final bottleneck to the actinides at N=126, many new visions open up: (i) for many mass formulas (e.g. [19]), there is a branch of the r-process leading to extremely long-lived superheavy elements beyond Z=110 with lifetimes of about $10^9$ years. If these predictions could be made more accurate, a search for these superheavy elements in nature would become more promising. (ii) At present the prediction for the formation of uranium and thorium elements in the r-process is rather difficult, because there are no nearby magic numbers and those nuclei are formed during a fast passage of the nuclidic area between shells. Such predictions could be improved, if the bottleneck of actinide formation would be more reliably known. (iii) Also the question could be clarified if fission fragments are recycled in many r-process loops or if only a small fraction is reprocessed.

This description of our present understanding of the r-process underlines the importance of the present project for nuclear physics and, particularly, for astrophysics.

### 2.1.1 RPA for heavy ions

In the proposal of a new nuclear reaction scenario proposed here, we envisage to exploit the Radiation Pressure Acceleration (RPA) mechanism for ion acceleration. It was first proposed theoretically [20-24]. Special emphasis has been given to RPA with circularly polarized laser pulses as this suppresses fast electron generation and leads to the interaction dominated by the radiation pressure [20, 21]. It has been shown that RPA operates in two modes. In the first one, called 'hole-boring', the laser pulses interact with targets thick enough to allow to drive target material ahead of it as a piston, but without interacting with the target rear surface [20]. The first experimental observation of RPA in the 'hole-boring' regime was achieved in experiments led by the Munich group [25, 26]. The RPA laser ion acceleration mechanism in general provides the highest achievable efficiency for the conversion from laser energy to ion energy and for circularly polarized laser light RPA holds promise of quasi-monoenergetic ion beams. Due to the circular polarization, electron heating is strongly suppressed. The electrons are compressed to a dense electron sheet in front of the laser pulse, which then via the Coulomb field accelerates the ions. This mechanism requires very thin targets and ultra-high contrast laser pulses to avoid the pre-heating and expansion of the target before the interaction with the main laser pulse. The RPA mechanism allows to produce ion bunches with solid-state density ($10^{22}$ - $10^{23}/cm^3$), which thus are $\sim 10^{14}$ times denser than ion bunches from classical accelerators. Correspondingly, the areal densities of these bunches are $\sim 10^7$ times larger. It is important to note that these ion bunches are accelerated as neutral ensembles together with the accompanying electrons and thus do not Coulomb explode.

### 2.1.2 Stopping power of very dense ion bunches

In nuclear physics, the Bethe-Bloch formula [27] is used to calculate the atomic stopping of energetic individual electrons [28] by ionization and atomic excitation. For relativistic electrons, the other important energy loss is bremsstrahlung. The radiation loss is dominant for high energy electrons, e.g. E≥ 100 MeV and Z=10. If, however (see below), the atomic stopping becomes orders of magnitude larger by collective effects, the radiation loss can be neglected. For laser acceleration, the electron and ion bunch densities reach solid state densities, which are about 14-15 orders of magnitude larger compared to beams from classical accelerators. Here collective effects become important. One can decompose the Bethe-Bloch equation according to Ref. [29] into a first contribution describing binary collisions and a second term describing long range collective contributions. Ref. [30] discusses the mechanism of collective deceleration of a dense particle bunch in a thin plasma, where the particle bunch fits into part of the plasma oscillation and is decelerated $10^5 - 10^6$ stronger than predicted by the classical Bethe-Bloch equation [27] due to the strong collective wakefield. For ion deceleration we want to use targets with suitably low density. These new laws of deceleration and stopping of charged particles have to be established to use them later in experiments in an optimum way.

In the following, the opposite effect with a strongly reduced atomic stopping power that occurs when sending an energetic, solid state density ion bunch into a solid target, will be discussed. For this target the plasma wavelength ($\lambda_p \approx 1$ nm) is much smaller than the ion bunch length (≈ 500 nm) and collective acceleration and deceleration effects cancel each other. Only the binary collisions are important.

Hence, we may consider the dense ion bunch as consisting (in a simplistic view) of 300 layers with Angstrom distances. Here the first layers of the bunch will attract the electrons from the target and – like a snow plough – will take up the decelerating electron momenta. The predominant part of the ion bunch is screened from electrons and we expect a significant (here assumed as $\approx 10^2$ fold) reduction in stopping power. The electron density $n_e$ is strongly reduced in the channel because many electrons are driven out by the ion bunch and the laser. Again, all these effects have to be studied in detail. It is expected that the resulting very dense ion bunches should have a time evolution and the reaction products are emitted at different times and angles. Therefore, for the characterization of the dense bunches and their time evolution, the detection system needs to capture the reaction products, emitted at different times (analogous to time of flight measurements), and measure their angular distributions. Of course, the temporal evolutions, which can be followed, vary greatly depending on the temporal resolution of the diagnosis system. In a preliminary phase, it is expected that electrons and ions are emitted due to the Coulomb explosion of a part of the initially formed bunch (pre-bunch emission). Then, the remaining bunch will have a slower temporal evolution, which can be followed as a function of its time of flight in free space. The experimental study of deceleration of dense, high speed bunches of electrons and ions will require:

• Bunch characterization in free space: its components, their energies and the ion charge states, their angular distribution and temporal evolution; due to the large number of particles, the detection solid angles must be small (of the order of $10^{-7}$ sr or less).

• Tracking the changes introduced by bunches passing through different materials (solid or gas) and their deceleration study. Studies will be carried out depending on laser power and target type and thickness and for deceleration - depending on material type and its thickness.

The same detection system could be used for both diagnosis in free space and diagnosis after passing through a material. A rapid characterization may be done with a Thomson parabola ion spectrometer, and an electron magnetic spectrometer, implying measurements of the emissions at different times and possibly their angular distribution; in relevant case. A more complete analysis will require a diagnosis system working in real-time, using magnetic spectrometers and detection systems with high granularity or with position sensitive readout in the focal plane (e.g., stacks of ΔE-E detectors, with ionization chambers and Si or scintillation detectors). Even if the laser pulse frequency is small, the nuclear electronics can be triggered in the usual way.

### 2.1.3 Fission-Fusion reaction mechanism

The basic concept of the fission-fusion reaction scenario draws on the ultra-high density of laser accelerated ion bunches. Choosing fissile isotopes as target material for a first target foil accelerated by an intense driver laser pulse will enable the interaction of a dense beam of fission fragments with a second target foil, also consisting of fissile isotopes. So finally in a second step of the reaction process, fusion between (neutron-rich) beam-like and target-like (light) fission products will become possible, generating extremely neutron-rich ion species.

For our discussion we choose $^{232}$Th (the only component of chemically pure Th) as fissile target material, primarily because of its long half-life of $1.4 \cdot 10^{10}$ years, which avoids extensive radioprotection precautions during handling and operation. Moreover, metallic thorium targets are rather stable in a typical laser vacuum of $10^{-6}$ mbar, whereas, e.g., metallic $^{238}$U targets would quickly oxidize. Nevertheless, in a later stage it may become advantageous to use also heavier actinide species in order to allow for the production of even more exotic fusion products. In general, the fission process of the two heavy Th nuclei from beam and target will be preceded by the deep inelastic transfer of neutrons between the inducing and the fissioning nuclei. Here the magic neutron number in the superdeformed fissile nucleus with N=146 [31, 32] may drive the process towards more neutron-rich fissioning nuclei, because the second potential minimum acts like a doorway state towards fission. Since in the subsequent fission process the heavy fission fragments keep their A and N values [33], these additional neutrons will show up in the light fission fragments and assist to reach more neutron-rich nuclei.

Fig. 3 shows a sketch of the proposed fission-fusion reaction scenario. The accelerated thorium ions will be fissioned in the $CH_2$ layer of the reaction target, whereas the accelerated carbon ions and deuterons from the production target generate thorium fragments in the thick thorium layer of the reaction target. This scenario is more efficient than the one where fission would be induced by the thorium ions only. In view of the available energy in the accelerating driver laser pulse, the optimized production target should have a thickness of about 0.5 μm for the thorium as well as for the $CD_2$ layers. The thorium layer of the reaction target would have a thickness of about 50 μm. Using a distance of 2.8 Å between atoms in solid layers of $CH_2$, the accelerated light ion bunch ($1.4 \cdot 10^{11}$ ions) corresponds to 1860 atomic layers in case of a 520 nm thick $CD_2$ target.

In order to allow for an optimized fission of the accelerated Th beam, the thicker Th layer of the reaction target, which is positioned behind the production target, is covered by about 70 μm of polyethylene. This layer serves a twofold purpose: Primarily it is used to induce fission of the impinging Th ion beam, generating the beam-like fission fragments. Here polyethylene is advantageous compared to a pure carbon layer because of the increased number of atoms able to induce fission on the impinging Th ions. In addition, the thickness of this $CH_2$ layer has been chosen such that the produced fission fragments will be decelerated to a kinetic energy which is suitable for optimized fusion with the target-like fission fragments generated by the light accelerated ions in the Th layer of the reaction target, minimizing the amount of evaporated neutrons. For practical reasons, we propose to place the reaction target about 0.1 mm behind the production target, as indicated in Fig. 3. After each laser shot, a new double-target has to be rotated into position.

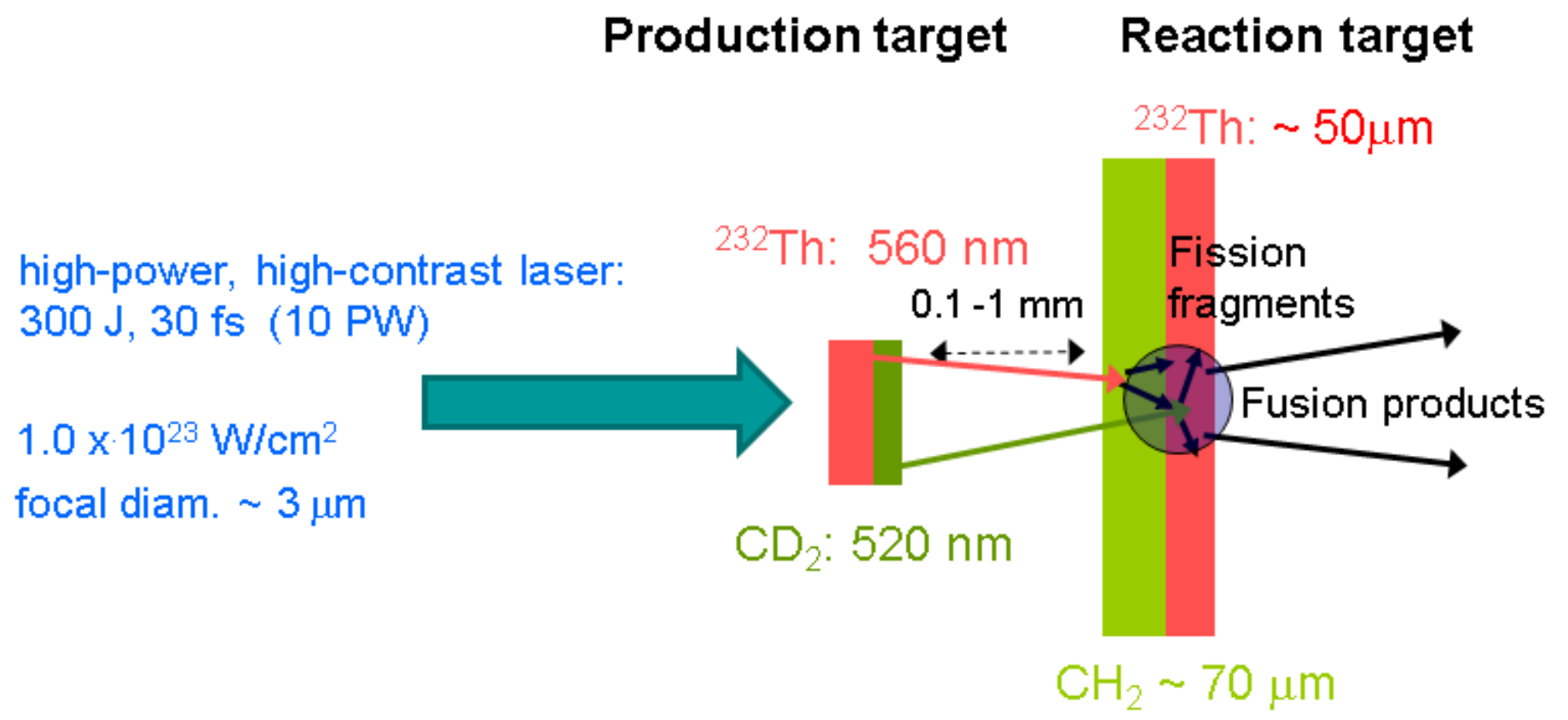

Figure 3 – Sketch of the target arrangement envisaged for the fission-fusion reaction process based on laser ion acceleration, consisting of a production and a reaction target from a fissile material (here $^{232}Th$), each of them covered by a layer of low-Z materials ($CD_2$ and $CH_2$, respectively). The thickness of the $CH_2$ layer as well as the second thorium reaction target have to be limited to 70 µm and 50 µm, respectively, in order to enable fission of beam and target nuclei. This will allow for fusion between their light fragments, as well as enable the fusion products to leave the second thorium reaction target.

In general, the fission process proceeds asymmetric [33]. The heavy fission fragment for $^{232}Th$ is centered at A=139.5 (approximately at Z=54.5 and N=84) close to the magic numbers Z=50 and N=82. Accordingly, the light fission fragment mass is adjusted to the mass of the fixed heavy fission fragment, thus resulting for $^{232}Th$ in $A_L$=91 with $Z_L \approx 37.5$. The width (FWHM) of the light fission fragment peak is typically $\Delta A_L$ = 14 mass units, the 1/10 maximum width about 22 mass units [33].

So far we have considered the fission process of beam-like Th nuclei in the $CH_2$ layer of the reaction target. Similar arguments can be invoked for the deuteron- (and carbon) induced generation of (target-like) fission products in the subsequent thicker thorium layer of the reaction target, where deuteron- and carbon-induced fission will occur in the $^{232}Th$ layer of the reaction target. Since we can consider the $2.8 \cdot 10^{11}$ laser-accelerated deuterons (plus $1.4 \cdot 10^{11}$ carbon ions) impinging on the second target per laser pulse as 1860 consecutive atomic layers, we conclude a corresponding fission probability in the Th layer of the reaction target of about $2.3\cdot10^{-5}$, corresponding to $3.2\cdot10^{6}$ target-like fission fragments per laser pulse. A thickness of the thorium layer of the reaction target of about 50 μm could be exploited, where the kinetic proton energy would be above the Coulomb barrier to induce fission over the full target depth. In a second step of the fission-fusion scenario, we consider the fusion between the light fission fragments of beam and target to a compound nucleus with a central value of A~182 and Z~75. Again we employ geometrical arguments for an order-of-magnitude estimate of the corresponding fusion cross section. For a typical light fission fragment with A = 90, the nuclear radius can be estimated as 5.4 fm. Considering a thickness of 50 μm for the Th layer of the reaction target that will be converted to fission fragments, equivalent to $1.6\cdot10^{5}$ atomic layers, this results in a fusion probability of about $1.8\cdot10^{-4}$. Very neutron-rich nuclei still have comparably small production cross sections, because weakly bound neutrons ($S_n$~3 MeV) will be evaporated easily. The optimum range of beam energies for fusion reactions resulting in neutron-rich fusion products amounts to about 2.8 MeV/u according to PACE4 [34] calculations. So, e.g., the fusion of two neutron-rich $^{98}_{35}Br$ fission products with a kinetic energy of the beam-like fragment of 275 MeV leads with excitation energy of about 60 MeV to a fusion cross section of 13 mb for $^{189}_{70}Yb_{119}$, which is already 8 neutrons away from the last presently known Yb isotope. One should note that the well-known hindrance of fusion for nearly symmetric systems (break-down of fusion) only sets in for projectile and target masses heavier than about 100 amu [35, 36]. Thus for the fusion of light fission fragments, we expect an unhindered fusion evaporation process. A detailed discussion of the achievable fission-fusion reaction yield is given in Ref. [37]. In addition to the scenario discussed above, the exceptionally high ion bunch density may lead to collective effects that do not occur with conventional ion beams: when sending the energetic, solid-state density ion bunch into a solid carbon or thorium target, the plasma wavelength ($\lambda_p \approx 5$ nm, driven by the ion bunch with a phase velocity corresponding to the thorium ion velocity) is much smaller than the ion bunch length (≈ 560 nm) and collective acceleration and deceleration effects cancel. As discussed already before, only the binary collisions remain and contribute to the stopping power. In this case the first layers of the impinging ion bunch will attract the electrons from the target and like a snow plough will take up the decelerating electron momenta. Hence the predominant part of the ion bunch is screened from electrons and we expect a drastic reduction of the stopping power. The electron density $n_e$ will be strongly reduced in the channel defined by the laser-accelerated ions, because many electrons are expelled by the ion bunch and the laser pulse. This effect requires detailed experimental investigations planned for the near future, aiming at verifying the perspective to use a significantly thicker reaction target, which in turn would significantly boost the achievable fusion yield.

Fig. 4 displays a closer view into the region of nuclides around the N=126 waiting point of the r-process, where nuclei on the r-process path are indicated by the green color, with dark green highlighting the key bottleneck r-process isotopes [38] at N=126 between Z=66 (Dy) and Z=70 (Yb). One should note that, e.g., for Yb the presently last known isotope is 15 neutrons away from the r-process path at N=126. The isotopes in light blue mark those nuclides, where recently beta half-lives could be measured following projectile fragmentation and in-flight separation at GSI [39]. Again the elliptical contour lines indicate the range of nuclei accessible with our new fission-fusion scenario on a level of 50%, 10% and $10^{-3}$ of the maximum fusion cross section between two neutron-rich light fission fragments in the energy range of about 2.8 MeV/u, respectively.

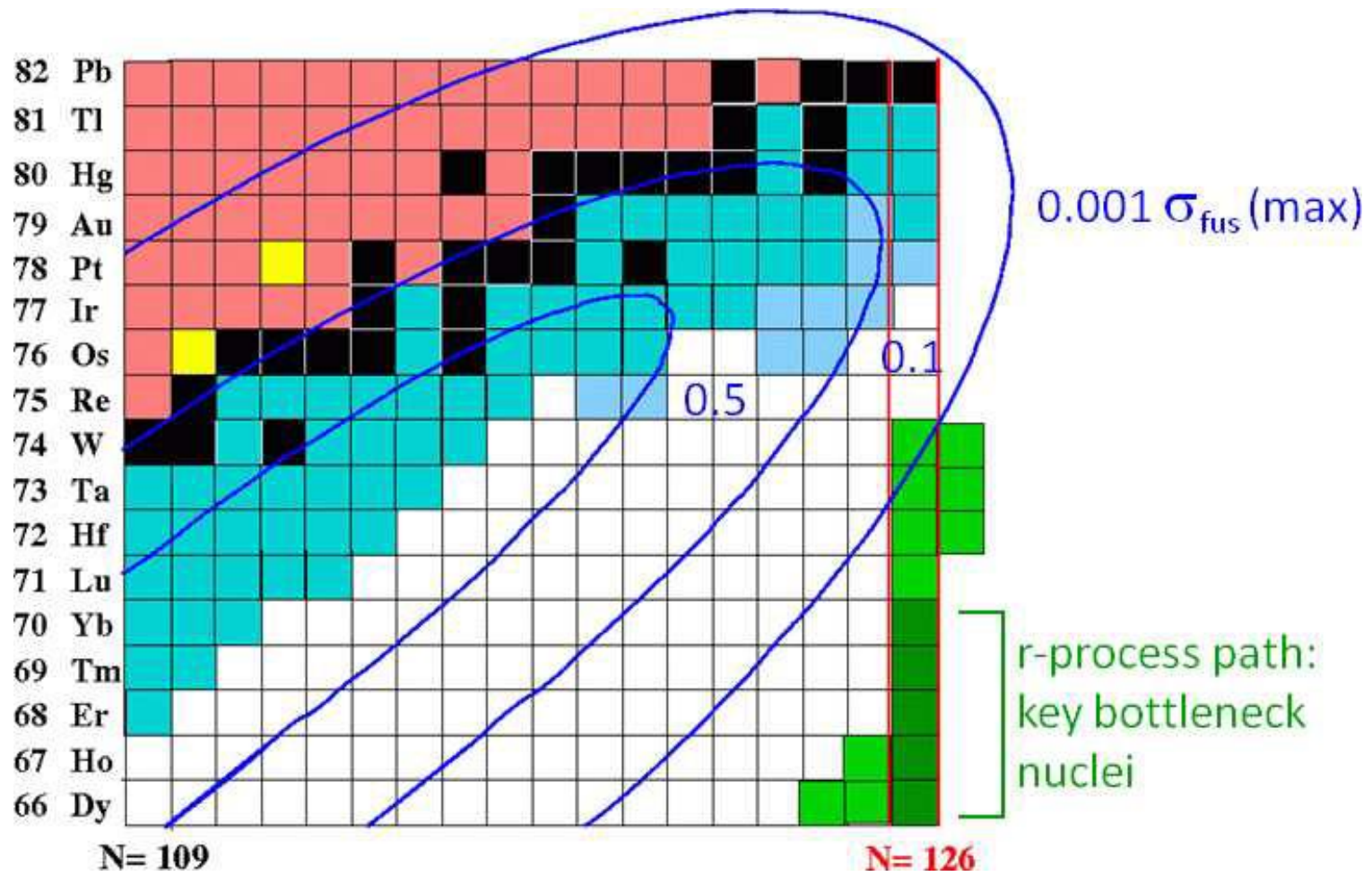

Figure 4 – Chart of nuclides around the N=126 waiting point of the r-process path. The blue ellipses denote the expected range of isotopes accessible via the novel fission-fusion process. The indicated lines represent 0.5, 0.1 and 0.001 of the maximum fusion cross section after neutron evaporation. In green the N=126 nuclides relevant for the r-process are marked, with the dark green color indicating the key bottleneck nuclei for the astrophysical r-process.

Besides the fusion of two light fission fragments, other reactions may happen. The fusion of a light fission fragment and a heavy fission fragment would lead back to the original Th nuclei, with large fission probabilities, thus we can neglect these fusion cross sections. The fusion of two heavy fission fragments would lead to nuclei with A~278, again nuclei with very high fission probability. Hence we have also neglected these rare fusion cross sections, although they may be of interest on their own. However, the multitude of reaction channels will require conclusive experimental precautions for a separation of the fusion reaction products of interest in the diagnostics and identification stage of the experimental setup.

*Requested beams*

For an estimate of the required laser intensities, focal spot area and target thickness, the 1-D RPA model as outlined in [20] is sufficient. It holds true for the relativistic 'hole-boring' regime of RPA. The following laser beam parameters have been assumed for the rate estimates:

- 2 laser beams, each with an energy of 150 J per pulse and a pulse length of 21 fs (corresponding to a power of ~7 PW). The focal spot on the thorium production target should have a diameter of 3 μm, leading to a focused intensity of approximately $10^{23}$ W/cm$^2$.

*Targets*

The target arrangement we want to use is depicted in Fig. 3 introduced before. It consists of two targets, termed production target and reaction target. The first is composed of a double layered, made from thorium and from deuterated polyethylene, $CD_2$. The two layers serve for the generation of a thorium ion beam and a beam containing carbon ions and deuterons. The reaction target has also a sandwich structure. The first layer is made from $CH_2$ and causes fission of the accelerated thorium nuclei. The second layer is a pure thorium film. The accelerated carbon ions and deuterons lead to fission of these thorium nuclei. Fusion of the fragments created in both layers generates neutron-rich nuclei in a mass range towards the waiting point N=126.

*Instrumentation and detectors*

Exploring this 'terra incognita' of yet unknown isotopes towards the r-process waiting point at N= 126 certainly calls for a staged experimental approach, starting with the development of laser ion acceleration of heavy ions (i.e. heavier than carbon as the presently heaviest species studied). Such preparatory studies will also be performed by the Munich group in Garching at the new CALA laser facility. Further studies should focus on the range and electronic stopping powers of dense laser-accelerated ion beams, followed by systematic optimizations of target properties in order to optimize the yield of fission fragments.

Also the yields for the fusion products should be measured in exploratory experiments, where it will be crucial to optimize the kinetic energy of the beam-like fission products. Subsequently the A, Z and N distributions of the light thorium fission fragments should be characterized, requiring detection setups for particle and decay studies. Fig. 5 shows a schematical view of the potential experimental setup of the presented reaction scenario. The high-intensity laser beam is tightly focused onto the target assembly in the target chamber (TC). Subsequent diagnostics and measurement devices can be added and operated according to successive project phases, where measurements of fusion products will be performed primarily in two stages. A first phase will aim at an identification of the produced isotopes via decay spectroscopy using a transport system (e.g. tape) directly behind the target chamber used to transport the reaction products to a remote, well-shielded detector system, where the characterization of the implanted fusion products could be performed either via β-γ-decay studies using, e.g., $LaBr_3$ scintillation detectors or spectroscopy with high-resolution germanium detectors (case a) in Figure 5 below), or after thermalization in a buffer gas stopping cell [40] and separation in a (multi-reflection) time-of-flight (MR-TOF) separator as developed by the Giessen group [41] (b). This device is particularly attractive when aiming at isotopic species with lifetimes shorter than 50 ms. Such a spectrometer could be operated either as an isobar separator or directly for mass measurements with a mass accuracy of up to $10^{-7}$. The probably most essential and also most demanding experimental task will be the separation of the reaction products. Fusion products with about 2-3 MeV/u will have to be separated from faster beam-like fission fragments with about 7 MeV/u, or target-like fragments with about 1 MeV/u, which could be achieved with a (2-stage) velocity filter. This separator has to accept a much broader momentum and charge-state range than typically requested from existing comparable devices operated at conventional accelerator facilities. This separator (e.g. 2-stage velocity filter) selects the ions of interest in order to prepare them (again after thermalization in the gas cell, followed by cooling and bunching in, e.g., a radiofrequency quadrupole ion guide, before then being transferred to perform either selective spectroscopic studies (c) or precision mass measurements either in the MR-TOF (d) or a Penning trap mass spectrometer (e), the latter potentially operated in an upgraded version with highly-charged ions to increase the performance (f). When using a Penning trap, such a setup would be similar to the SHIPTRAP facility at GSI [42] or ISOLTRAP at ISOLDE/CERN [43] for mass measurements with an accuracy of $\Delta m/m \approx 10^{-8}$, (corresponding to about 10 keV/$c^2$ [44]) for isotopes with half-lives longer than circa 100 ms, while the MR-TOF would grant complementary access also to shorter-lived species with half-lives longer than about 1 ms.

However, priority should be given to the design study of the separation stage forming the indispensable prerequisite for any unambiguous investigation of specific isotopes produced during the laser-driven reaction process. Here, as soon as possible, personnel resources should be allocated to start the design process. Here the Giessen/GSI-team brings in longstanding expertise in design, construction and operation of various types of separators. In case of length problems of the setup to be integrated into the available floor space, also a (90°) bent RFQ for extraction and bunching could be foreseen (g).

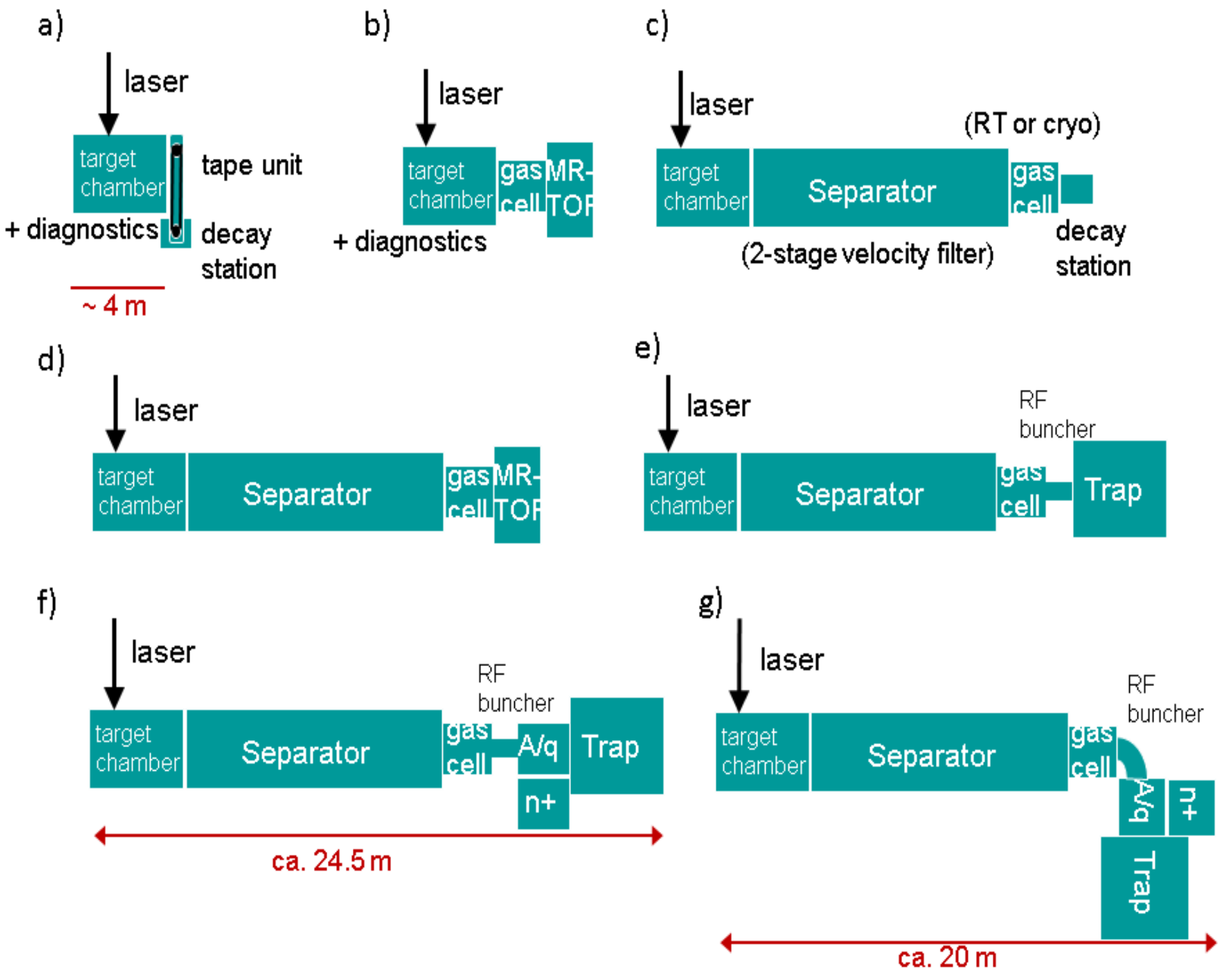

Figure 5 – Schematical view of the experimental arrangement for fission-fusion studies for different phases of the experimental development process. Measurements of fusion products will be performed primarily in two stages, first aiming at an identification of the produced isotopes via decay spectroscopy using a transport system (e.g. tape) directly behind the target chamber (a),or after thermalization in a buffer gas stopping cell and separation in a (multi-reflection) time-of-flight separator (MR-TOF) as developed by the Giessen group [41] (b) while later on a separator (e.g. 2-stage velocity filter) selects the ions of interest in order to prepare them (again after thermalization in the gas cell) either for selective spectroscopic studies (c) or for precision mass measurements either in the MR-TOF (d) or a Penning trap mass spectrometer (e), the latter potentially operated with highly-charged ions to increase the performance (f). In case of length problems of the setup to be integrated into the available floor space, also a (90 deg.) bent RFQ for extraction and bunching could be foreseen.

*Theoretical support*

Theoretical calculations and simulations will be needed at different stages of the present proposal:

- Theoretical guidance during the development of the laser ion acceleration of heavy species is a necessary and important ingredient for the success of the present proposal. Here we draw on the support by the LMU theory group of H. Ruhl.

- Detailed simulations of the acceleration process of heavy ions and the subsequent nuclear interactions in the fission and fusion stage of the proposed novel reaction scheme will be required to specify the properties of the produced reaction products during the fission-fusion process and to quantify the expected range of neutron-rich fusion products.

*Implementation scheme*

The proposed project exploits unique properties of laser-driven ion beams, not accessible elsewhere at conventional accelerator facilities and at present unrivalled at existing high-power laser facilities. Therefore, it should be pursued at ELI-NP with high priority from the start of the facility, in particular in view of the progressive stages required to reach its final goals. In its first phase the development and optimization of the laser acceleration process as well as of the corresponding targetry will addressed. This stage can go along with investigations of potential collective effects in the stopping behavior of laser-accelerator dense ion bunches. In order to reach these goals, initially a postdoc position and a PhD position will be needed, where the postdoc assumes responsibility of working on the laser-ion acceleration process, while the PhD candidate focuses on the collective stopping effects. In this context, LMU Munich has already granted a PhD position by the German federal funding agency to start the development of this topic at the local Garching high-power laser facility LEX/CALA. Thus, ELI-NP should contribute with postdoc position, primarily based at Magurele, but flexible to temporarily join also other experimental facilities to acquire practical expertise and perform exploratory investigations prior to the start of the ELI-NP experimental program. In view of the central importance of the recoil separator described before, design work on this central piece of equipment for the E1 area besides the interaction chamber (IC) should start as early as possible, since it will require extensive ion optical simulation studies, including the most recent results of laser-driven heavy ion acceleration. A postdoc position will be required for this task, potentially hired via a research contract between ELI-NP and a partner institution carrying long term expertise in designing and building of separators, e.g., GSI Darmstadt/Univ. Giessen. This postdoc should initially collect all required input data for the separator design from various high-power laser facilities, which exceed the parameters of ‘conventional’ recoil separators (e.g. via their large charge and momentum spread). This work should go along with the local ion acceleration studies thus that, once the prerequisites of the target interactions are under control, a profound layout of the separator, including its construction timeline and costing, is available, allowing the ELI-NP management to decide upon its construction.

In hot plasma various mechanisms of nuclear excitation and de-excitation may appear. Beside direct interaction with free electrons and X-rays/γ-rays through mechanisms such as:

- photoexcitation,
- electron inelastic scattering,
- stimulated gamma ray emission,

other excitation/de-excitation mechanisms involving the bound states of electron cloud:

- Internal Conversion (IC): nuclear de-excitation resulting in the emission of an orbital electron to the continuum,
- Bound Internal Conversion (BIC): same as IC, but the electron is promoted to a bound state,
- NEEC (Nuclear Excitation by Electron Capture from continuum): inverse of IC,
- NEET (Nuclear Excitation by an Electronic Transition): inverse of BIC

will occur with different rates compared to isolated atoms or materials in normal conditions. Significant changes in nuclear life-times are predicted in hot and dense plasmas [45], as in the case of the 6.85 h isomeric state of $^{93}Mo$ shown in Figure 6. Here, the 4.85 keV photoexcitation from the $21/2^+$ to the $17/2^+$ state, followed by the decay of the last one through a much faster transition towards the $13/2^+$ state, corresponds to an effective lifetime decrease of the $21/2^+$ isomeric state in plasma conditions. Such a mechanism of induced energy release (2.5 MeV in case $^{93}Mo$) as a result of an excitation of much lower energy (500 time less in the case of $^{93}Mo$) has potential applications for energy storage with much higher energy density compared to electrochemical processes in batteries.

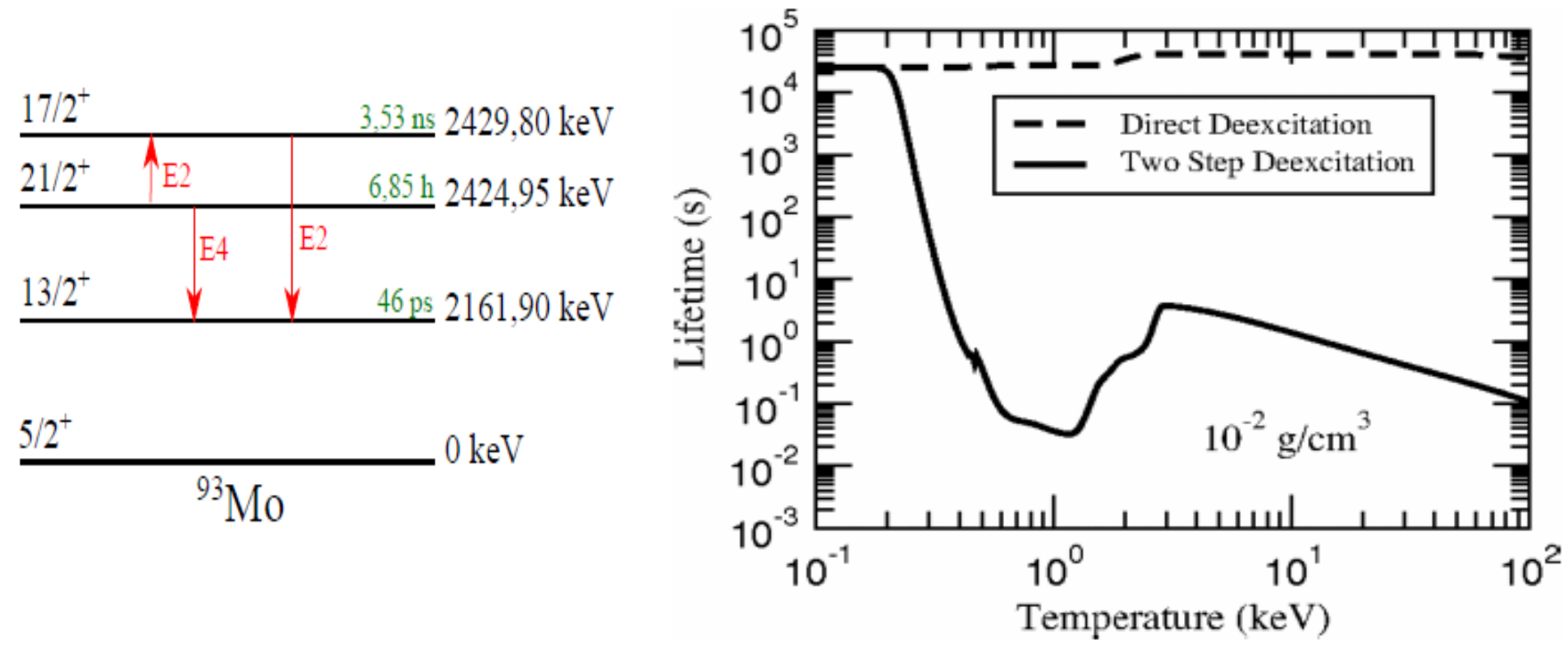

Figure 6 – The partial level scheme of $^{93}Mo$ and the lifetime of the $21/2^+$ isomeric state as a function of the plasma temperature.

The possible changes of lifetimes of nuclei having beta-decaying isomeric states at low energy above the ground state, as $^{176}$Lu, $^{26}$Al and $^{34}$Cl, are also highly relevant for astrophysical nucleosynthesis processes.

The NEET and NEEC mechanisms are schematically presented in Figure 7. The NEET has been observed [46-48] in normal (cold) target conditions in several heavy nuclei: $^{197}$Au, $^{189}$Os and $^{193}$Ir with probabilities $P_{NEET} \sim 10^{-8}$ or lower. These probabilities are in good agreement with theoretical predictions. Higher probabilities for the NEET process are reported [49] in $^{237}$Np, however, the experimental value of $(2.1\pm0.6)\times10^{-4}$ is much larger than the theoretical predictions [50]. We note also that the BIC process, i.e. the reverse of NEET, has been observed in $^{125}$Te [51]. The NEEC process was never observed.

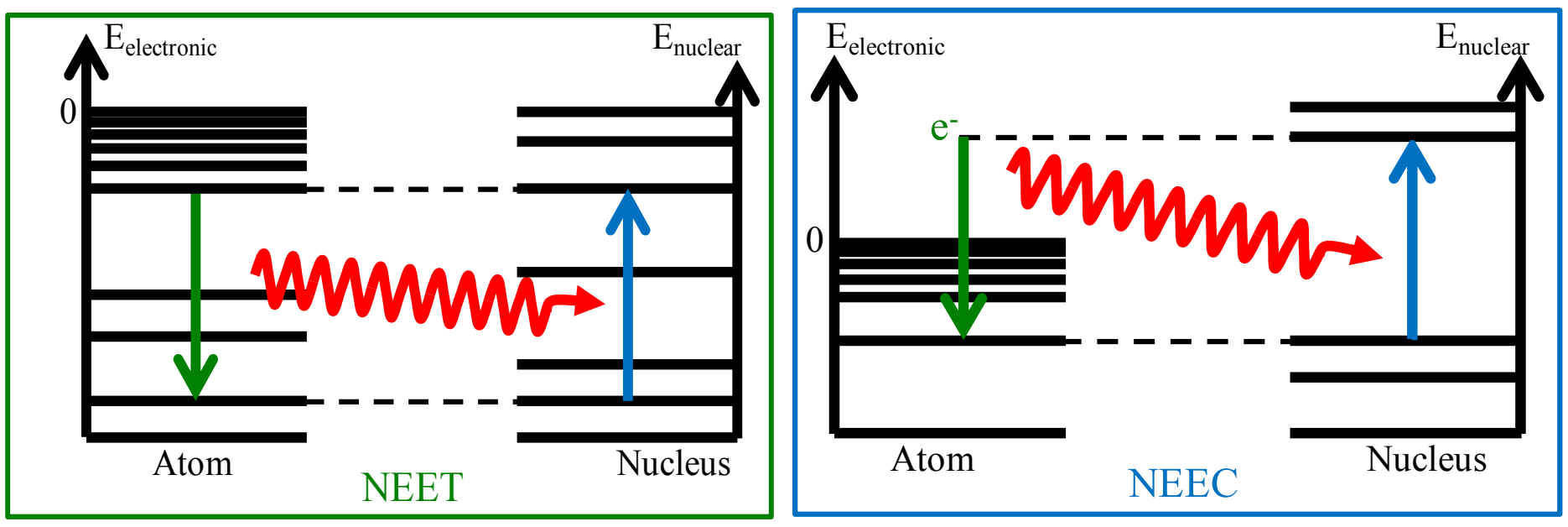

Figure 7 – Schematic representation of the NEET and NEEC processes of nuclear excitations.

The possibilities to create hot plasma offered by lasers opened new opportunities to study these phenomena. The presence of ions in different charge states, each of them with various electron configurations, enhances the chances for existence of an electronic transition, corresponding to an X-ray line, of "equal" energy and multipolarity with the nuclear transition. However, none of NEET, NEEC or BIC processes have been observed in plasmas. Moreover, in each ion the ionization state and electron configurations will change at high rate in a complex interplay of collision/excitation/deexcitation mechanisms, depending on the rapidly evolving properties of the plasma. This makes a theoretical description very difficult. Understanding and optimizing plasma formation using various targets and irradiation conditions are the keys for observation of nuclear excitations, while the possibilities for extending the plasma lifetime by trapping it using very high pulsed magnetic fields, as suggested in section 2.4, has also to be explored.

Several attempts have been done to evidence nuclear excitation in laser plasma in:

- $^{235}$U, having an isomeric state of $T_{1/2}$=26.8 min at only 76.8 eV above the ground state and

- $^{181}Ta$, which has an isomer of $T_{1/2}$=6.05 μs at an excitation energy of 6.237 keV.

The positive results for NEET reported by Andreev et al. [52] in $^{181}Ta$ have not been confirmed by more recent experiments [53] of the ENL (Excitations Nucléaires par Laser) group from CENBG, who has under study two other candidates for NEET: $^{201}Hg$ and $^{84}Rb$.

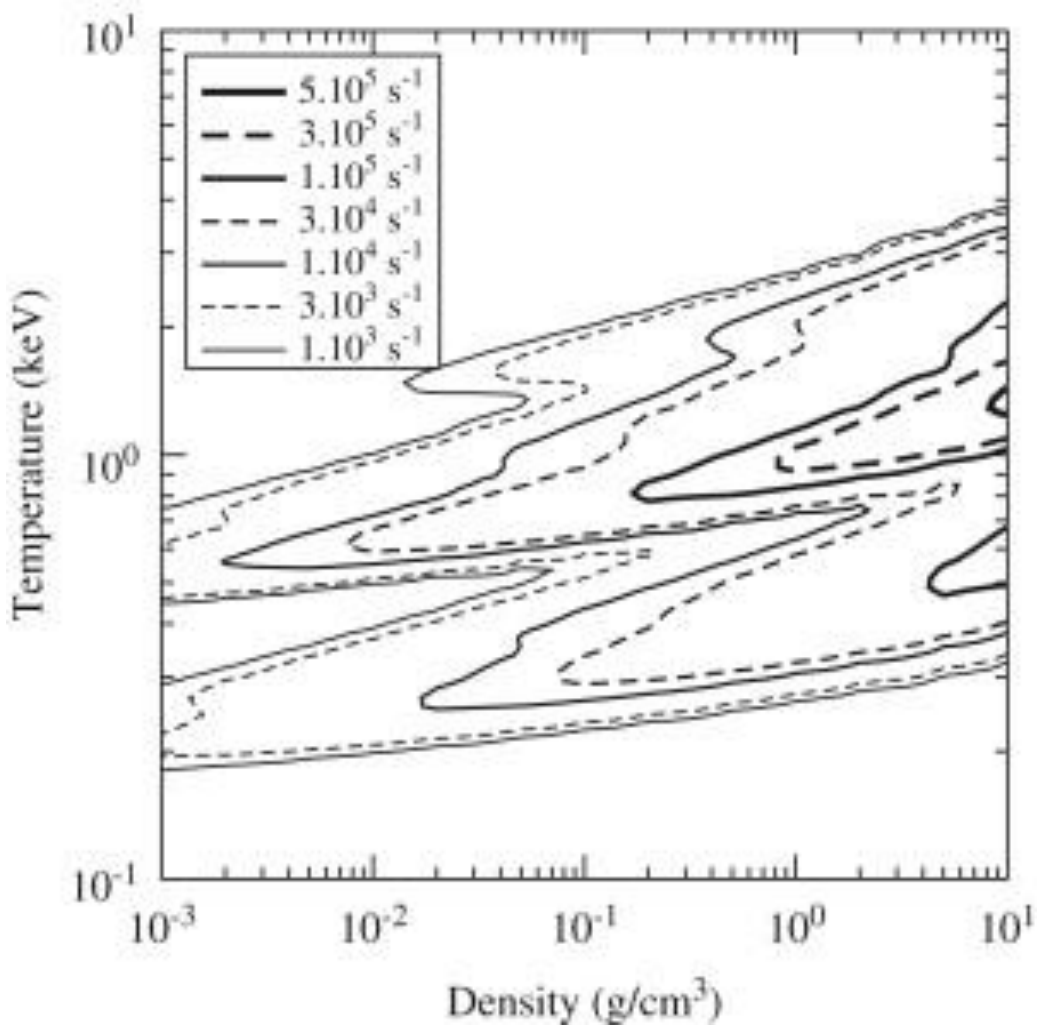

Figure 8 – The calculated NEET rate in $^{201}Hg$ is plotted as function of plasma density and temperature [54].

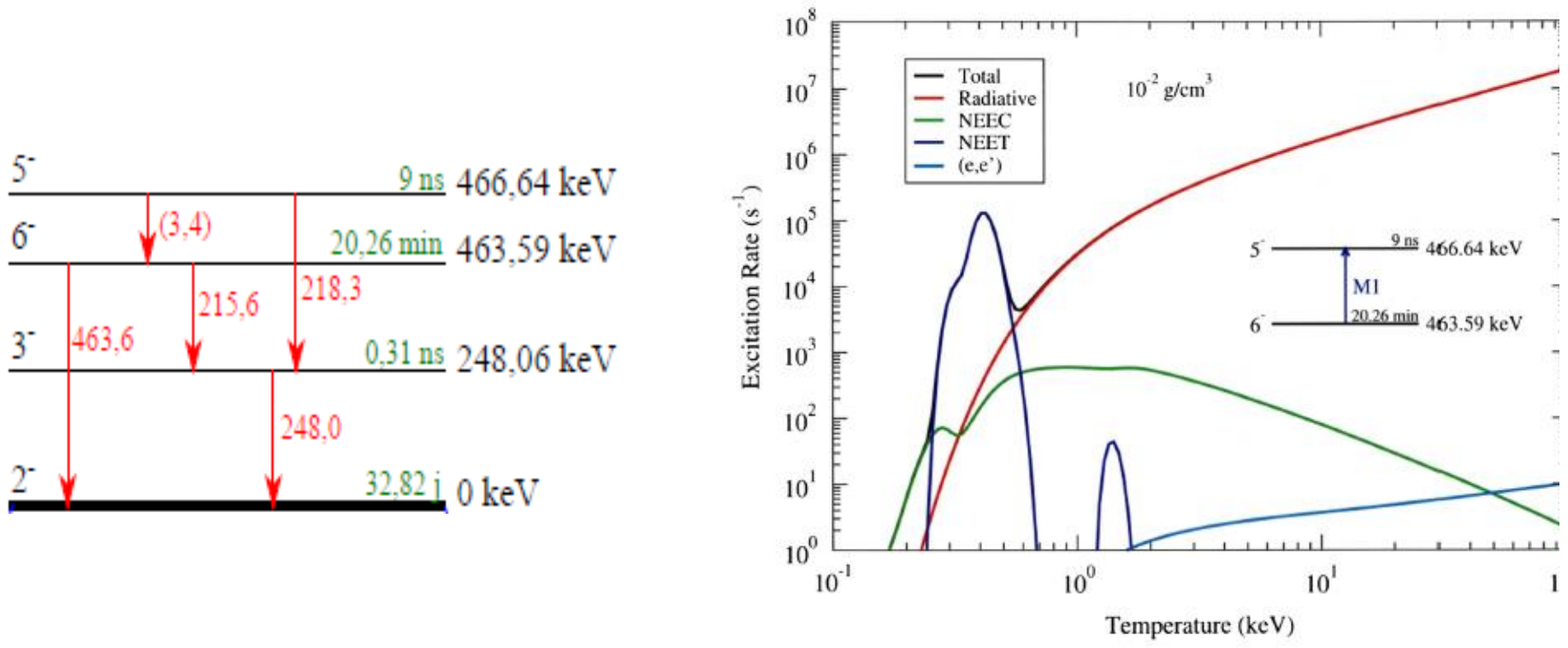

Figure 9 – Partial level scheme of $^{84}Rb$ and the rate for excitation of the $5^-$ state through different processes [55].

In Figure 8, the calculated NEET rate [53] in $^{201}$Hg (excited state at 1.56 keV with $T_{1/2}$=81 ns) is plotted as a function of plasma density and temperature. Values larger than $10^4$ $s^{-1}$ are obtained for temperatures of 500 – 700 eV and densities down to $2\times10^{-3}$ g/cm$^2$ achievable with uncompressed (0.9 ns) ELI-NP laser pulses. Needed intensities are of order of $10^{15}$ W/cm$^2$, corresponding to a focal spot of ~100 μm.

In Figure 9, the partial level scheme of $^{84}$Rb (left panel) and the excitation rates of the $5^-$ state at 3.498 keV above the isomeric state of $T_{1/2}$=20.26 min at 463.6 keV, are shown [53, 55]. According to the ISOMEX code based on the relativistic average atom model, the NEET process is expected to dominate for temperatures around 400 eV with a rate around $3.5\times10^3$ $s^{-1}$.

To study this case, before plasma creation with the laser, the desired isomeric state has to be created through nuclear reactions. The 1 PW-class PHELIX laser installed at GSI near the UNILAC heavy ion accelerator is well suited for such experiments. Otherwise, at ELI-NP, the presence of two high power high repetition laser pulses allows to use one for particle acceleration and isomeric state production while the other one, uncompressed, can be used for plasma formation and heating. The most appropriate production mechanism has to be defined for each isomer under study: besides the proton or heavy ion acceleration, the electron-to-gamma conversion in high Z target should be considered as well. In any case, the needed energies of accelerated particles are not very high, the optimization should to be targeted towards highest possible particle fluxes and low divergence, such as to keep the high density of isomers in a spot of ~250 μm on the secondary target. For the study of $^{84m}$Rb, taking into account its lifetime, a sequence of 60 minutes of 100TW laser @10 Hz is needed to reach the maximum production. The isomer will be populated by the $^{76}$Ge($^{12}$C, p+3n)$^{84m}$Rb reaction. Afterwards, a high energy long duration laser shot will produce the hot and dense plasma in which the nuclear excitations take place.

The isomeric production yields and the problem of measuring γ-rays from short lived isomers produced by high power lasers has been addressed [56] in a recent experiment at the ELFIE-100 TW facility at LULI on the $^{90}$Nb nucleus, produced in the $^{90}$Zr + p reaction. The level scheme of $^{90}$Nb (see Figure 10) presents three isomeric states with half-lives of 18.8 s, 6 ms and 63 μs, respectively. The 2.3 keV transition is a candidate for both NEET and BIC processes, if the plasma is generated (by an ELI-NP uncompressed pulse) shortly after the high power pulse, which is expected to produce both isomeric states involved in the transition (at 122.37 keV and 124.67 keV) in similar quantity. For long enough delays between the laser pulses, only the state at 124.67 keV will survive and the BIC process can be studied alone.

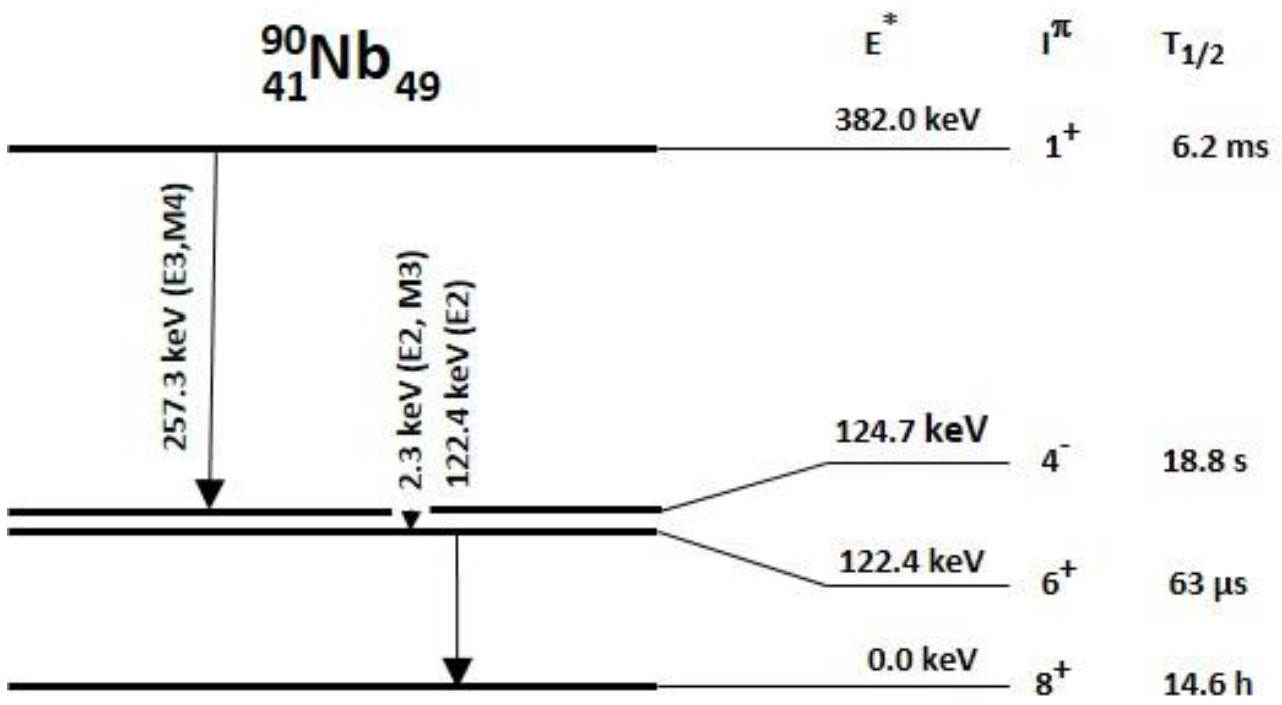

Figure 10 – Partial level scheme of $^{90}$Nb with spin assignments and half-lives.

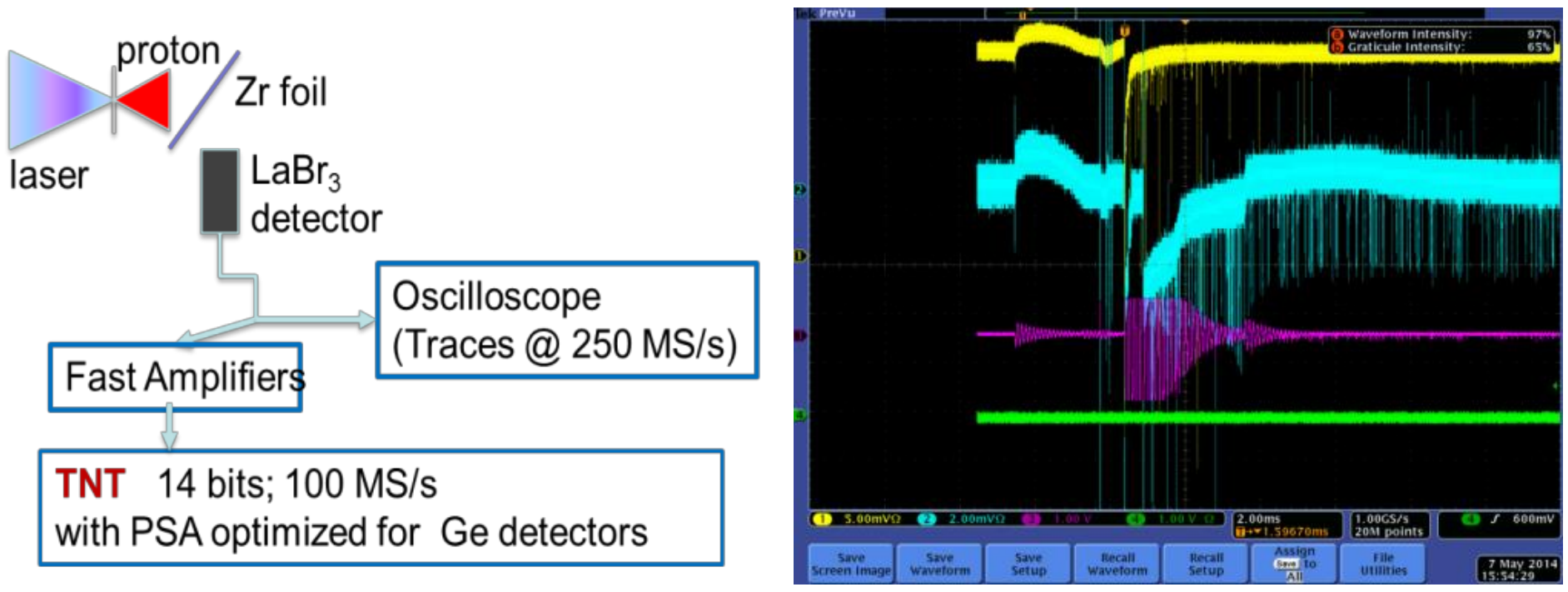

Figure 11 – Left: Diagram of the experimental setup. Right: signal traces measured with an oscilloscope. The blue trace corresponds to the $LaBr_3$ detector placed inside the interaction chamber, the yellow trace corresponds to a similar detector (not gated), placed as a reference outside the interaction chamber.

For the detection of γ-rays, a $LaBr_3$ scintillator coupled to a gated-photomultiplier tube (PMT) has been installed at ~10 cm from the Zr target. The properties of $LaBr_3$ scintillators (165% photon yield compared to NaI and 16 ns decay time) are well suited for on-line measurements of isomers or unstable nuclei with very short life-times. However, the strong X-ray flash is generating a huge amount of scintillation. The recovery from saturation effects takes several milliseconds as shown in Figure 11. Nevertheless, the gamma signals (represented by thin lines due to the 1 ms/division scale of the oscilloscope) are visible even below 1 ms after the laser pulse, with reduced amplitude. The signal from the detector was split and sent also to a digitizer with on board pulse processing that was able to provide the energy and arrival time of each detected gamma ray. This scheme was used because very long acquisition times are needed to measure the yield of the 18.8 s isomer. The off-line processing of traces from the oscilloscope has shown that the digitizer gives good results also during the first millisecond. The obtained bi-dimensional spectra are shown in Figure 12. One can observe well separated signatures from the isomers of interest, together with (i) the 511 keV line corresponding to $\beta^+$ decay of $^{27}Si$ populated by the proton reaction in the Al holder of the Zr target and (ii) the 206 keV isomer in $^{79}Br$ due to photonuclear reactions in the scintillator itself. We remark also the very low background in the range of tens of millisecond after the main pulse.

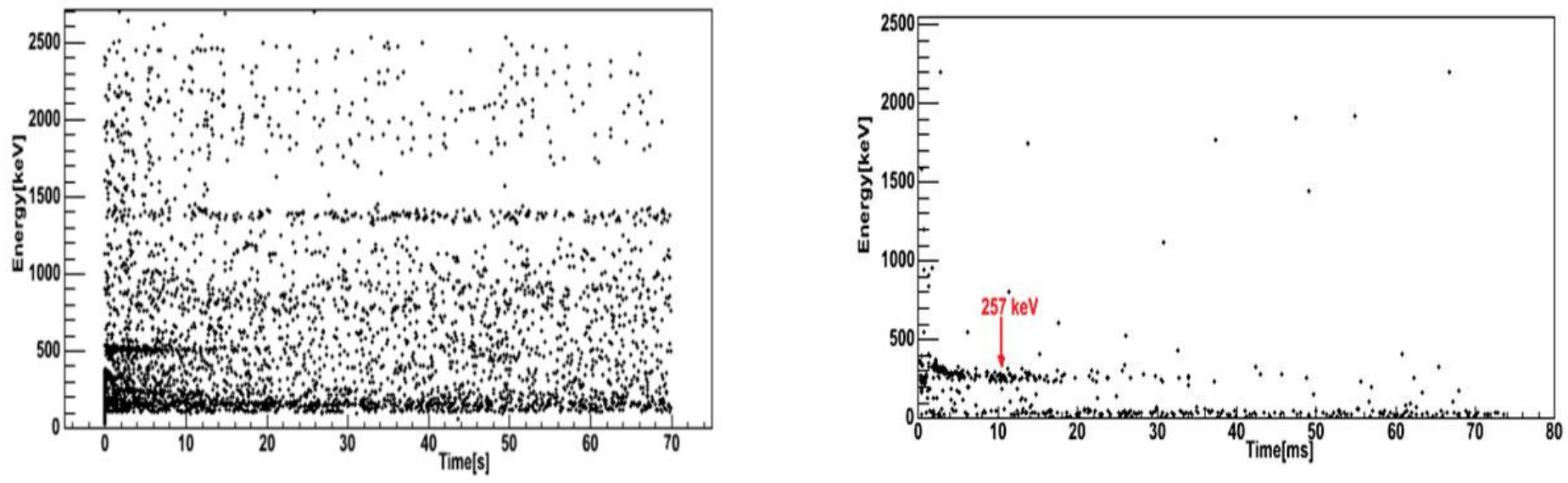

Figure 12 – In-situ single shot bi-dimensional energy-time gamma spectra obtained with a Zr secondary (isomeric) target and a $LaBr_3$ detector.

Based on these results, with several improvements in progress (increasing the lead shield around detector and optimizing its shape, using a calibrated LED signal to correct the pulse height during the recovery period etc.) it seems possible to measure lifetimes of the order of 100 µs or even below in high power laser experiments. The technique can be used as on-line diagnostic method, well adapted for high-repetition lasers, complementing the activation technique that requires transport of irradiated the sample in front of gamma detectors placed outside the experimental hall.

The $10^{15}$ W/cm$^2$ laser intensity used for plasma creation does not generate very energetic electrons or hard X-rays, such that much shorter lifetimes will be possible to observe. With adequate shielding and filtering, the 218.3 keV transition of only 9 ns in case of $^{84m}Rb$ might be measured.

The results on yields are also encouraging: the number of isomers produced per shot was around $10^6$ in the 50 µm Zr target for the 18.8 s and 6 ms isomers. The ELI-NP laser parameters promise to increase this number to ~$10^8$ isomers per shot.

### $^{26}$Al case

Properties of unstable nuclei, which play a key role in explosive stellar environments, have been the paramount interest of astrophysical nuclear research since its emergence more than 50 years ago [57]. With the projected ELI intensities, a new world of possibilities opens up to study their behavior for the first time under the extreme temperature and pressure conditions present in the inner cores of planets and stars. The quest to study nuclear astrophysics with ELI should focus on the most prominent puzzling systems. Hence the SUPA collaboration proposes to study the possible enhancement of the decay of the long-lived $^{26}$Al radioisotope in astrophysical environments with ELI. This endeavor would be a complementary effort to already established successful experimental research projects of current SUPA physicists (S.D. Pain) at the Holifield Radioactive Ion Beam Facility at Oak Ridge.

The $\gamma$-ray mapping of the $^{26}$Al decay across the galaxy provides one of the most interesting constraints on nuclear physics parameters in astrophysical environments. The $^{26}$Al nucleus was the first radioisotope detected in the interstellar medium, by the observation of the characteristic 1809 keV $\gamma$-emission associated with the decay of its ground state [58]. As the half-life $^{26gs}$Al ($5^+$) state is $7.2\times10^5$ years, the presence of this nucleus provides evidence of ongoing galactic nucleosynthesis. Wolf-Rayet stars and Asymptotic Giant Branch (AGB) stars and novae [59] have been suggested as possible sources of the origin of $^{26}$Al. At a temperature of T=0.03GK, the $^{26gs}$Al(p,$\gamma$)$^{27}$Si reaction is expected to be the main destruction mechanism for the $^{26}$Al isotope. However, at these hot stellar temperatures, the dominant contribution to the $^{26gs}$Al(p,$\gamma$)$^{27}$Si reaction rate is capture through low-lying resonances, for which the strengths have not been measured and an experimental benchmarking of theoretical studies, such as Hauser-Feshbach based calculations [60], remains elusive. The disintegration process of $^{26}$Al is further intricated by the presence of a $0^+$ isomer at 228 keV above the ground state. This isomer, which originates like the ground state from the coupling of the two unpaired nucleons in the odd-odd $^{26}$Al system, is prohibited to decay into $^{26gs}$Al due to the large spin difference $^{26m}$Al decays via $\beta$+ emission with $T_{1/2}$ = 6.35 s directly to $^{26gs}$Mg ($0^+$). This is a very specific and complicated scenario.

Equilibration between $^{26gs}$Al and $^{26m}$Al can only proceed via the coupling through a sequence of intermediate states (IS), for which no branching ratios are experimentally established. Theoretical work [61] based on shell-model calculations predicts a dramatic reduction of the effective life time $\tau_{eff}$ ($^{26gs}$Al) by a factor of $10^9$ within the temperature range from 0.15 to 0.4 GK, superseding previous estimates by Ward and Fowler [62] by orders of magnitude. This significant decrement of $\tau_{eff}$ is due to a variety of physical processes triggered and influenced by hot plasma environments, which will gradually become accessible with the emerging ELI project. At high densities, the increasing Fermi energy of the electron opens up electron capture channels otherwise energetically forbidden. Moreover, hot bremsstrahlung radiation will lead to an enhancement of the coupling of ground and isomeric states via the manifold of known as well as hitherto unresolved IS at several MeV, where the nuclear level density is high. The population of these states, and thus their contribution to the true astrophysical disintegration rate, will reflect an overlap of Boltzmann distributions from ground *and* excited state in the hot and dense environmental conditions provided [63]. The ELI laser system will deliver energetic particle and radiation bursts of sufficient intensity to create planet and stellar-like environmental conditions. Most importantly, these radiation pulses are ultrashort in time and synchronous, thus providing ideal conditions for an 'astrophysical laboratory' capable of resolving ps time scales. In a first instance, we want to expose a miniature $^{26}$Al target specimen to an isochorically heated environment with ELI. Work by Patel *et al.* shows that isochorical heating by laser induced thermally distributed proton beams with end-energies of only a few MeV can be used to create very localized ($\varnothing$=50μm) high energy-density plasma states [64]. In this study a 'modest' 10 J, 100 fs high intensity laser system was able to produce several tens of eV within the ps time domain. The ELI system, even in the first phase, will be able to surpass these values by several orders of magnitude, especially once the onset of the pressure dominant acceleration regime is established as predicted by Esirkepov [65]. For increasing laser intensity, the electromagnetic field will eventually start to directly interact with the nucleus, thus presumably contributing further to an enhancement of the decay probability. In all instances, the spatial confinement of particles and radiation emerging from laser acceleration will help this particular investigation tremendously. The isotope $^{26}$Al is only available in minute quantities, which will just allow the production of miniature pellet targets or thin layers on backing or radiator materials. The onset of an enhanced transition rate and the coupling of ground and isomeric state via IS can be deciphered via the 511 keV annihilation radiation following the β+ decay of $^{26m}$Al. The coincident 511 keV photons are measurable with semiconductor or scintillation detector systems and would exhibit a characteristic temporal behavior with $T_{1/2}$ = 6.35 s. Ideally, a fast target transportation would need to be developed to retrieve the target probe from the interaction zone after irradiation.

We are aware of the many conceptual and technical aspects that need to be addressed prior to such an experimental engagement with ELI. Most importantly, once ELI parameters are firmly established, precise yield estimates have to be undertaken. Furthermore, we have to consider the reaction yield for the $^{26}$Al(γ,n)$^{25}$Al channel with $S_n$($^{26}$Al)=11.4 MeV, which also causes the emergence of 511 keV annihilation radiation with $T_{1/2}$=7.18 s, as $^{25}$Al is a $\beta^+$ emitter. This suggests, e.g., the use of neutron detectors for discrimination. Moreover, as the decay of $^{25}$Al also produces a coincident 1612 keV γ-ray with low branching, intensity measurements with a high resolution germanium detector will allow to estimate the background contribution from this intruding reaction channel. To achieve isochorical heating, a series of conceptual studies have to be performed to derive an ideal setup for the miniature aluminum targets, which will include fabrication, alignment and the encapsulation of the tiny probes. Additionally, as particle reaction channel yields have to be estimated, Hauser-Feshbach calculations have to be performed for increasing temperatures [66]. Besides theoretical codes, GEANT4, SRIM [67] and LASNEX [68] simulations need to be undertaken. Furthermore, there may be a need to development of a special target chamber due to the radioactivity of the target probe. We also propose to implement prima facie experiments on bulk targets of stable isotopes that have low-lying isomeric states with similar life-times as proof of concept studies (e.g. $^{107,109}$Ag). Results will be first and foremost interpreted in light of the theoretical evaluations shown in [61]. The study of $^{26}$Al could become a benchmark experiment, as it would manifest ELI as a novel accelerator system, providing environments of astrophysical interest. It will align and allow a further development of existing projects with radioactive beam facilities that will deliver a lot of interesting results for nuclei of pronounced astrophysical interest in the next years.

### 2.3 NUCLEAR REACTIONS IN LASER PLASMAS

Plasma is by far the most common form of matter known. Plasma in the stars and in the tenuous space between them makes up over 99% of the visible universe and perhaps most of that which is not visible. On earth we live upon an island of "ordinary" matter. Given its nature, the plasma state is characterized by a complexity that vastly exceeds that exhibited in the solid, liquid, and gaseous states. Correspondingly, the physical properties of nuclear matter (structure, life times, reaction mechanisms etc.) could be drastically changed inside the plasma. These studies represent one of the most far ranging, difficult and challenging research areas today, implications could cover others fields, from quantum physics to cosmology, astrophysics etc.

In this context, one of the most crucial aspects concerns the role of electron screening.

Direct and indirect measurements of the relevant cross sections have been performed over the years. Direct measurements using accelerated beams show that, at very low energies, the electrons in the target's atoms partially screen the Coulomb barrier between the projectile and the target [69], resulting in an enhancement of the measured cross section compared with the bare nucleus cross section [70]. The electron screening effect is significantly affected by the target conditions and composition [71], it is of particular importance for the measurement of cross-sections at extremely low energetic domains including plasma effects, i.e. in an environment that under some circumstances and assumptions can be considered as "stellar-like" (for example, for the study of the role played by free/bounded electrons on the Coulombian screening can be done in dense and warm plasmas).

Electron screening prevents a direct measurement of the bare nucleus cross section at the energies of astrophysical interest. In the last decade, the bare cross section has been successfully measured in certain cases by using several indirect methods [72].

Usually, astrophysically relevant reactions are performed in the laboratories with both target and projectile in their ground state. However, in high temperatures plasmas ($10^8$ K), an important role can be also played by the excited states, as already deeply discussed in the pioneering theoretical work of Bahcall and Fowler [73]. In that case, the authors studied the influence of low lying excited $^{19}F$ states on the final $^{19}F$(p,alpha) reaction, predicting an increase of a factor of about 3 in reaction rate at temperatures of about 1-5 GK.

Thus determining the appropriate experimental conditions that allow to evaluate the role of the excited states in the stellar environment could strongly contribute to the development of nuclear astrophysics. The study of direct measurements of reaction rates in plasma offers this chance. In addition, other new topics can be conveniently explored, such as three body fusion reactions as those predicted by Hoyle [74], lifetime changes of unstable elements [75] or nuclear and atomic levels [76] in different plasma environments; other fundamental physics aspects like non-extensive statistical thermodynamics [77] can be investigated in order to validate/confute the general assumption of local thermal equilibrium that is traditionally done for plasmas.

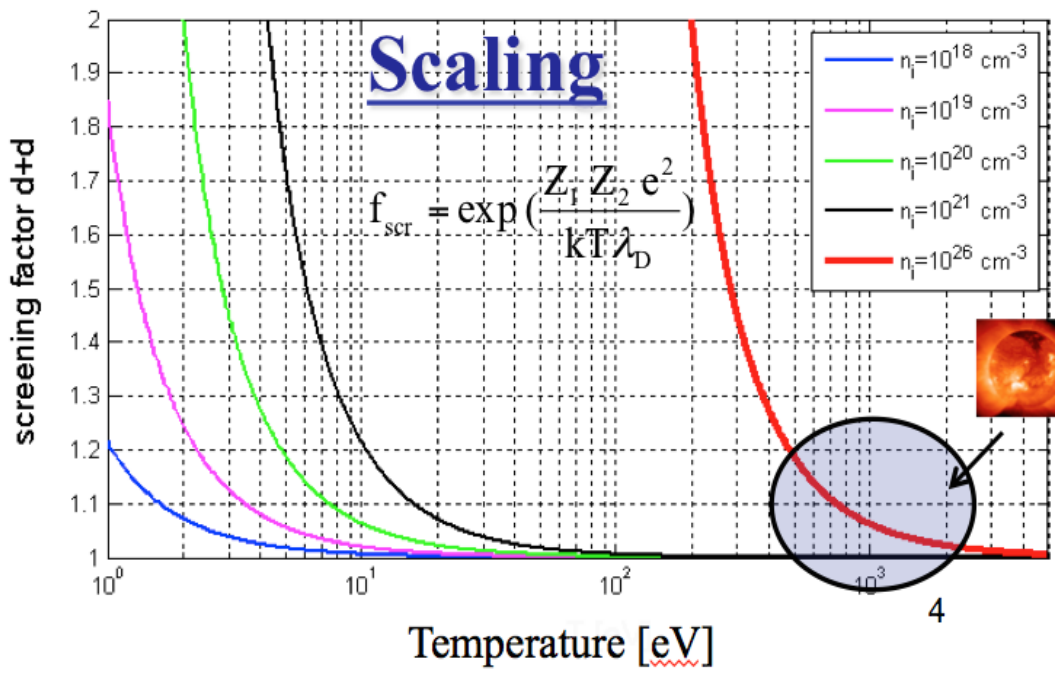

Figure 13 – Calculated screening factor for the D+D reaction as a function of electron temperature and density. Highlighted are the typical solar values.

Although it seems practically impossible to reproduce in the laboratory the extreme properties of stellar matter, according to a method commonly used in other fields of plasma physics, it is possible to rescale the plasma parameters (temperature and density) in order to make the laboratory conditions similar to the ones of an astrophysical plasma. As an example, Figure 13 shows the calculation of the screening factor for the *D+D* reaction as a function of electron temperature and density. It can be noticed that the typical values of solar screening can be reproduced in alternative plasma conditions of temperature and density.

The future availability of high-intensity laser facilities capable of delivering tens of petawatts of power (e.g. ELI-NP) into small volumes of matter at high repetition rates will give the unique opportunity to investigate nuclear reactions and fundamental interactions under extreme plasma conditions [78], including also the influence of huge magnetic and electric fields, shock waves, intense fluxes of X and $\gamma$-rays originating during plasma formation and expansion stages.

### 2.3.1 First cases of study

To investigate these research topics, we are proposing the construction of a general purpose experimental setup, where it will be possible to study the electronic screening problem in a wide variety of cases and configurations with different purposes. In particular, we propose to study the screening effects on low energy fusion reactions and on weakly bound nuclear states (Hoyle, Efimov [79] etc.). Concerning the first question, among the various nuclear reactions which have attracted relevant attention also for astrophysical or cosmological reasons, we would select the $^{13}C(^{4}He,n)^{16}O$ and $^{7}Li(d,n)^{4}He$-$^{4}He$ reactions: the former for its relevance in the frame of stellar nucleosynthesis, the latter for the role played in Big Bang primordial nucleosynthesis. Through the laser-target interaction, we aim at producing plasmas containing mixtures of $^{13}C$ + $^{4}He$ and $^{7}Li$ + deuterons in order to investigate inner-plasma thermo-nuclear reactions.

The $^{13}C$+$^{4}He$ reaction is of key interest for the investigation of the helium burning process in advanced stellar phases [80]. In particular, it can be activated at the base of AGB stars, thus constituting one of the most interesting neutron sources in stellar conditions. These are in turn important for the so-called “slow-process”, i.e. the neutron induced reactions responsible of the heavy elements production. Thus, by gaining further knowledge about the $^{13}C$+alpha reaction, it will be possible to evaluate more carefully the available neutron flux for the following s-process nucleosynthesis. For the astrophysical factor S(E) of $^{13}C(alpha, n)^{16}O$ reaction no experimental data are available in the region below 270 keV, but only model predictions [81].

The $^{7}Li(d,n)^{4}He$-$^{4}He$ reaction was recently addressed by Coc et al. [82] as one of the most important reactions affecting the CNO abundances produced during the primordial nucleosynthesis (BBN). From such an analysis, it was found that the $^{7}Li$ nucleosynthesis is strongly influenced by the $^{7}Li(d,n)$ $^{4}He$-$^{4}He$ reaction rate. Data collected by these authors give a variation of two orders of magnitude on the $^{7}Li$ abundance during the BBN epoch, around 1 GK of temperature, with respect to the reaction rate measured by Boyd et al. [83]; the latter is usually adopted for the BBN evaluation. These discrepancies can be explained if one considers that very few experimental data exist, and authors consequently assume a constant S-factor ranging between two extreme hypotheses from 5 to 150 MeV·b. Providing new experimental data focused on the determination of the outgoing neutron flux is essential in order to up-grade our knowledge of this process and consequently of the BBN at a temperature of about 1 GK. This critical temperature domain will be affordable by the petawatts laser facility of ELI-NP [84], including the configuration based on two laser beams producing colliding plasmas [85, 86].

In relation to the weakly bound nuclear states, as a first case of study we propose to investigate the $^{11}B(^{3}He, d)^{12}C^{*}$ reaction in a plasma. Nucleonic matter displays a quantum-liquid structure, but in some cases finite nuclei behave like molecules composed of clusters of protons and neutrons. Clustering is a recurrent feature in light nuclei, from beryllium to nickel. Cluster structures are typically observed as excited states close to the corresponding decay threshold; the origin of this phenomenon lies in the effective nuclear interaction, but the detailed mechanism of clustering in nuclei has not yet been fully understood. The second $J = 0^{+}$ state at 7.654 MeV in $^{12}C$, first predicted by Hoyle [74] in 1953 and thus called the Hoyle state, plays a central role in nuclear physics. It is a well-known fundamental testing ground of models of the clustering phenomena in light nuclei, which is highlighted by recent developments of *ab initio* theoretical calculations that are able to calculate light nuclei such as $^{12}C$. The Hoyle state plays a central role in stellar helium burning by enhancing the production of $^{12}C$ in the Universe, allowing for life as we know it. It is the first and quite possibly still the best example of an application of the anthropic principle in physics. Early on after the discovery of the Hoyle state, it was suggested by Morinaga [87] that we can learn more about the structure of the Hoyle state by studying the rotational band built on top of it, which led to a 50-yr long search for the second $2^{+}$ state in $^{12}C$. Recently, the existence of the second $2^{+}$ state in $^{12}C$ has been the subject of much debate.

The current evaluation of the triple-$\alpha$ reaction rate assumes that the $\alpha$ decay of the 7.65 MeV $0^{+}$ state in $^{12}C$, proceeds sequentially via the ground state of $^{8}Be$. This assumption has been sustained also by a new upper limit of $5x10^{-3}$ on the direct $\alpha$ decay of the Hoyle state at 95% of C.L. [88] extracted from the study of the $^{11}B(^{3}He, d)$ reaction. This assumption is challenged by the recent identification of two direct $\alpha$-decay branches with a combined branching ratio of 17.5% [89]. If correct, this would imply a corresponding reduction in the triple-$\alpha$ reaction rate with important astrophysical consequences. This data has been extracted by the fragmentation of quasi-projectiles from the nuclear reaction $^{40}Ca$ + $^{12}C$ at 25 MeV/nucleon, used to produce excited states candidates to $\alpha$-particle condensation. This approach differs from the previous one for the presence of nuclear medium. In the $^{11}B(^{3}He, d)$, carbon Hoyle state is populated and decay in vacuum while the fragmentation approach is populated and decay in presence of nuclear matter. For the important astrophysical consequence is mandatory to study these topics in a plasma environment.

## 2.3.2 Methodology

To perform the proposed experiments, providing relevant data concerning the aforementioned reactions and others, we aim to take advantage from the excellent and unique performance of the ELI-NP facility and realize an experimental setup where two laser beams generate two colliding plasmas. The reaction products (neutrons and charged particles) will be detected through a new generation of plastic scintillators wall and through a new silicon carbides wall. The sketch of this configuration is drawn in Figure 14.

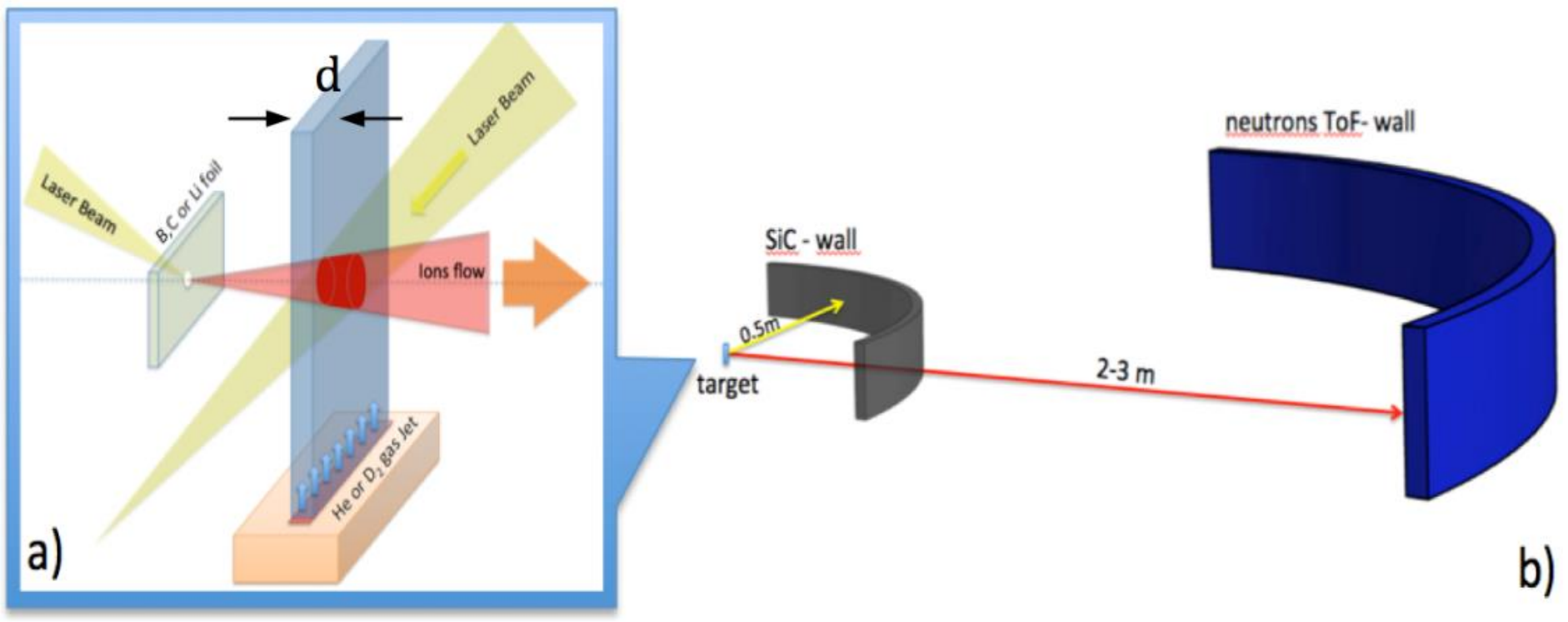

Figure 14 – Layout of the experimental setup. a) Target configuration, the main laser pulse impinging on B, C or Li thin foil generates a primary plasma which impacts on a second plasma slab produced through the interaction of a secondary laser pulse on a He or $D_2$ gas jet target. b) Layout of the detectors configuration; the setup combine high granularity SiC charged particles detectors (in vacuum) and a new generation of neutrons time-of-flight detectors (in air).

*Target configuration*

The use of colliding plasma plumes suitable for nuclear physics studies was proposed few years ago by some of us [85] and recently adopted also by other research teams [86]. The basic principle is the following: a first laser pulse imping on a $^{13}C$, $^{7}Li$ or $^{11}B$ solid thin target (few micro-meters) producing, through the well-known TNSA (Target Normal Sheath Acceleration) acceleration scheme, boron, carbon or lithium plasma. The rapidly streaming plasma impacts on a secondary plasma, prepared through the interaction of a second laser pulse on a gas jet target (made by $^{4}He$, $D_2$ or $^{3}He$). TNSA was intensively studied in the last years; experiments [90] and models [91] show that this acceleration scheme works very well in the intensity domain between $10^{18}$-$10^{20}$ W/cm$^2$. The produced ions expand along a cone, whose axis is normal to the target surface, with a low emittance [92]. The observed ion energy distributions have an exponential shape [93] with a high-energy cut-off, linearly depending on the laser intensity [90] and scaling with the atomic number $E_{max}^{i} \propto Z$. Figure 15(a) shows some carbon ions energy distributions [94] measured in a TNSA regime at 6-7×$10^{20}$ W/cm$^2$. These experimental observations are well described and predicted by theoretical models (see [91] and reference there in). A further fine-tuning can be done acting on other parameters: i.e. the laser incident angle or polarization [94], the structure of the target surface [95], or the target thickness [93, 96]. The total number of accelerated ions obviously depends on the target composition [90]; in particular, for a single component target we can estimate that it roughly corresponds to the removed mass. In this last condition also a fraction of protons was experimentally observed due to the presence of hydrogenated contaminants on the target surface. This component can be in any case reduced through a preliminary heating of the target surface.

Due to the wide possibility of ions properties tuning (energy and number, especially), the idea underlying this proposal is to take advantage of the unique opportunities provided by ELI-NP (high rep. rate and petawatts laser) to operate in the TNSA domain (few $10^{18}$ W/cm$^2$) in order to ensure, by using large focal spots, the production of a *very large flux* of ions (some estimates are shown in Figure 15(b)) with energy distributions optimized for our purpose (lower high energy cut-off) in order to make possible the study of nuclear reactions at very low cross-section in a plasma environment.

As already mentioned before, after the production, B, C or Li ions forward-streaming towards the gas-jet made by $^4$He, $D_2$ or $^3$He. There, a second laser pulse synchronized with the first one, can be used to obtain a helium or deuterium (depending on the reaction under analysis) plasma with a low center of mass velocity, but with densities ranging in the $10^{18}$ - $10^{20}$ ions/cm$^3$ domain [97]. The properties of the secondary plasma (working as a “plasma target”) can be modified or tuned, depending on the energetic domains one wants to explore. By using femtosecond pulses, secondary plasma temperatures lie in the tens of eV range. For reactions with fully-thermalized plasmas at medium-high ion temperatures, the duration of the secondary laser beam can be extended in the nanosecond domain: temperatures or few keV for deuterons or alpha particles can be obtained in this case.

Specific simulations (see sec. 4.3) have been done in order to describe and tune the experimental conditions under these assumptions. To optimize the experimental setup (e.g. number of ions, energy etc.) we foresee a further R&D activity on targets. The goal is the manufacturing of targets with high light absorbance, tuned for ELI-NP laser wavelengths, by using nanostructured surfaces or materials [98]. Such structured materials have been very well manufactured [99, 100], as an ordered array of metallic nanowires, by using nano-porous alumina as a template. Our goal is to replace the alumina substrate with bulk carbon (or lithium) and the metallic nanowires with carbon nanotubes [101] (or lithium nanowires). Moreover, the development of these materials could lead to the implementation of a third, alternative setup like the one shown in Figure 16(a), where two identical laser pulses impinge on a thick target with micro cells, filled with gaseous elements (He, D) and enclosed from both sides by thin nanostructured carbon or lithium foils. In such a configuration the “in-cell” gas is self-ionized by the impact with plasmas generated on the two surfaces and can be further compressed by the shock waves developed during the laser–matter interaction. High plasma densities are expected in this case, however, contaminants due to nanotechnological and nanofabrication processes can play an important role, which has to be investigated.

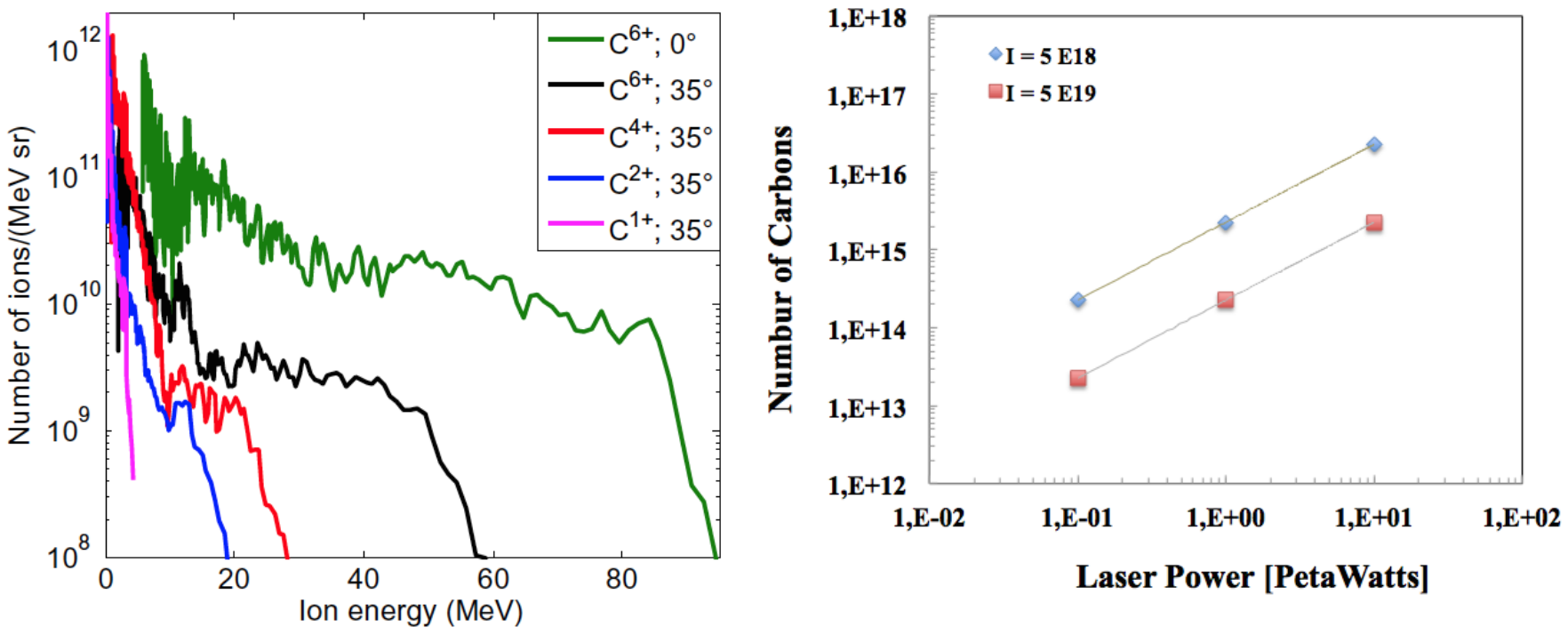

Figure 15 – a) Carbon ions energy distributions measured along the target normal axis at the rear side for laser incident angles of 0° and 35° with respect to the target normal axis and intensities of ~ $5\times10^{20}$ W/cm$^2$ (data taken from Ref. [93]). b) Maximum number of carbon ions expected at ELI-NP as a function of the laser power. The estimation has been obtained by using laser pulses focalized on a 1 μm fully drilled target [93], in order to achieve intensity (working with two focal spot radii of 160 and 50 μm respectively) of $5\times10^{18}$ W/cm$^2$ and $5\times10^{19}$ W/cm$^2$.

*Detectors*

The proposed activity requires also the construction of a highly segmented detection system for neutrons and charged particles. The segmentation is required for the reconstruction of the reaction’s kinematic. The “ideal” neutron detection module for these studies must have: high efficiency, good discrimination of gammas from neutrons, good timing performance for TOF neutron energies reconstruction. In addition, it must be able to work in hard environmental conditions, like the ones established in the laser-matter interaction area. All these aspects may be met by configuration based on 50x50x50 mm PPO-Plastic scintillator plus a SiPM read-out and a totally digital acquisition of the multi-hit signals (Figure 16(b)).

Moreover, also an R&D activity is planned on SiC detectors in collaboration with CNR-IMM Catania, in order to realize a wall device to detect charged particles in coincidence with neutrons. The SiC detectors have been proven recently to have excellent properties [101]: high energy and time resolution, resistance to radiation, insensible to visible light etc. It is fundamental for the study nuclear reaction such as the $^{11}B(^{3}He, d)^{12}C^{*}$ ($Q_{reac}$=10.46 MeV), where only the position and energy measurement of light charged particles can give access to the desired information. A sketch of the overall setup is shown in Figure 14(b).

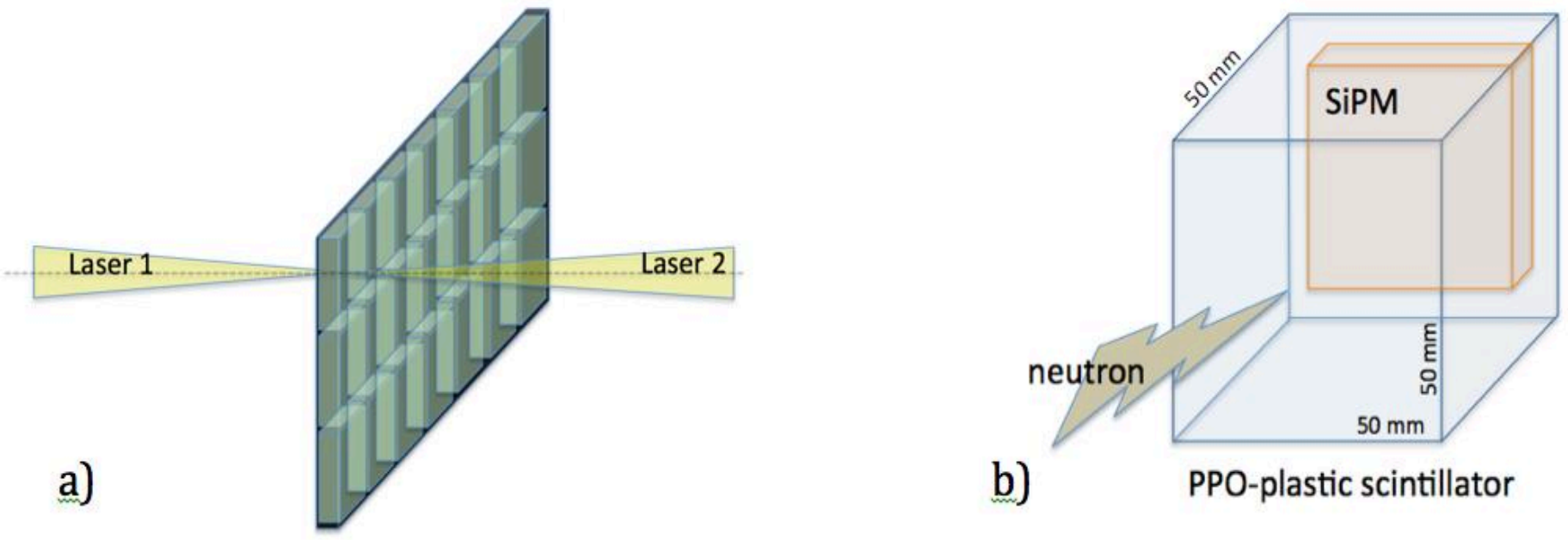

Figure 16 – An alternative two laser configuration, based on micro-cells gas targets enclosed by two thin carbon or lithium foils with nano-structured surfaces.

### 2.3.3 Hot and dense plasma trapping in high magnetic field for fusion and astrophysics studies

The main aim of the proposal is the development and operation of a Laser Based Compact Magnetic Photo-Fusion (LB-CMPF) device (see Figure 17) in open magnetic topology for study burning process in different types of fusion fuels (D-D, D-T and p-$^{11}$B) and other nuclear processes in hot and dense plasma trapped by the high external magnetic field. For the burning process of high density ($10^{18}$ cm$^{-3}$ – $10^{19}$ cm$^{-3}$) and high temperature (tens of keV) magnetized plasmas, the trapping by a high (about 120 T) external applied mirror-like magnetic field is a challenging objective. The high magnetic field generated by a recently developed pulsed magnetic driver and the initial plasma density is produced by high intensity laser beam interaction with clusters or thin foils.

The use of a multi-fluid code allows to simulate: the spatio-temporal evolution of the plasma in different external applied magnetic field topologies, the trapping time of the plasma, the burning process of the fuels, the neutron and alpha production in the fuels, the effect of the initial spatial profile of the magnetic field on the efficiency of the burning process and the optimization of the device operation to improve the reaction rates. A 2-D single-fluid resistive MHD code allows studying the spatio-temporal evolution of the plasma in different magnetic configurations and evaluating the axial and radial plasma losses. A number of diagnostics on particle and plasma measurements was developed and improved over the last few years.

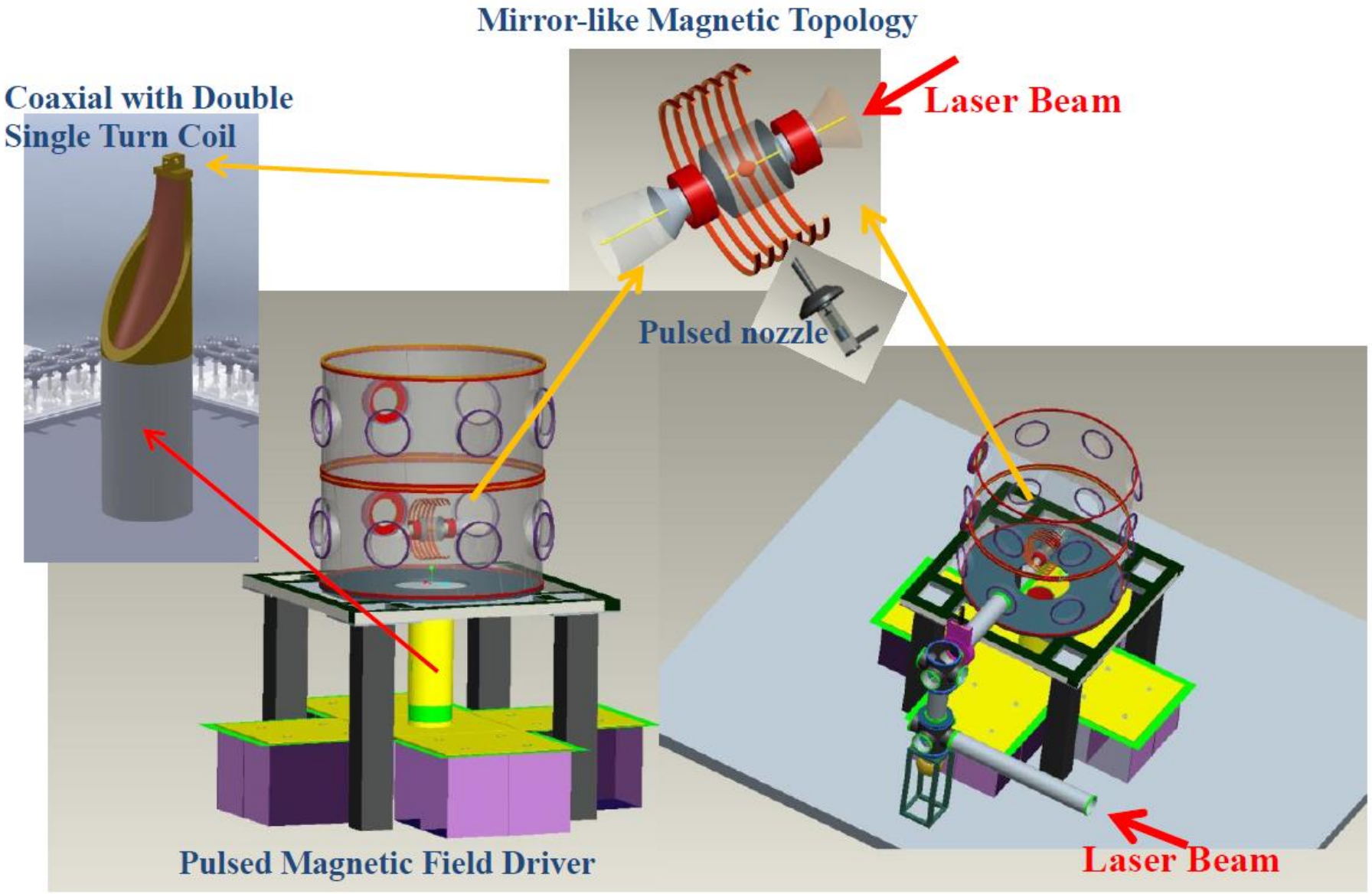

Figure 17 – Main elements of the proposed Laser Based Compact Magnetic Photo-Fusion (LB-CMPF) device.

The last few years there have been an increased interest to develop experimental setups or laboratory prototypes of compact fusion devices working with intermediate plasma densities ($10^{16}$ -$10^{18}$ $cm^{-3}$), to be compared to Tokamak machines, which operate at lower plasma density or to ICF machines, which operate at much higher plasma density. These compact fusion devices are proposed for important industrial applications, such as magnetic fusion plasma studies, fusion energy production, space propulsion and blanket material studies for the future Tokomak or ICF machines using the produced neutrons from the fusion nuclear reactions.

A laser based new scheme and methodology for the production, trapping and refueling of high-density and high-temperature plasma in high externally-applied magnetic field is proposed for research, development and operation in laser facilities providing petawatt laser beams and above, such as the ELI-NP pillar. The initial high-density and high temperature plasma is produced by ultra-short, high-intensity laser beam interaction with clusters or thin solid targets (thin disc), and different fuels such as D-D, D-T and p-$^{11}$B can be investigated. The term photo-fusion is used due to the laser beam induced plasmas in the proposed LB-(CMPF) device.

The proposed development is based on different experimentally well-established technologies such as:

a) the production of a high density and high temperature plasma by laser beam interaction with clusters or thin solid targets (thin disc),

b) laser filamentation and non-linear propagation of ultrashort, high intensity laser beam in a plasma,

c) the development of a pulsed high magnetic field driver in mirror-like or other topologies,

d) the coupling and the trapping of the high-beta plasma in different high magnetic field configurations produced externally by the magnetic field driver,

e) numerical simulations using multifluid codes, describing the spatio-temporal evolution of the state parameters of each plasma species and the reaction rates of the nuclear fusion reactions.

*2.3.3.A. Plasma production*

*(a) Laser-Cluster interaction*

The interaction of a high-intensity ultra-short laser pulse with a molecular beam of neutral deuterium clusters produces high-density and high-temperature plasma [103–105]. The used clusters are composed by a relatively big number of molecules up to 200.000 or higher [107] and are formed during the adiabatic expansion in vacuum of the deuterium gas from a pulsed high pressure nozzle [106, 107]. The interaction of these big clusters with the ultra-short laser pulse ionizes the molecules of the cluster and forms an electron cloud around the clusters. A high electric field is formed from this charge separation and high energy D ions are produced due to the Coulomb explosion of the remaining, positive, big clusters. The collisions between the D ions of the plasma produce neutrons by fusion nuclear reactions [103, 104, 107]. But the produced plasma expands very fast in the vacuum, decreasing the local plasma density and consequently the number of D-D ions nuclear fusion reactions, because the rate of the nuclear fusion reactions depends on the square of the local density. The above description is compatible with experimental data [103–105] concerning laser-cluster and/or laser-micro-droplets interaction. The produced plasma have a high density up to few $10^{18}$ cm$^{-3}$ and the energy measurements of the D ions using a Thomson parabola verify the production of D ions with kinetic energy up to 70 keV for laser beam energy up to 700-800 mJ and pulse duration up to 30 fs. The laser beam intensity corresponds to $5\times10^{16}$ W/cm$^2$ [104, 105]. Numerical simulations [106, 107] based on a 2-D MHD resistive code [108] (see next paragraphs) confirm that the application of an external applied high magnetic field in mirror-like topology enables to decrease the plasma expansion velocity, increase the trapping time of the plasma and improve the neutron production [109, 110]. In a recent experiment we observed similar results but with a lower laser beam energy of 200 mJ (December 2012, CELIA laser facility in Bordeaux).

The observed effect is due to high contrast ratio for the laser beam, because in the case of relatively low contrast ratio the pre-pulse of the laser pulse destroys the big cluster before the arrival of the main pulse and reduces the Coulomb explosion effect.

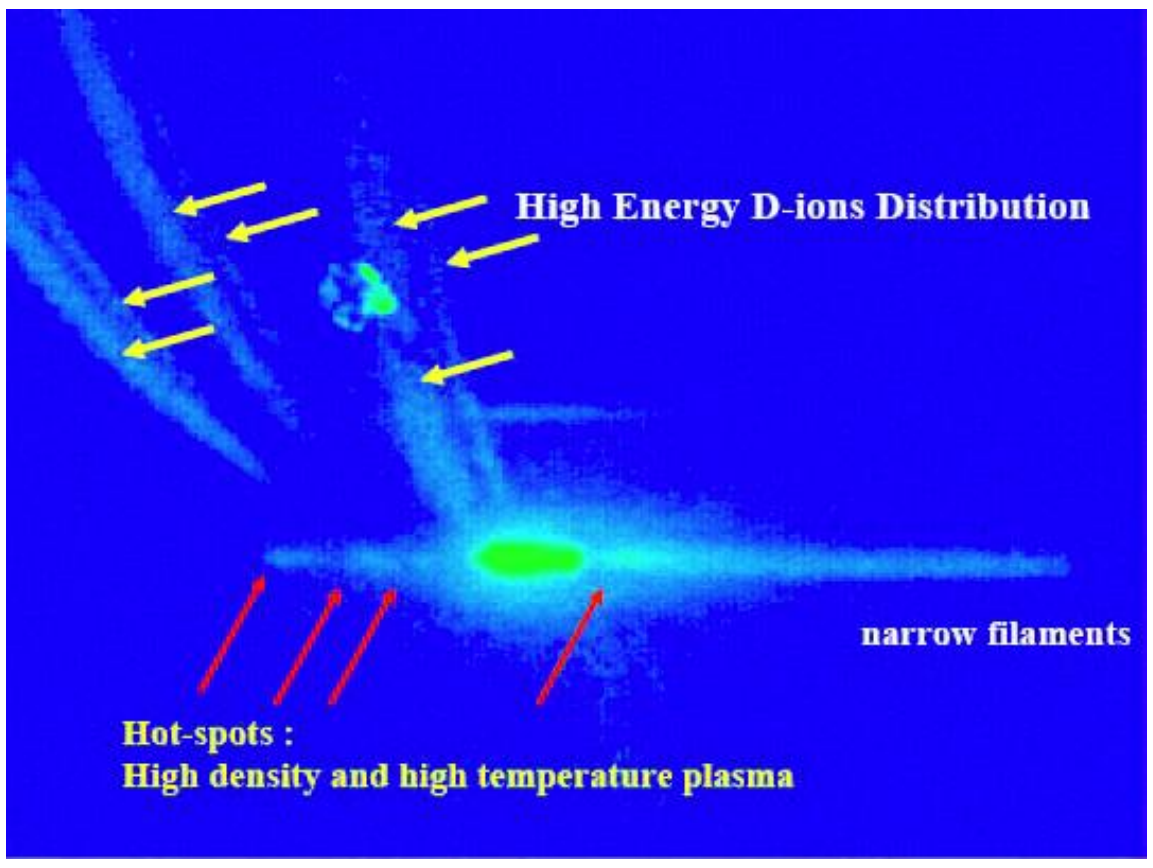

Figure 18 – Distribution of D ions measured with a Thomson parabolas mass spectrometer. The formation of multi-plasma spots and multi high density plasma regions is due to the non-linear propagation of the laser beam in the cluster volume.

*(b) Non-linear propagation and filamentation in deuterated clusters*

The initial plasma volume corresponding to the laser-cluster interaction is limited by the optical focusing system in the most of the experiments. In fact, the interaction volume in most of the experiments is relatively small, corresponding to the focal dimension of the ultra-short laser beam in the clusters. The fusion reaction rates and consequently the number of the produced neutrons is proportional to the interaction volume and the time interval for which the plasma density remains high (few tens of ps in the case without magnetic trapping and few μs with an external applied magnetic field up to 110 T). A new idea to increase the interaction volume is based on the effect of the non-linear propagation effects of the ultrashort laser beam during the propagation in the cluster volume [110-112] (see Figure 18). Under the conditions of non-linear propagation [112] of the ultra-short laser beam the interaction volume increases considerably, preserving all the plasma parameters as in the case of a single focus. Figure 18 shows the output data of the Thomson parabola mass spectrometer, which presents a number of hot spots (multi-plasma spots), corresponding to the filamentation formation of the laser beam in the clusters. The non-linear propagation produces a relatively long filament [110, 111] with an important number of multi-focal spots (or hot spots) of high density and high temperature plasma. Each parabola in Figure 18 corresponds to the energetic D-ions produced by each hot-plasma spot formed during the laser beam non-linear propagation in the clusters. The multi-plasma spots preserve the same density and temperature values as was for the case of the single focus spot, described in the previous paragraph. This new experimental condition confirms the volume increase of the high density and high temperature plasma and improves the efficiency of the energy transfer from the laser beam to the plasma (cluster volume).

*(c) Non-linear ponderomotive force for plasma block acceleration and production of energetic particle beams*

The main advantage of the proposed LB-CMPF device is that we can study experimentally and numerically the burning process of different fusion fuels and investigate on the reaction rate efficiency of fuels in various external applied high magnetic field topologies. An important fusion nuclear reaction is $p{-}^{11}B$, because it is neutron-less reaction and produces 3 alpha particles with a total kinetic energy of 8.2 MeV. The disadvantage is that the reaction cross section is significant for energies higher than 400 keV, which is difficult to achieve by laser-plasma interactions in magnetic configurations, because the plasma temperature is relative low up to few tens of keV. Recently, important experiments were performed at the LULI laser facility in the École Polytechnique [113] and in the PALS laser facility at the University of Prague [114], employing the ultrashort laser beam impact on a solid target to produce a high energy proton beam, then to interact with a solid Boron target and generate $>10^8$ alphas per laser shot [114]. The efficiency of the process remains relatively low, due to both the relatively small interaction volume and the interaction time of the protons during their penetration in the solid Boron target. We plan to use a new experimental scheme for the p and B ion interaction in the proposed LB-CMPF device with a bigger interaction volume and a longer plasma interaction time by trapping the produced (p, B) plasma in the high magnetic field of the device. Theoretical [115], experimental [116, 117] and numerical work [118, 119] show that the non-linear (ponderomotive) force can accelerate plasma blocks [118, 119, 120] when high-contrast, ultrashort laser pulses up to a few $10^{17}$ W/cm$^2$ interact with a solid target. Our recent numerical simulations confirm experimental investigations and show that the interaction of high contrast laser pulses with thin foils enables producing energetic ion beams [121] with densities close to the solid density (i.e. up to $10^{23}$ m$^{-3}$). A new experimental setup using the LB-CMPF device can be employed in order to study the $p{-}^{11}B$ fusion process in the mirror-like magnetic configuration of the proposed device. A brief description of the proposed experimental setup allows to evaluate the advantage of the LB-CMPF device. A thin solid disc will be placed in the vicinity of each magnetic mirror of the device in order to produce high density and high energy proton and $^{11}B$ beams by the interaction of thin discs, when each solid target is irradiated by a PW laser beam.

The high contrast PW laser beam is necessary in order to keep the laser beam intensity up to $5\times10^{17}$ W/cm$^2$ for the ion beam production, but with a relatively larger irradiated surface on the target by using the high energy of the PW laser beam and consequently enlarging the initial section of the produced ion (p and $^{11}B$) beams. Both particle beams with energies up to 400 – 600 keV will be injected inwards the LB-CMPF device (from the magnetic mirrors to the center of the device) and in few nanoseconds will fill up the volume between the magnetic mirrors of the device. The magnetic field topology, the relatively high trapping time up to μs and the relatively large volume (about 1 $cm^3$) allow optimizing the reaction rate process of the p–$^{11}$B fusion reaction. The numerical results of the simulation for the proposed experimental configuration will be presented in the section 4.3.3.

*2.3.3.B. Development of a Pulsed High-Value Magnetic Field Driver*

During the last few years, we have investigated on the development of a pulsed high magnetic field driver [107, 122] designed for trapping high density and high temperature plasmas in a high magnetic field with mirror-like topology. The magnetic driver operation is based on the fast discharge of a high voltage capacitor-bank storing around 8 kJ into a slotted single-turn coil (Figure 19). The aim is to operate in non-destructive conditions, thus the coil withstands the magnetic forces without modification. The targeted magnetized volume is of the order of 2-3 $cm^3$. A low-impedance flat transmission line was used for efficient transfer of the stored electrical energy into the coil. For a good coupling between the flat line and the coil, the switch was chosen first as a multi-channel surface discharge gap in atmospheric air [123]. This triggered switch has a minimum insertion inductance in the main circuit and improves dramatically the peak current in the coil. A new solution was adopted for the design of a fast spark-gap, based on a series of 15 multi-gap switches installed in parallel [124]. This multi-gap multi-channel switch (MGMCS, see Figure 20) has been used successfully by the team to feed an X-pinch at the level of 250 kA with 750 J storage [125].

Figure 19 shows a global view of the device with the capacitor bank, the main flat transmission line, the MGMCS and the coil at the end of a tapered section of the flat transmission line connected with the spark-gap. Different types of slotted or bored single-turn coils were tested in order to measure the value of the high magnetic field produced by the driver and the spatial profile of the magnetic field along the z-axis in order to determine the mirror-like profile of the B-field inside the coil. Examples of coils are presented in Figure 21(a) with dimensions of interest (Figure 21(b)). Figure 22 shows the magnetic field profile measured along the z-axis inside the coil for the case of the slotted single-turn coil #1. The results are in good agreement with the expected mirror-like topology. Similar results were obtained from all types of the used coils. The coils are massive metallic structures made off brass, in order to withstand the high magnetic forces produced by the high current up to 1 MA.

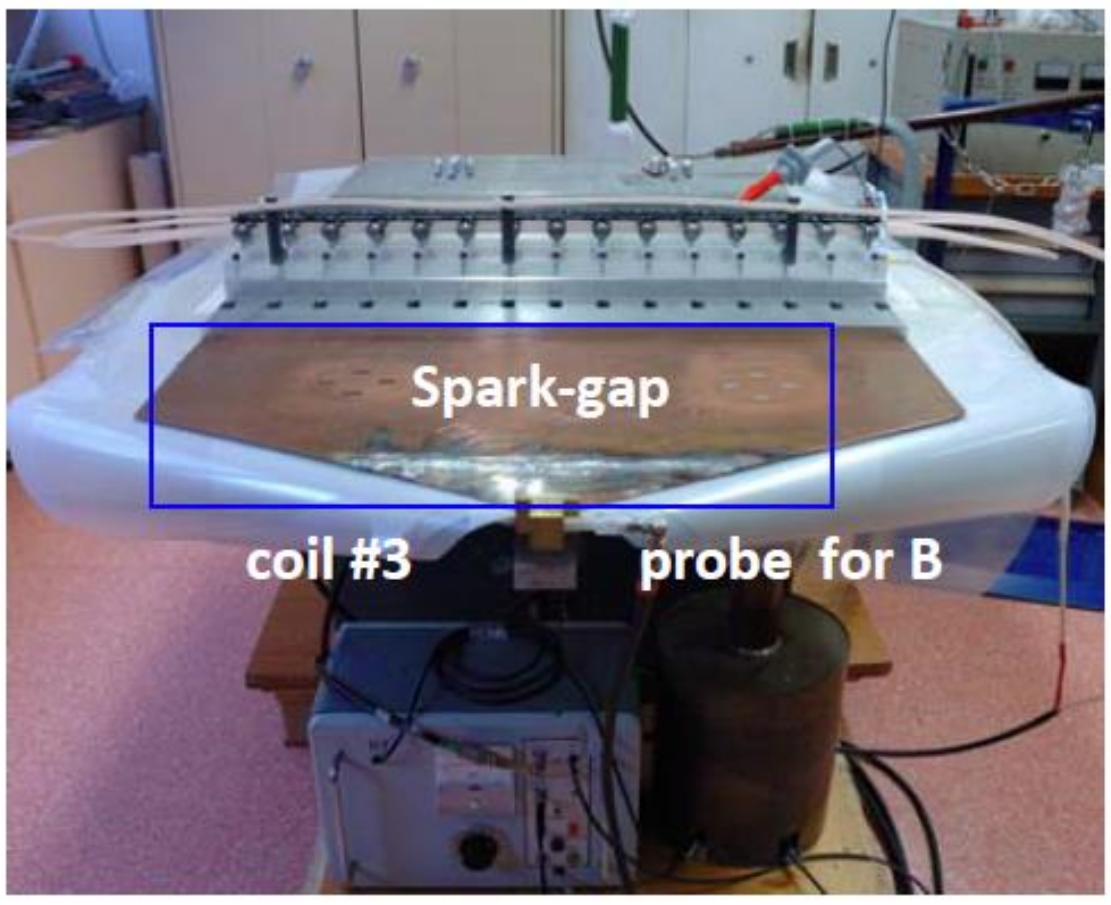

Figure 19 – From rear to front, the flat transmission line on top of the capacitor bank, the MGMCS and the coil at the end of an adapted flat transmission line connected to the switch.

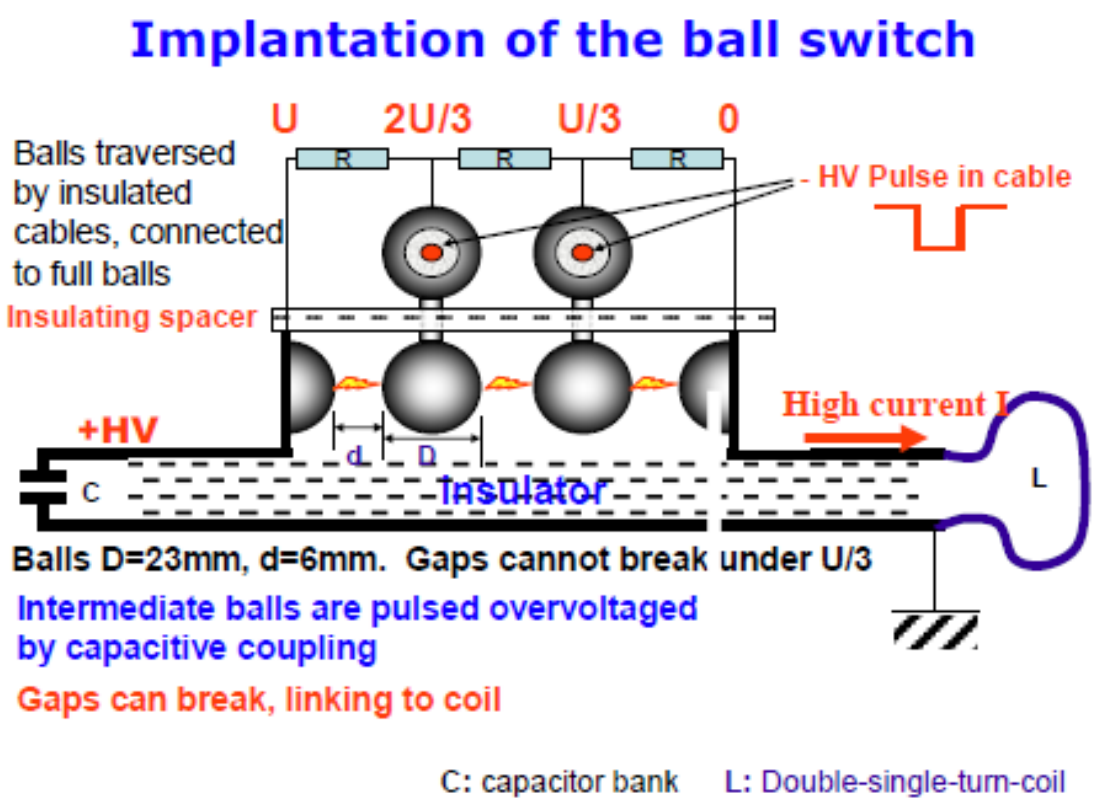

Figure 20 – Sketch of the principle of operation of the new spark-gap switch with multi gaps and multi channels (MGMCS). The iron balls have 16 mm diameter and the spacing is 5 to 6 mm.

From measurements using a 6-mm$^2$ pickup coil installed on axis, there is a linear dependence of B on the pulsed current (3.4 μs pseudo period) maximum value of the magnetic field as a function of the current for the different coils. A top value of approximately 33 T was measured for a current around 930 kA. The correct operation of the used fast spark-gap is very important, because it determines the total value of the inductance of the circuit and consequently the maximum value of the current in the mirror-like coil (see Figure 23). The inductance of the spark-gap depends on the number and location of the lighted channels in the MCMGS. Better conditions are obtained with a sharp triggering pulse, when channels are operating on the whole width of the transmission line.

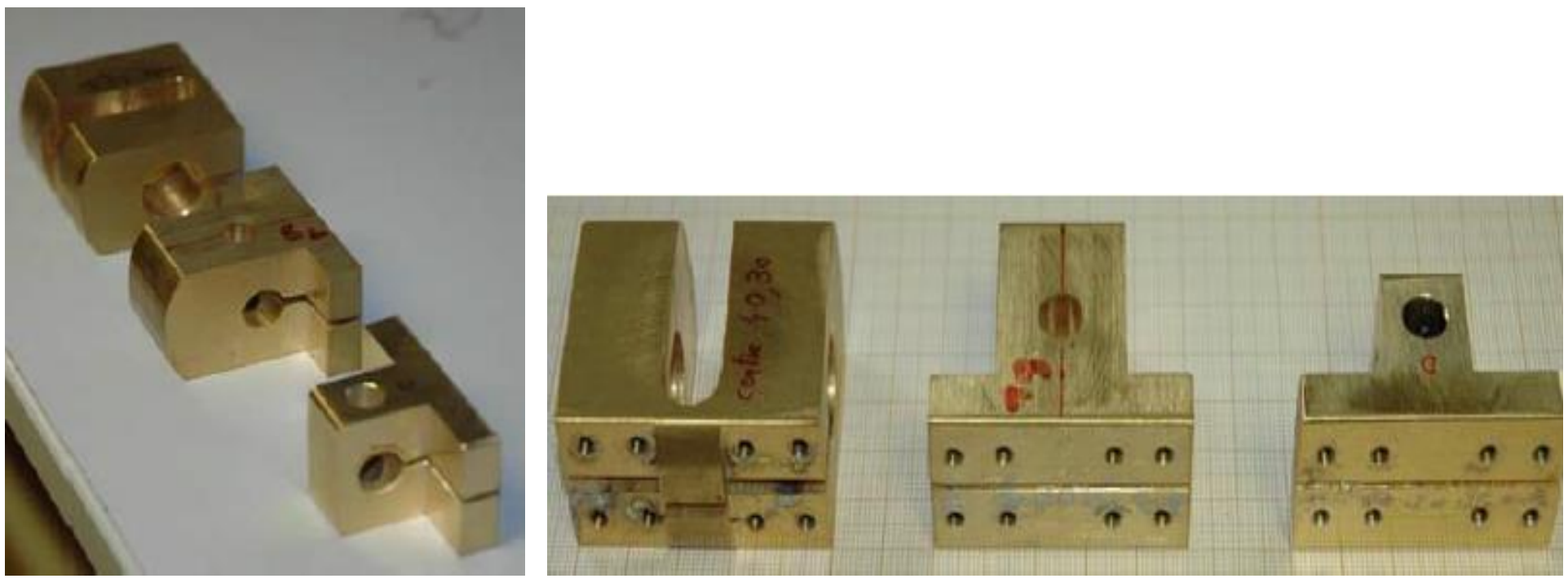

(a) (b)

Figure 21 – (a) The different types of slotted and bored single turn coils tested for the mirror-like topology of the magnetic field. (b) Spatial dimensions of the different coil used to test the mirror like topology at the end of the transmission line of the high magnetic field driver. The coils are identified by numbers from 1 to 3 from the left to the right, respectively. The graphical paper main pitch is 1 cm.

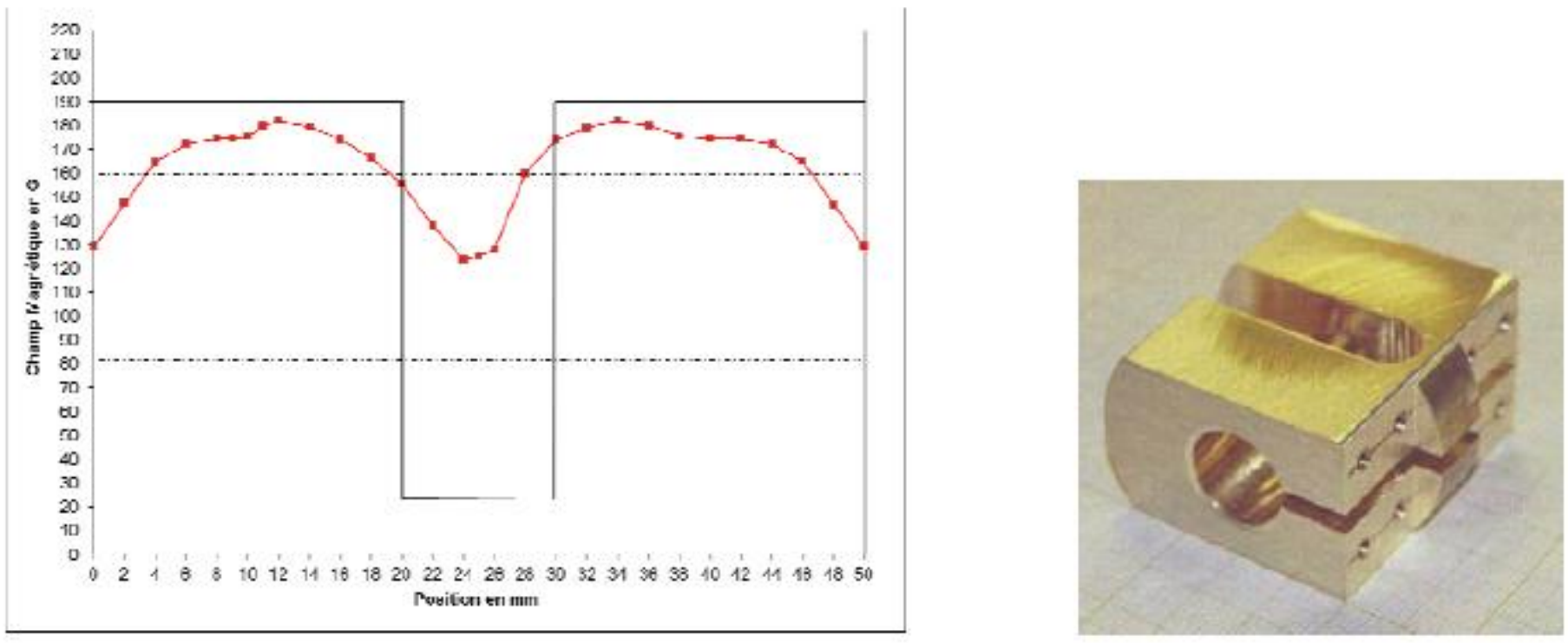

Figure 22 – Magnetic field profile measured along the z-axis of the slotted single-turn coil, presented on the right. The low values of B (180 gauss maximum) correspond to calibration shots with a ringing frequency less than 1 kHz, using a 2000 μF capacitor at 300V and a mechanical switch.

The experimental results show that the proposed configuration allows the development of a relatively compact driver for pulsed high magnetic field generation up to 30-35 Tesla (see Figure 23). But the numerical simulation from the 2-D MHD resistive code shows that for high neutron production, a longer trapping time of the plasma and a higher magnetic field up to 90-110 T are necessary (see also the numerical simulations using the multi-fluid code). With these constraints, a new configuration of four modules composes the capacitor bank in order to achieve the requested value of the magnetic field. A co-axial transmission line is connected between the fast spark-gap (switch) and the coil (see Figure 24). A new design for a common fast spark-gap switch, based on the one previously used, will connect the four modules of the capacitor bank to a coaxial convolute feeding the mirror-like coil. The switching performance is linked to a perfect synchronization of the discharged currents from the four modules in the coil. The proposed configuration with a square MCMGS and a common triggering will facilitate the coupling. The coaxial section of the transmission line will allow the transition to experimental chamber under vacuum whereas the tulip transition (commonly used in high power microwaves for coupling a guide to an antenna) will adapt the line to the coil.

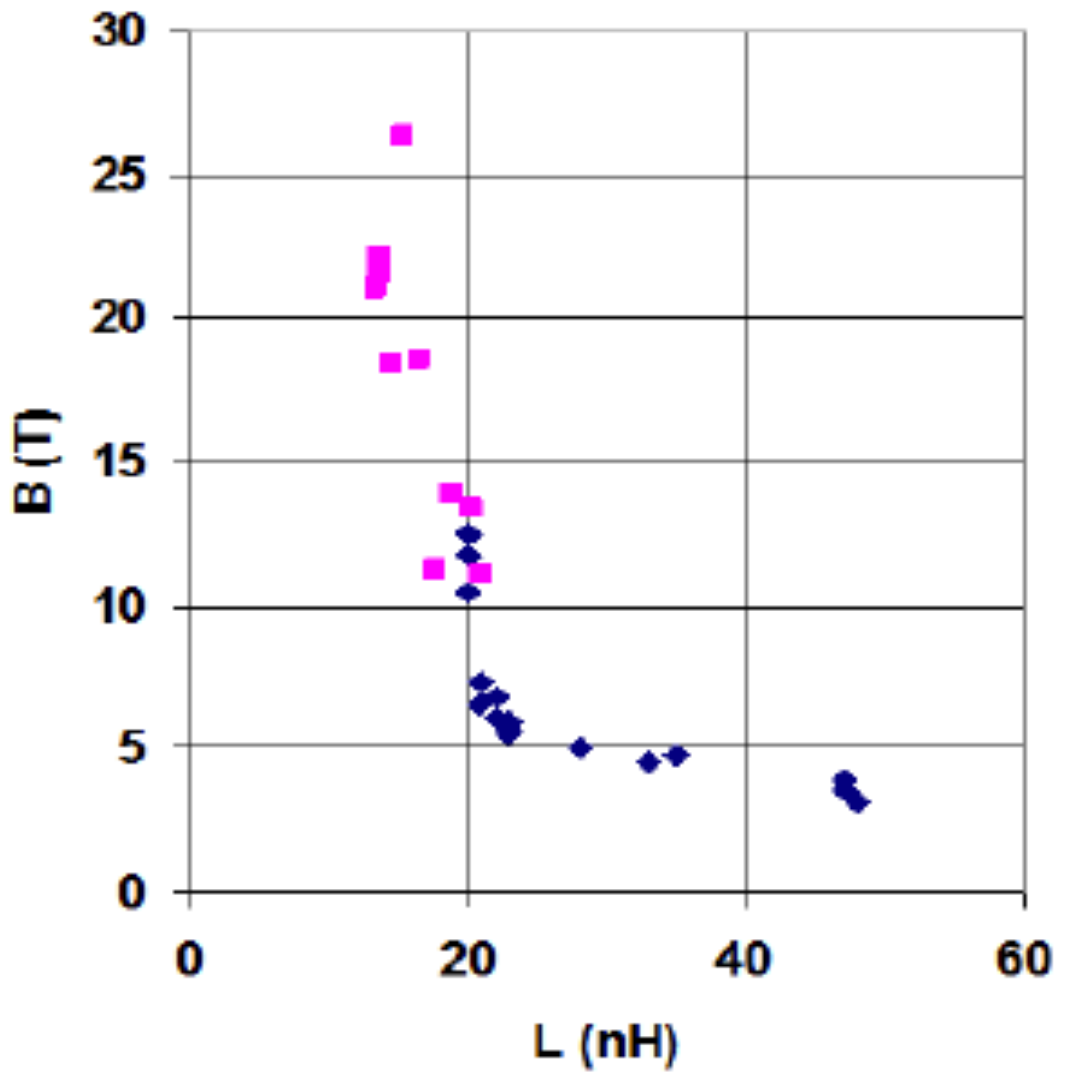

Figure 23 – Example of dramatic increase of the magnetic field measured on the z-axis of the mirror-like coil 3 as a function of the whole circuit inductance. The reduction of L is correlated to an increased number of lighted channels, up to coverage over the whole width.

*Remark:* The value of the 110 T is the upper limit for operating with non-destroyed single double-turn coils. For higher magnetic fields, the coils will be destroyed after every shot. The existing magnetic driver, as described previously, could be used to test different high magnetic field topologies using non-destroyed or destroyed coil configurations. For the PW laser system of the ELI-NP laser facility, which operates with one shot every one minute, it is very convenient to use the non-distractive coil configurations. But if the application requires higher magnetic fields, it is possible to test and evaluate the reproducibility of the magnetic field topology using the existing magnetic driver before the realization of the final experimental setup. Following the OMEGA laser facility magnetic field amplification scheme [126], we can increase by a factor of 3 the value of the high magnetic field, but using a destroyed coil configuration.

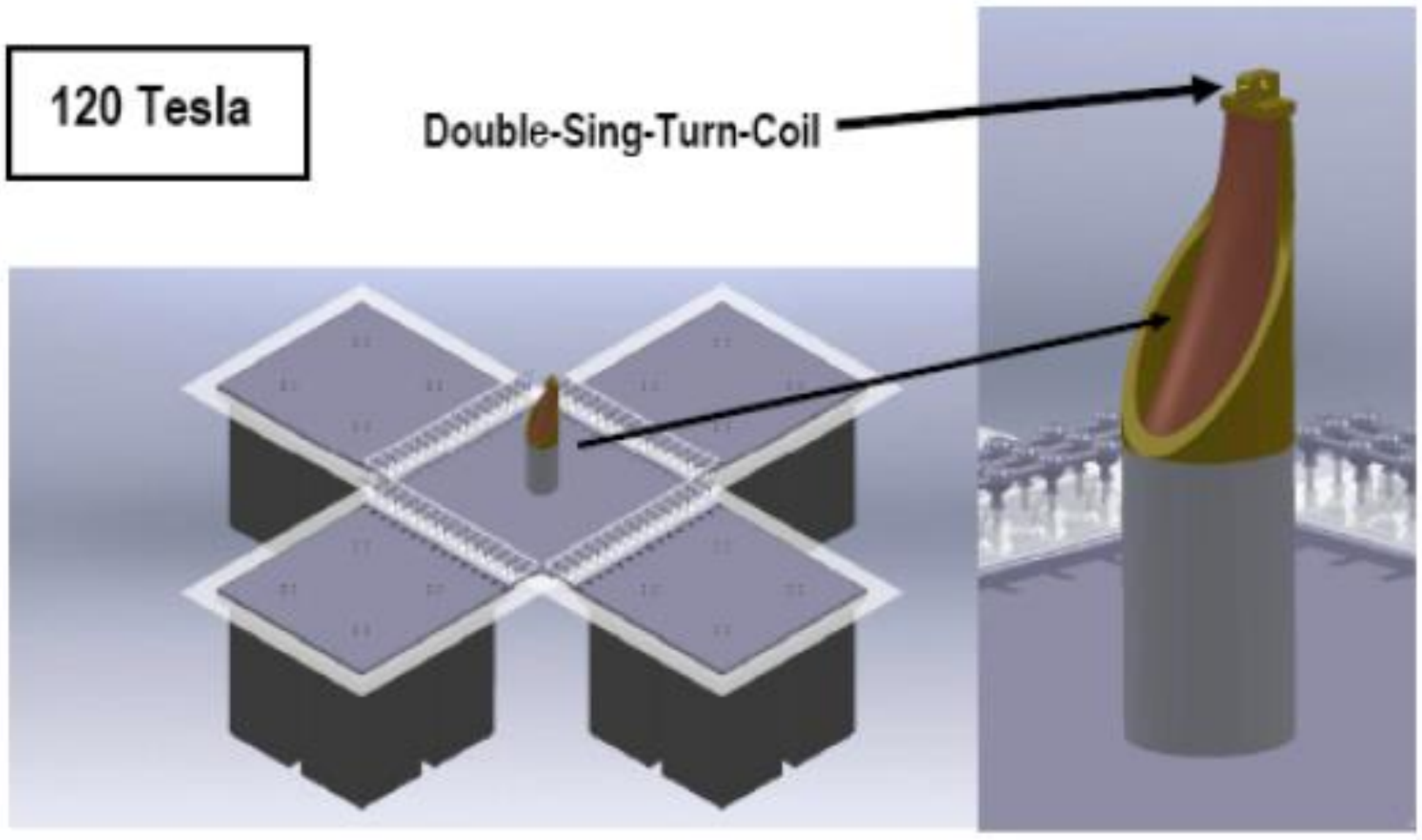

Figure 24 – The proposed 4 x 4 capacitor bank to generate up to 110-120 Tesla and the coaxial transmission line charged with the slotted single-turn coil.

### *2.3.3.C. Applications*

Beside the above mentioned applications on D-D and p-$^{11}$B fusion, another important application concerns the study of nuclear reactions of high density plasma composed by heavy nucleus in high magnetic field. The production of high density (near to solid density) plasma jets by laser impact on thin solid (disc) targets allow to study the acceleration process and investigate research on the interaction of counter propagated plasma jets in high magnetic topologies (see Fig. 59(a) and 59(b) and the related text of the section 4.3.3) with astrophysical applications. The proposed magnetic driver could be used to generate high magnetic field (110 T – 300 T) in small volumes (about 1 cm$^3$) or relatively smaller magnetic field (30 T – 35 T) but in larger volumes (tens of cm$^3$) in different magnetic topologies. This possibility allows to adapt the magnetic driver for applications concerning the NEET in a hot plasma when an electronic transition in the atom excites its nucleus, as proposed by the team of ENL/CENBG/CNRS of University of Bordeaux. The magnetic trapping for hundreds of ns of a reasonable high density of Rb plasma with 1.4 keV electronic temperature allow to study the effect of the induced NEET process. The main advantage of the proposed LB-CMPF device is the possibility to work with different targets producing high velocity and high density (values close to the solid density) jets (beams) of different particles (see the text and the simulations in the next paragraph). The nature of the particles in the plasma jet could be selected from the teams working on shock interactions and are of interest to study related astrophysical phenomena in the presence of high magnetic fields, as proposed by the team of J. Fuchs from LULI (École Polytechnique, Palaiseau, France). The use of the multifluid code allows calculating the particle density in the jets, the kinetic energy of the beams and the spatiotemporal evolution of the plasma in the magnetic tubes inside the magnetic topology.

The existing multi-fluid numerical code allows simulating high density and high temperature plasmas of different composition in the presence of extremely high magnetic fields up to 10 kT [127, 121]. The multifluid code could be used to describe more complex systems, such as the processes involving the muon production by high energy particle beam (see section 2.4.3 below), their trapping in a high magnetic field and the use of muons in a DT plasma for catalyzed fusion studies in a compact magnetic device. Another scenario that can be study is the effect of heating low temperature plasma trapped in a mirror like magnetic topology using the high energy gamma and/or neutron pulses produced by laser that will induce fission on a mixed Th D target. The proposed multifluid code could be used to describe spatio-temporal evolution of the state parameters of the fission fluid fragments (ions), of the D plasma in the mirror-like topology and the heating effect of the D ions of the plasma due to collisions with the fission fragments. The code allows estimating the production of the mono-energetic neutrons (2.4 MeV) from the fusion and comparing with the neutron measurements from the experiment.

## 2.4 NEUTRON PRODUCTION AND OTHER APPLICATIONS

Intense neutron generators serve an important role in many research fields including engineering material science [128], life sciences [129], and condensed matter physics [130]. Until recently, the experimental access to a high neutron flux was exclusive to reactor and accelerator-based facilities. For the past few years, the availability of tabletop particle sources based on high intensity lasers has enabled the realization of high flux neutron generators [131, 132].

The neutron interactions with matter are unique in many respects, offering specific capabilities to probe materials and processes, complementary to X-ray or charged particle based techniques. While the low energy neutrons (thermal neutrons), having a wavelength comparable with interatomic spacing in solids, are widely used for structural characterization of samples from many fields, the application range of fast neutron sources encompasses active interrogation of sensitive material, nuclear waste transmutation, material testing in fission and fusion reactor research and others. In this chapter, various methods for fast neutron production using high power laser are proposed to be studied at ELI-NP. The experimental techniques described have applications well beyond the neutron generation, which will be emphasized through the presentation.

### 2.4.1. Neutron production through fast light ions nuclear reactions

The method of choice for efficient production of well-collimated, fast neutron beams has been established to be the bombardment of targets by light ions. This method is implemented in the form of light-ion accelerators in which tens of MeV ions are impinged on low Z-number targets, or in the form of spallation sources in which hundreds of MeV protons impinged on high Z-number targets release many neutrons per incident particle.

Several mechanisms for laser acceleration of ions off solid foil targets were identified over the past two decades. These schemes are under extensive investigation and are a major research focus on ELI and elsewhere. High-quality, brilliant laser-generated ion beams are required for several studies portrayed in this document, to include neutron generation. For a recent analysis of the different acceleration mechanisms at extreme laser intensities see Ref. [133].

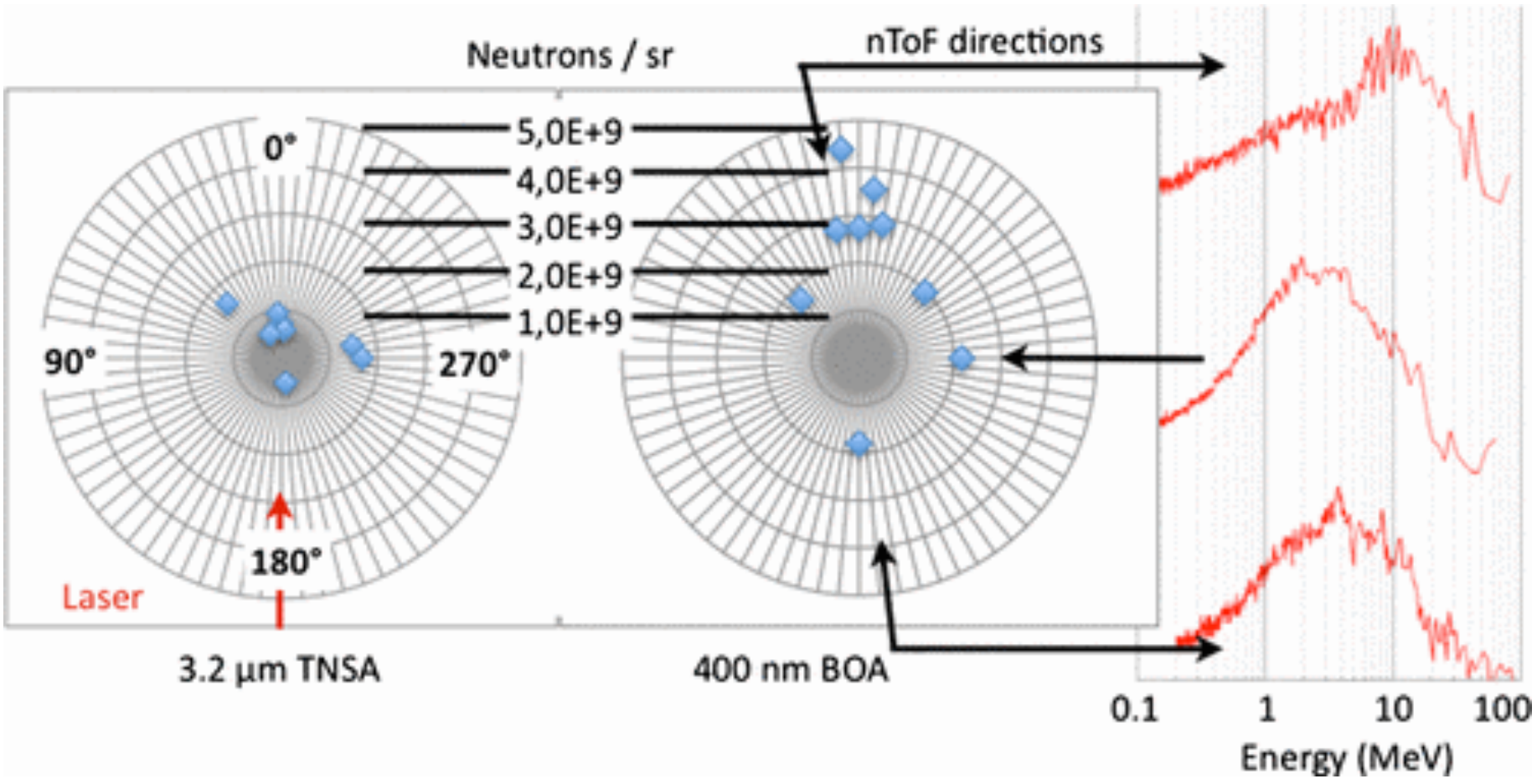

Figure 25 – Measured yield with a 3.2 µm target (left) compared to a 400 nm target (center) [131].

For the laser intensities at ELI-NP, two acceleration mechanisms, the Radiation Pressure Acceleration (RPA) [23] and the Break-out Afterburner (BOA) [135], are predicted to dominate the emitted ion spectra. Both of these mechanisms require similar experimental conditions, i.e. high temporal contrast and thin (<µm) targets.

On the first steps of our experimental campaign, we will characterize the ion spectra emitted from thin targets. This will be done using a suite of ion diagnostics (Thomson parabolas, activation stacks etc.) and optical diagnostics (Single-shot 3rd order Auto-correlator, interferometry). We will scan the target parameters (e.g. thickness and material) to compare the measured ion spectra with available models. The goal of this study is to establish a well-characterized ion beam for the relevant studies portrayed in this document.

The second step of this campaign, once the accelerated ion beams are characterized, is to establish a bright, laser-ion driven neutron source. Based on the ion beam properties, we will optimize an experimental setup that includes the ion target, a neutron converter, and neutron diagnostics.

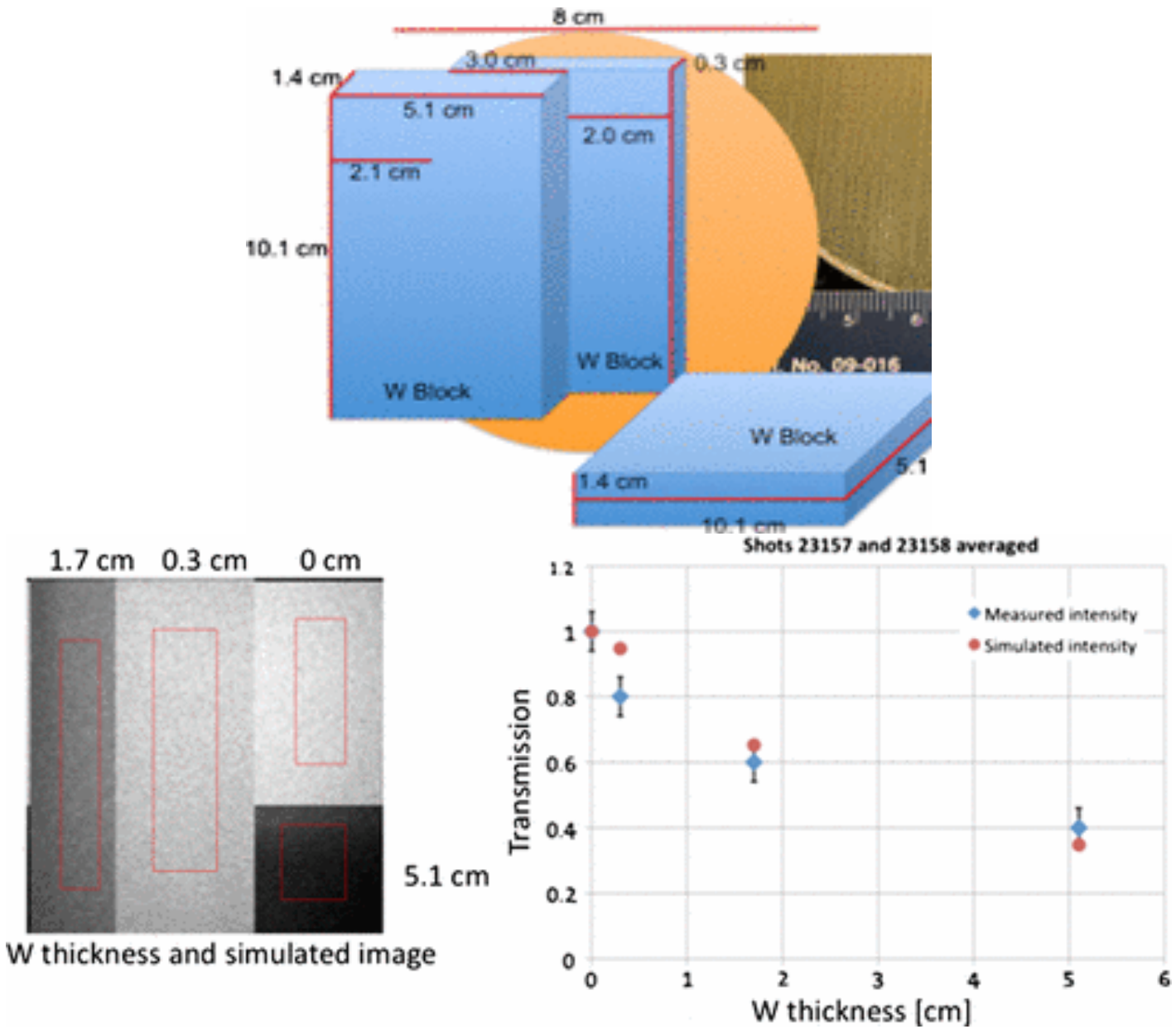

Figure 26 – Comparison between the measured transmissions for different tungsten objects to MCNPX calculations using the measured neutron energy distribution

Neutron generation using laser-accelerated ions was successfully demonstrated in few recent studies [131, 132, 136-138]. Some major achievements in this field are reported by Roth et al. [131]. Neutrons with mean energy above 10 MeV and yield of $5\times10^9$ n/sr are reported (see Figure 25) at forward angles following the high-energy deuteron reactions in a thick Be converter. The experiment took place at Los Alamos National Laboratory 200 TW Trident facility using laser pulses of, typically, 80 J / 600 fs reaching focus intensity in the range of $10^{20}$–$10^{21}$ W/cm$^2$. The ultra-high contrast of the TRIDENT laser enabled acceleration of deuterons through the BOA mechanism, which is described in detail in [135, 139], on 400 nm thick deuterated plastic foils.

The capability of this neutron source to perform radiography of thick objects was demonstrated using 3 blocks of tungsten, arranged as depicted in Figure 26 above.

Coupled with techniques of ion focusing and energy selection driven by cones or cylinders attached to the target [140] or micro-lenses triggered by a laser split pulse [141], this method promises high brilliance, nanosecond duration, micrometric dimensions, variable energy neutron sources with applications in fast evolving processes probing. At ELI-NP, these methods of neutron production can be studied taking full advantage of new accelerated mechanisms, such as BOA and RPA, expected to dominate at laser intensities larger than $10^{22}$ W/cm$^2$.

### 2.4.2 Neutron production through photonuclear reactions

We will pursue the production of intense neutron bursts using laser-driven electron jets. This effort will run in parallel to the ion-driven campaign portrayed in 2.4.1. The scheme is depicted in Figure 27 and follows the method described in Pomerantz et al. [142].

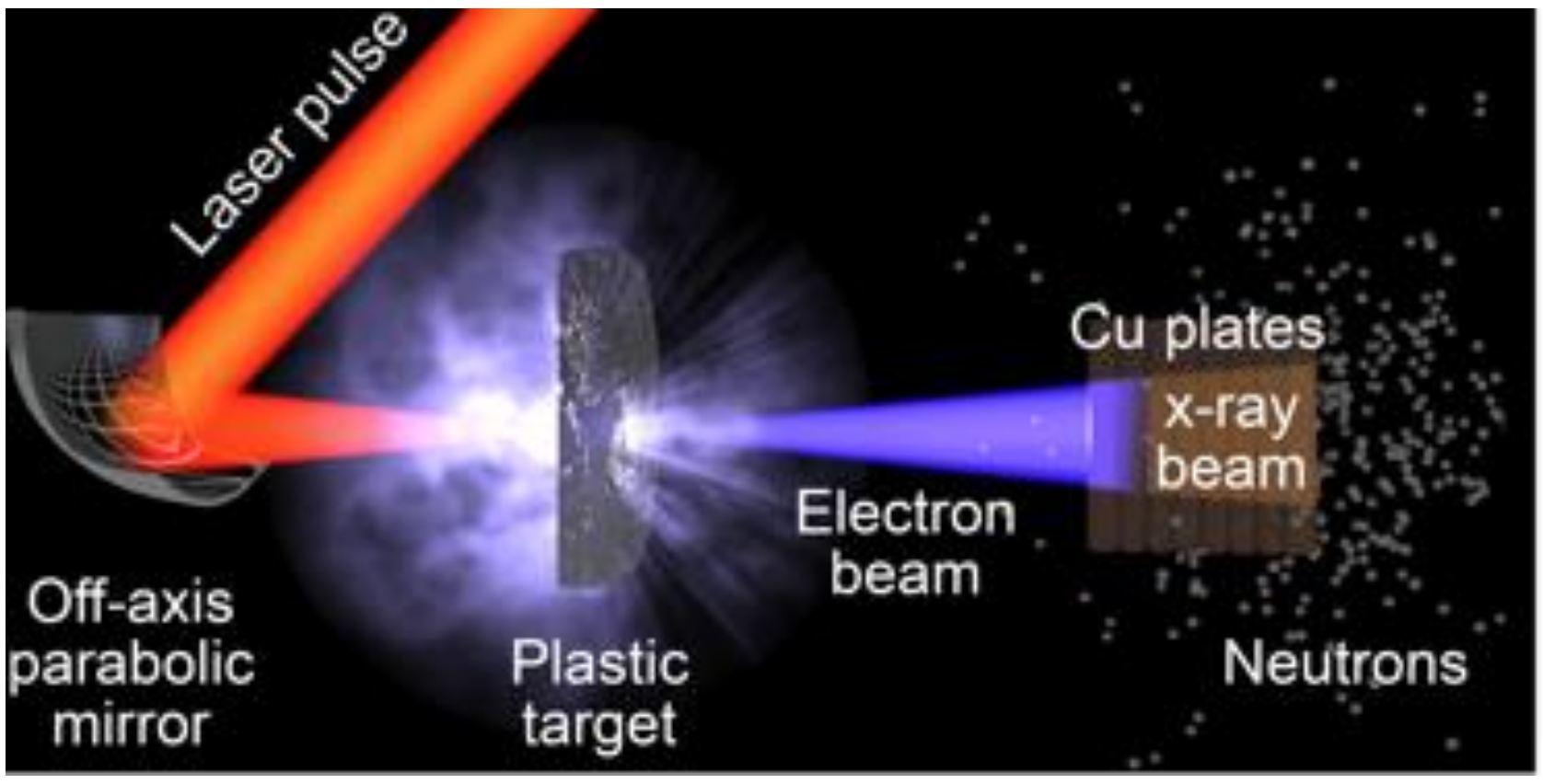

Figure 27 – Depiction of the electron-driven laser-neutron generation setup.

The laser pulse is focused onto a thin plastic target. Energy deposited in the target from low-level light arriving 10s of ns prior to the arrival of the main pulse is sufficient to turn the target into an exploding plume of plasma. The interaction of the main laser pulse with the plasma accelerates electrons and forms a high-energy electron jet that co-propagates with the laser beam. The electron jet impinges on a secondary high-Z converter positioned downstream of the plastic target. The electrons are stopped in the converter and radiate high-energy (Bremsstrahlung) gamma rays. These gamma rays interact with the copper nuclei and release photo-neutrons.

The experiments in Ref. [142] were performed on the Texas Petawatt Laser facility at the University of Texas at Austin [143]. Ultra-short laser pulses of 150 fs (FWHM), with 90 J of energy on target were employed to realize a neutron source with unprecedented short pulse duration (<50 ps) and high peak flux (>$10^{18}$ neutrons/$cm^2$/s), an order of magnitude higher than any existing source. The measured neutron energy spectrum is peaks at about 0.5 MeV, as expected from an evaporation spectrum of photo-nuclear reactions [144].

We will implement this method using the 10 PW beam at ELI-NP with the goal to achieve record values of the peak neutron flux. The experimental setup is very similar to the ion-driven method described above. This will enable running these two campaigns in parallel. The setup benefits from low sensitivity to the temporal quality of the laser-pulse, which makes it optimal for day-one experimental conditions.

One application of pulsed neutron beams is Fast Neutron Resonance Radiography (FNRR) [145]. This technique is used in various research, industry, and security applications. Studies include two-phase flow [146] and contraband detection of explosives [147], narcotics [148] and special nuclear materials [149]. FNRR takes advantage of the resonance absorption of neutrons at specific neutron energies in the few-MeV range (shown in Figure 28) to identify different elements.

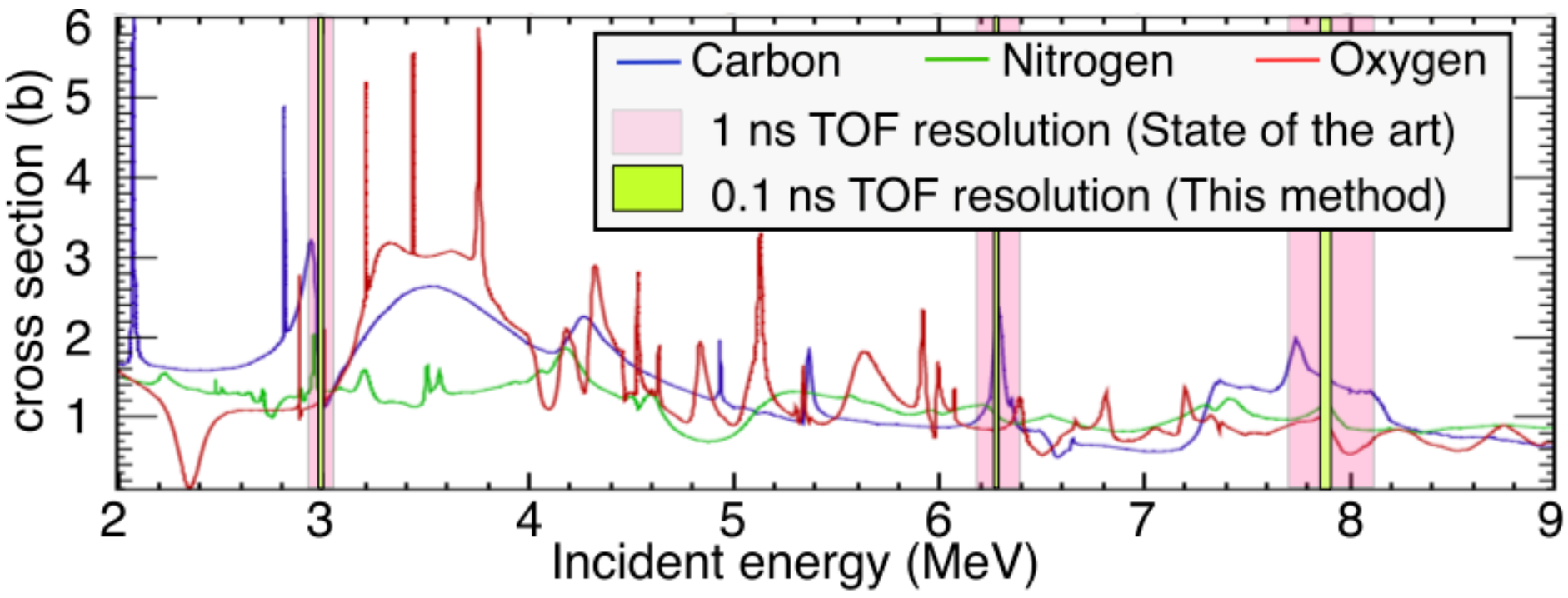

Figure 28 – The total absorption cross-sections for neutrons in carbon (blue), nitrogen (green) and oxygen (red) are shown. The energy resolution for a TOF distance of 3.5 m is shown for three example energy values, for the case of 1 ns (pink) and 100 ps (bright green) timing resolutions. The figure is taken from Ref. [142].

The method is presented in Figure 29. Neutrons generated from a pulsed source with a wide energy range are transmitted through the sample (a). Multiple neutron radiographs (b) are taken at a few time windows, each corresponding to a different neutron TOF and therefore to a different neutron energy bin. Setting these time windows to correspond to specific absorption energy resonances yields radiographs that shows not only the geometry of the imaged sample, but also its material composition (c). Thus it is possible, for example, to distinguish between harmless materials and explosives. The radiograph's contrast and its resolution for distinguishing between different compounds is a product of the detector's energy resolution, which in this limit can only be as good as the uncertainty of the neutron emission time, i.e. the neutron pulse duration. To-date, these studies are conducted at accelerator facilities and are limited to ns time resolution. The vertical bands in Figure 28 indicate the uncertainty in resolving the resonances for a few example energies using a state-of-the-art 1 ns temporal resolution [145, 150] (pink bands) vs. the 100 ps resolution (green bands) that may be achieved with the scheme proposed here.

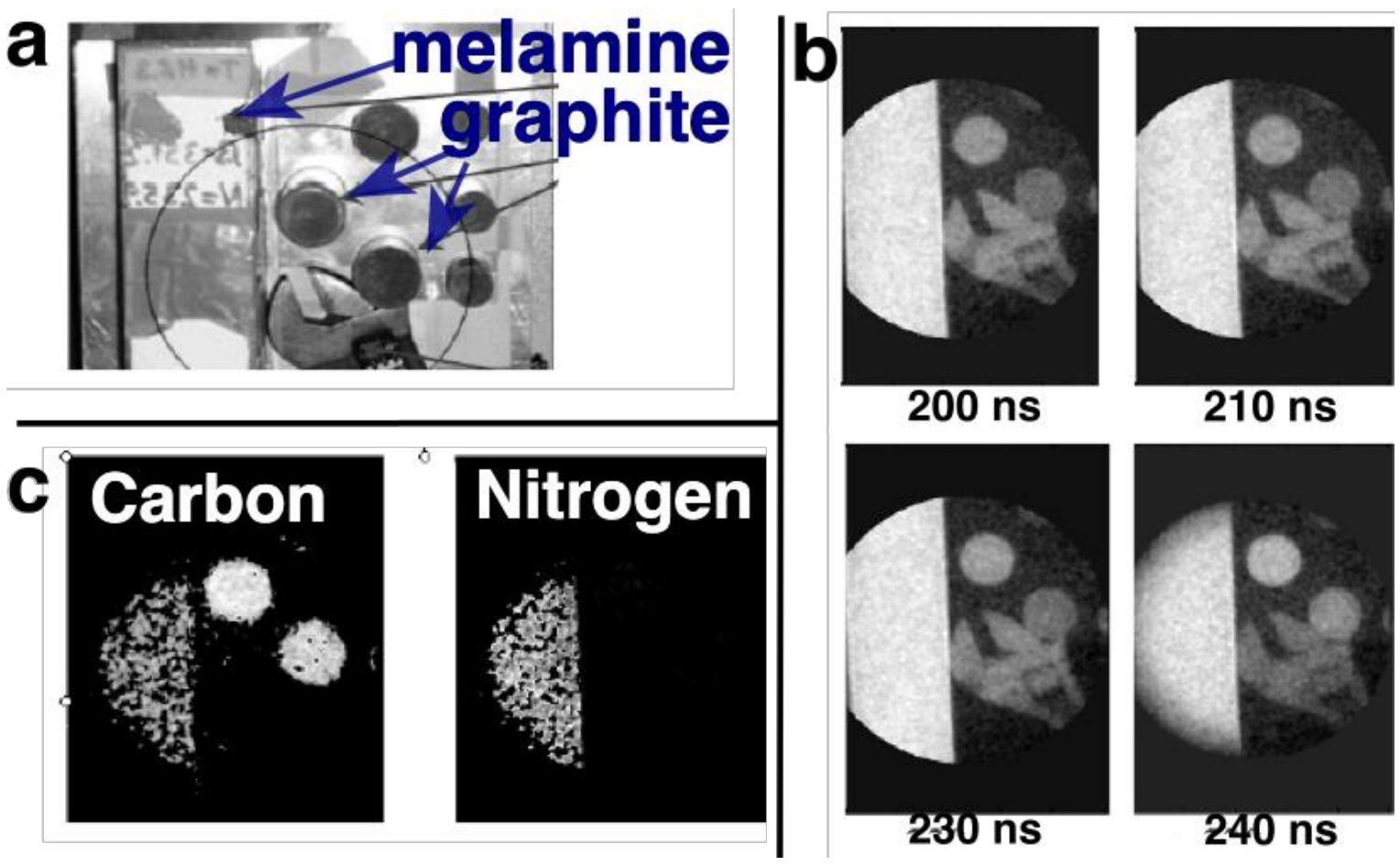

Figure 29 – Accelerator based, ns-level resolution FNRR. Figures adopted from [152]. (a) The sample containing carbon rich and nitrogen rich materials. The circle indicates the region imaged by a neutron detector. (b) 4 neutron radiographs taken at different TOF windows. (c) The reconstructed carbon and nitrogen distributions in the sample derived from these radiographs.

### 2.4.3. Muon-source and muon catalyzed fusion

The muon (μ) is an elementary particle in the lepton family with a short life time of 2.2 μs, until it decays to a positron and two neutrinos. Due to its small Bohr radius of $2.55927\times10^{-11}$ cm and large mass (105.6595 MeV/$c^2$), muons easily overlap with the nucleus and provide valuable information on the structural and polarization features of the host atom [134]. Muons can also play an important role in nuclear fusion [153]. The investigation of nuclear fusion from muonic molecules is of great importance in determining the properties of various exotic nuclear systems. In addition, due to their catalyzing effect, the muonic fusion process can lead to the development of intense neutron sources [154] and nuclear fuel breed systems [155].

A short burst of muons can be produced by bombarding short burst of protons on low Z material. Current facilities around the world (only five – ISIS, UK; TRIUMF, Canada; Paul Scherrer Institute, Switzerland; J-PARC, Japan; and Joint Institute for Nuclear Research, Russia) employ proton beams close to GeV energy for generating muons. However, the muon yield peaks for a proton energy of about 600 MeV and can be significant for lower proton energies as shown in Figure 30.

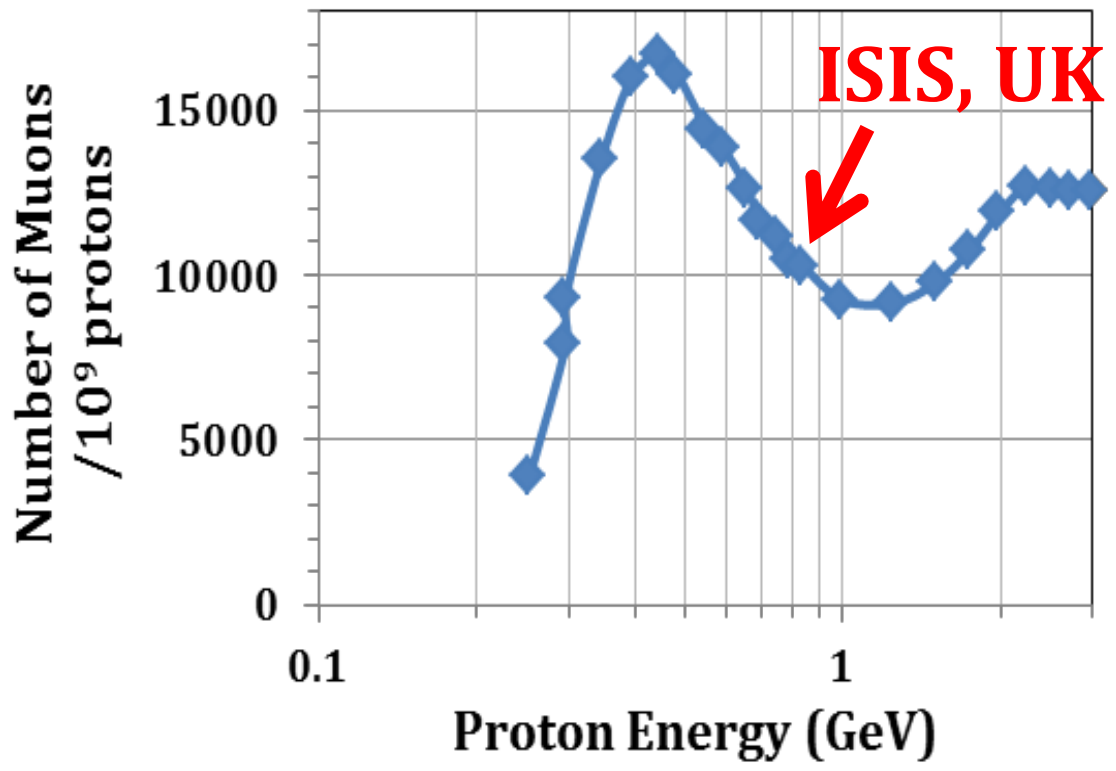

Figure 30 – Total muon yield for different energy protons incident on 5 cm thick graphite target, as reported by A. Bungau et al. [156].

On the basis of current understanding of the acceleration mechanisms and its scaling laws, it may be possible to produce protons in excess of 300 MeV using the planned 10 PW beamline at ELI-NP. Impinging these protons on a low Z secondary target ($^7$Li or graphite), muons can be efficiently produced via natural decay of pions. Figure 31 shows the angular flux distribution and spectra of muons, obtained by FLUKA [157] simulations, carried out for graphite targets irradiated by protons with energies relevant to the E1 target area [158]. The $\mu^+$ are usually produced in the backward direction, with respect to the incident protons, whereas a $\mu^-$ beam is primarily emitted in the forward direction of the target.

The muons created through this technique can be used in different applications, such as Muon Catalyzed Fusion (μCF) and as a probe to study exotic nuclear systems. In the μCF, the negatively charged μ is captured by deuterons producing muonic deuterium atoms (dμ), followed by the formation of ddμ molecules. Since the size of a ddμ molecule is about 200 times less than that of normal molecules, the two deuterons will be enclosed in a small volume within a distance of less than 500 fm (strongly reduced width of repulsive Coulomb barrier [159]) which would dramatically improve the probability of fusion, producing an intense neutron burst.

Muon spin rotation, relaxation and resonance techniques (μSR) provide information on the chemical and physical properties of matter, through a correlation between the emission of the $e^+$ and the spin moment of the decay. This is achieved by measuring the positron emission from muonic atoms created by short μ bursts. The incoming muon burst triggers the clock and the decay positron stops the clock. Studying time-differential positron distribution allows for a detailed analysis of the muon spin ($S_\mu$) dynamics under the influence of the local magnetic dipole at the interstitial site where the muon is captured [160]. Information obtained by μSR studies can help developing applications related to security threat surveys and material studies.

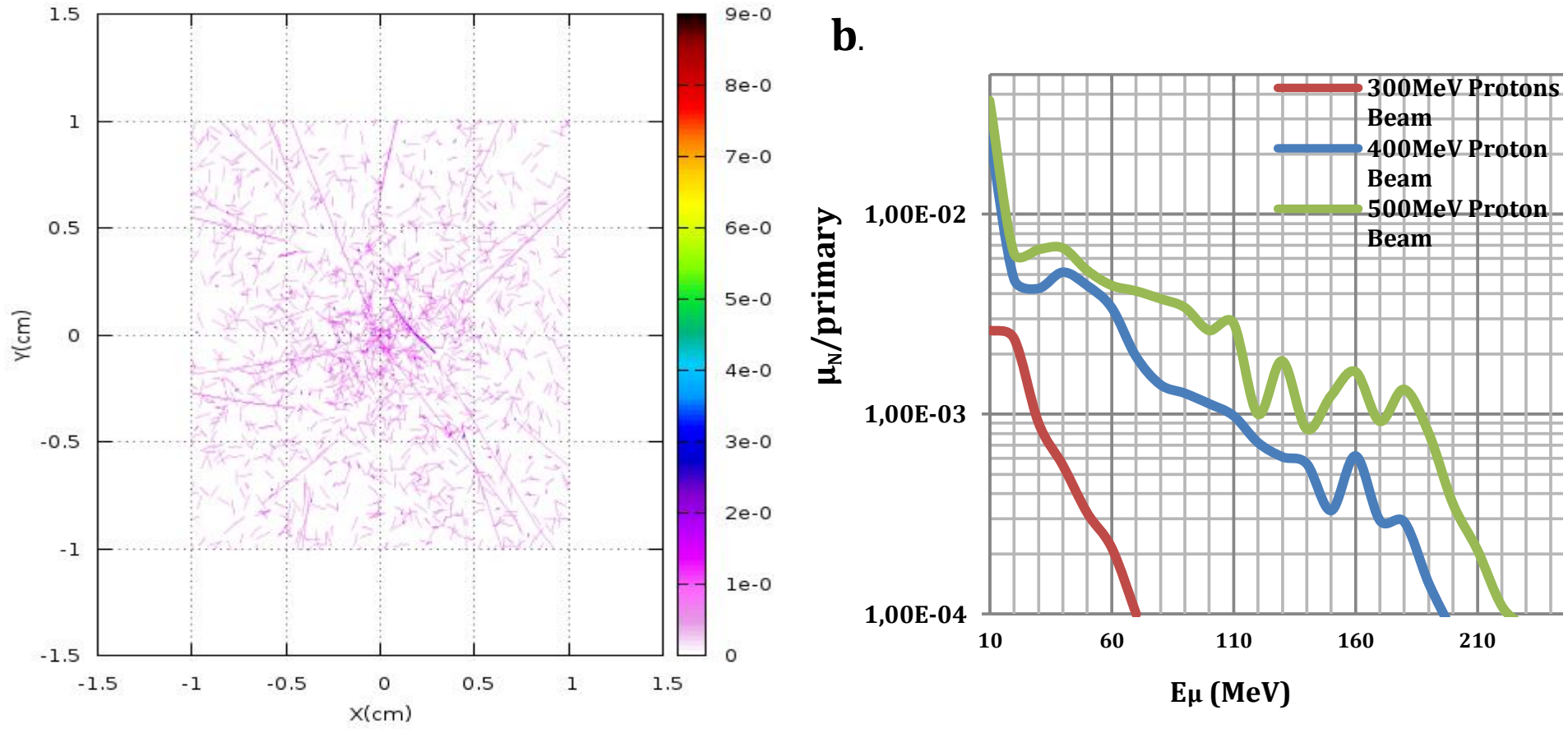

Figure 31 – FLUKA simulation: a) shows the distribution of μ generated in the graphite target (what thickness) with 400 MeV proton beam. b) shows the number versus energy of muons generated at three different beam energies of 300, 400 and 500MeV. The simulation was done for 10E7 protons.

## 3. TECHNICAL PROPOSAL

### 3.1 REQUIREMENTS FOR LASER BEAM CONFIGURATION AND PARAMETERS

The laser configurations described below are the result of a combined analysis of requests from experiments, risk mitigations and cost optimization. A number of 3 major configurations have been defined at this stage, each of them allowing a range of changes in angles and/or focal length of the parabola that will be still compatible with the constraints imposed by the interaction chamber design and the condition of keeping the ions acceleration line along the beam-dump axis.

They are shown in Figure 32. The values for focal length and deflection angles mentioned in this figure have been chosen taking into account the needs of experiments in other areas, trying to minimize the number of different parabolas and allowing interchange of components between experimental setups.

The three configurations can be described and justified as follows:

(a) Perpendicular direction of the focusing configuration means that one 10 PW beam is used for ion acceleration tightly focused onto a solid target, while the second 10 PW beam, that can also have a long duration (uncompressed) pulse of about 300 J and 1 ns, will be used for gas plasma heating, or microlens charging, or a high B-field production. The distance between the two foci can be varied (by at least 0.5 m) as well as the timing. The acceleration beam will be available with circular polarization.

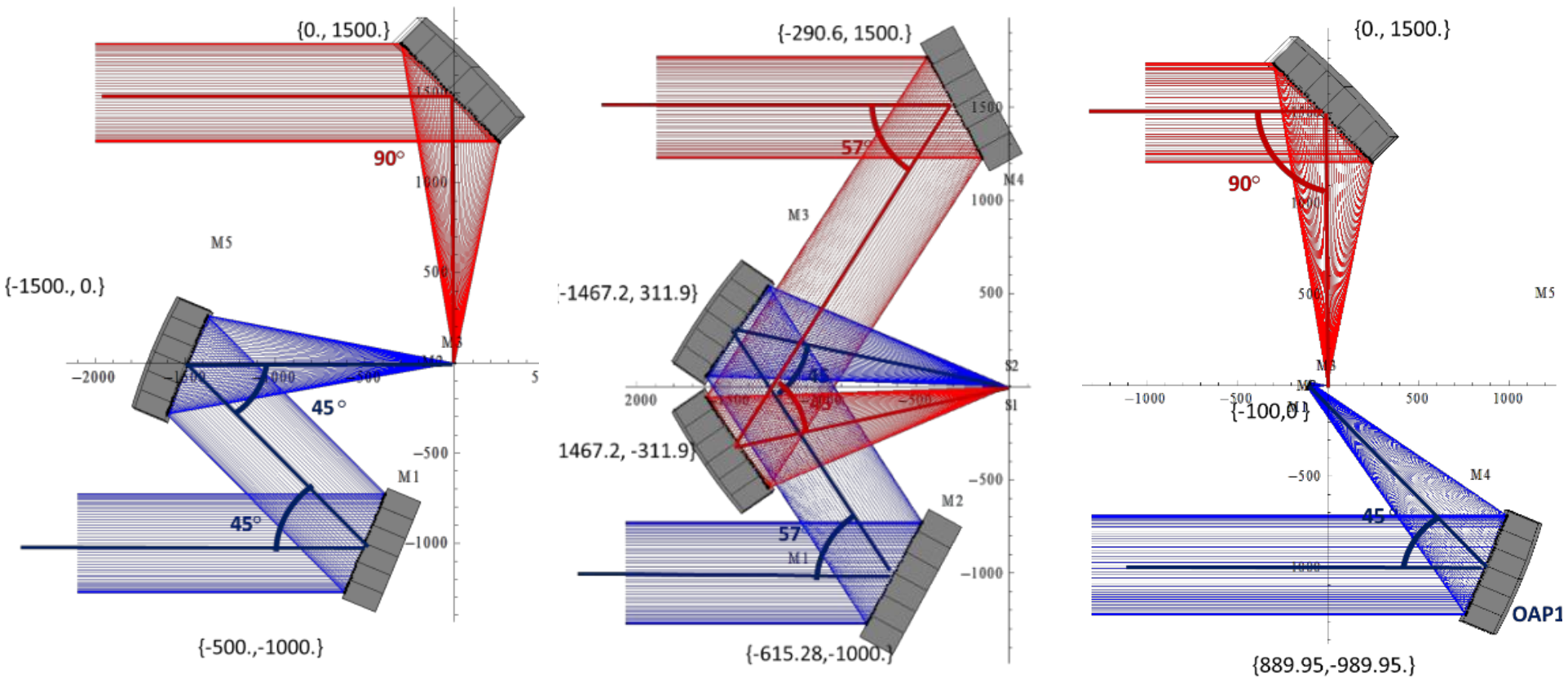

Figure 32 – Laser beam configurations proposed for experiments in the E1 area (see text for details) shown in a top view and target position placed in the origin of the axes. Axis units are mm. The diameter of 10 PW beams is 550 mm and the off-axis parabola focal length is 1500 mm.

(b) The small angle focusing configuration is meant to provide maximum possible intensity on target by combining the two 10 PW pulses. Circular polarization for both beams is requested. However, the Laser Beam Delivery (LBD) TDR is proposing to install in the first stage the polarization control system only to one 10 PW arm. This system integrates also a deformable mirror, such that in the first stage the second 10 PW arm will have neither polarization control nor adaptive optics after the compressor.

(c) The configuration with a plasma mirror (PM) before the target will be probably the first to be implemented, in order to diminish the pollution of the off-axis parabola (OAP) with target debris through a shielding (not shown in the Figure). A second laser pulse of 10 PW or uncompressed of 300 J and 1 ns, perpendicular to the acceleration direction will be available in this configuration. The PM may be requested also by experiments where very high temporal contrast is mandatory. The HPLS is supposed to provide contrast levels of $1{:}10^{12}$ in the ns and ps domain, however, the rising of beam power in the range below picoseconds in front of the main peak can be steepened only through a PM. However, the diameter of the (elliptical) beam spot on the PM will be of order of few centimeters, implying large costs and large dead times for replacement and realignment. We mention also that all HPLS outputs provide pulses in *p* polarization, which is less adapted for a PM compared to *s* polarization. Small angles of incidence are needed in *p* polarization in order to increase the reflectivity of the PM. We mention that the diameter of the (elliptical) beam spot on the PM will be of the order of few centimeters, implying large costs and large dead times for replacement and realignment.

### 3.2 E1 INTERACTION CHAMBER

Several options for the interaction chamber design have been considered. Due to the size of the beam and of the mirrors, the option of a small chamber with focusing mirrors outside was considered to limit too much the flexibility. A polygonal chamber, following the example of 2-m diameter Titan target chamber, adopted also by Apollon/Cilex, will lose the advantage of easy access to the TC center (TCC), when scaled up to allow implementation of all the three configuration in Figure 32. The second advantage of the polygonal chamber is that diagnostics installed on the lateral flanges are directing towards the TCC where the radiation is emitted from.

However, this advantage is not of big importance for E1 experiments because:

- due to the IC dimensions, it will be possible to install many diagnostics inside the chamber
- the nuclear reaction of interest will occur in a secondary target (e.g. plasma target in the focus of a second laser pulse) that will be installed sometimes far downstream the primary (acceleration) target, in order to perform some beam purification or energy selection. In other cases, enlarging this distance will be needed in order to allow for shielding the detectors from radiation against radiation emitted from the primary target.

Rectangular chambers are used in RAL for Vulcan and Astra Gemini laser systems.

The favored option for E1 is therefore a rectangular chamber with dimensions of about $3\times4$ $m^2$, to be optimized when the OAP motorized supports will be available. Still, these dimensions imply that the configuration with a plasma mirror (shown in subfigure *c* of Figure 32) will work shifting the interaction point backwards by a minimum of 0.5 m. The height of the chamber will be about 2 m, leading therefore to a volume of about 24 $m^3$.

The main functional requirements for the E1 IC are:

- working vacuum level $10^{-5} - 10^{-6}$ mbar
- pump down time to $10^{-6}$: max. 60 minutes (in the conditions of an empty chamber)
- Primary (acceleration) target position is not in the center of the rectangle. Direction of acceleration is perpendicular to the faces with longest dimensions (chamber front face toward beam-dump). Secondary target can be located downstream on acceleration direction at a distance of up to 50 cm.
- coupling of the two 10 PW laser beam lines on the back side with DN800 flanges
- for access inside IC: several door-flanges will be installed. However, the big E1+E6 area is not a clean room (actually due to radioprotection restriction this area will be in under-pressure compared to adjacent corridors or experimental areas), which imposed a local soft-wall clean room attached to one the doors of the IC, such that only this door will be used (including for large OAPs).
- modularity: the IC will be equipped with few large rectangular flanges each having 2 or 3 intermediate dimension rectangular flanges, holding a number of standard dimension (from DN25 to DN250) ports, possibly having inclinations such as to point towards the target position. When other configuration of ports will be needed, only some of these rectangular flanges will be changed. The possibility to exchange these intermediate dimension flanges with other ICs will be foreseen.
- At forward angles a square flange with dimensions of about $1\times1$ $m^2$ will be installed (Nuclear Technologies, for radioprotection considerations, suggested it to be made as thin as possible from Inconel-718) centered on the acceleration direction that could be replaced by an extension box of about 1 $m^3$, allowing for the placement in vacuum of additional diagnostics in forward directions
- access on top of the IC should be provided because the exchange systems (manipulators) for targets and diagnostics will have a load-locked box located above the primary target (and/or above secondary target in direction of acceleration)
- the optical table supporting the mirrors, targets and some of diagnostics will be decoupled from the chamber. The distance between the optical table and the chamber wall should be about 10 cm. Expected height of the optical table is 800 mm while the laser beam axis is 1500 mm.
- connection with the primary vacuum system could be done on the bottom side or back side of the chamber. The high vacuum pumps (turbo molecular and cryopumps) should be placed under the chamber (as first option) or on top of the chamber (second option) taking into the need to access the load-locked box for target exchange.

Materials and design should consider activation minimization. Basic choice is Aluminum. EMP containment inside the IC has to be taken into account thoroughly in design.

Compatibility of flanges and ports with other laser ICs has to be assured.

A proposed design of the E1 IC is shown in Figure 33. The lid is intended for removal only for the installation of the optical table inside the chamber. The intended vacuum seal for all large openings is a differentially pumped dual o-ring. The chamber, lid, and removable walls construction are of aluminum alloy, ribbed to be lighter with high rigidity. The lid provides ports for the high vacuum pumping and safe access for targets handling.

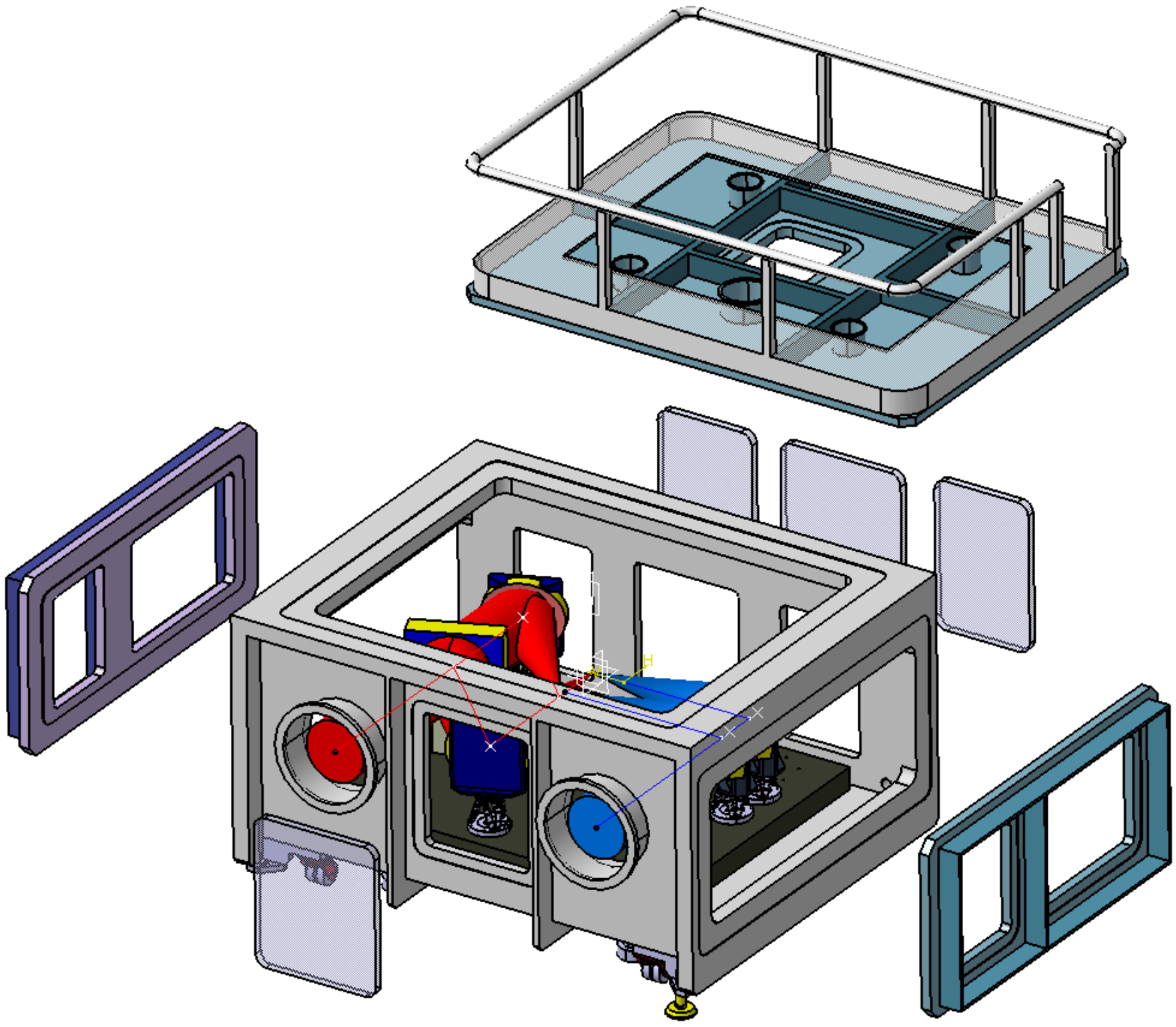

Figure 33 – Proposed design of the E1 Interaction Chamber

The requirements on the interaction chamber for the *nuclear reaction in plasma* experiments should be ideally a spherical Aluminum-ERGAL chamber, with a wall thickness of about 1 cm to minimize neutron absorption. This is difficult to be achieved. Instead, thin windows in the direction of neutron detectors can be foreseen for the E1 chamber described above.

### 3.3 TARGETS AND TARGET SYSTEMS

Thin and ultra-thin solid targets are requested by most of the experiments. Some experiments may ask for structured solid targets (such as cylinders, cones, coils attached to targets) to control the acceleration and optimize the yields. Secondary solid targets are also requested to produce the desired nuclear reactions or nuclear states.

Primary gas targets (for ion shock acceleration) or secondary gas targets (for plasma-target formation) have to be also considered in a wide range of densities. The absence of target debris and replacements/realignments are important advantages of gas targets. However, the radioactivity induced in gases has to be evaluated, monitoring implemented and gas collection inside pressure vessels for temporary storage before release might be needed.

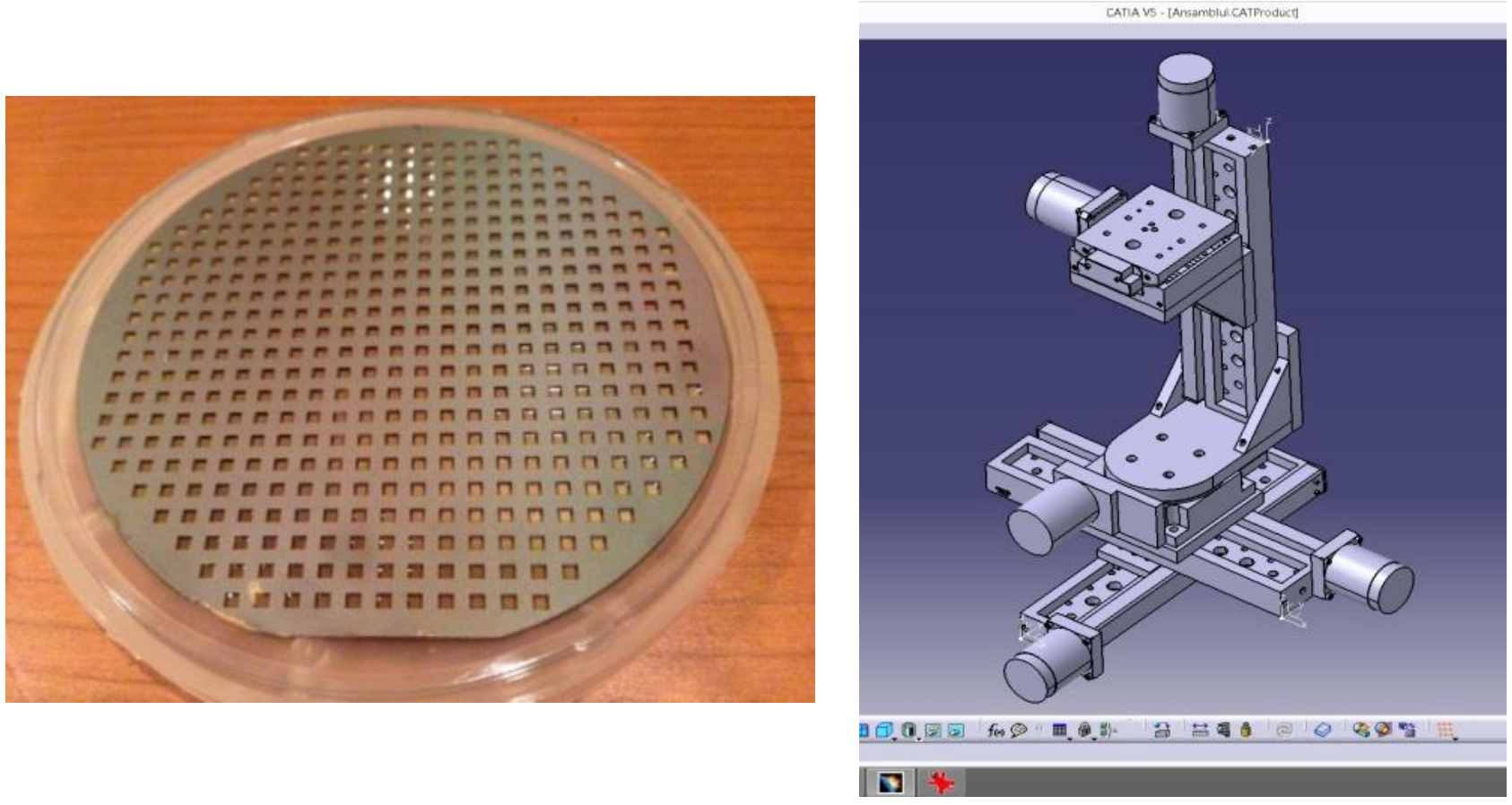

Figure 34 – Si wafer with ~1000 ultrathin SiN windows and a 5-axis micro-positioning system

The requirements for targets are very similar with those expressed by the High Field QED TDR with two complementary remarks:

- Proton removal from the target foil is essential for heavy ions and for the reduction of activation levels. It has to be done in-situ, before the laser shot (e.g. by heating or by a ~1 W laser irradiation), according to a procedure to be defined.
- The number of needed targets will be ~300/day for experiments in E1 and $>10^3$/day in the E5/E4 areas. A raster target with dimensions up to 150 x 150 mm, as shown in Figure 34, mounted on a 5-axes micro-positioning system with automatic alignment should be considered as standard choice for most of the experiments. More than $10^3$ multi-targets (with μm – nm thickness) can be accommodated.

➢ A tape target solution as depicted below has to be considered as cost effective and satisfactory for experiments requiring above micrometer thick target (commercial solutions are available)

➢ The liquid crystal target produced in-situ [161] seems to be an adequate solution for high repetition rate lasers, however, they are useful mainly for proton acceleration

➢ The same remark is valid in the case of cryogenic targets (H and D)

The target exchange system should allow for the change of a primary or secondary target, or even some diagnostics such as RCF stacks, without opening the large volume E1 IC. This will be achieved by a load-lock vacuum chamber placed on top of the IC and equipped with a motorized vertical motion linear stage. A prototype is under design, to be followed by its realization and delivery for test at the CETAL facility.

### 3.4 ACTIVE DETECTORS AND FAST DIAGNOSTICS. EMP DAMAGE MITIGATION

Due to the high laser pulse repetition, active detectors/diagnostics are of high priority for laser driven experiments at ELI-NP. In this category we include:

- Thomson parabolas with plastic scintillators (or microchannel plates (MCP) followed by phosphorous screen) viewed by gated charged-coupled devices (CCD) or photodiodes arrays such as the Radeye1 sensor
- $LaBr_3$ and liquid or plastic scintillators for gamma ray detection, respective, neutron spectroscopy
- Electron spectroscopy is also important to understand the heavy ion acceleration mechanism
- Plasma diagnostic is fundamental to characterize the plasma conditions (temperature and density). The main requirements are the X-ray diagnostic (imaging and spectroscopy), Interferometry, VUV imaging and spectroscopy. A low energy probe beam is also considered for plasma diagnostics

Because these detectors are active during the laser pulse, they and their associated (analogue and digital) electronics are subject to potential damages produced by an EMP. Some of them are placed in vacuum, where the EMP can be very high. Since the EMP amplitude at ELI-NP cannot be estimated a priori, it was proposed to implement strong filtering and shielding for the building, because here it will be difficult to add something at later stages, while at the level of experiments initially a reasonable starting solution will be chosen, which could be reinforced if it would appear necessary during first experiments. We will learn also from first experiments at CETAL in 2015 and at Apollon/CILEX in 2015/2016. Enclosing the detectors in metal cases, use double shielding coaxial cables for signal transmission and filters that cut frequencies higher than those used by detectors, along with the installation of electronics in EMP-shielded racks, are the main solution to be applied when optical transmission is not possible, as depicted in

Figure 35. Reduction of the EMP has to be implemented through various methods, as detailed in the ELI-NP "EMP Shielding and Damage Mitigation" Report.

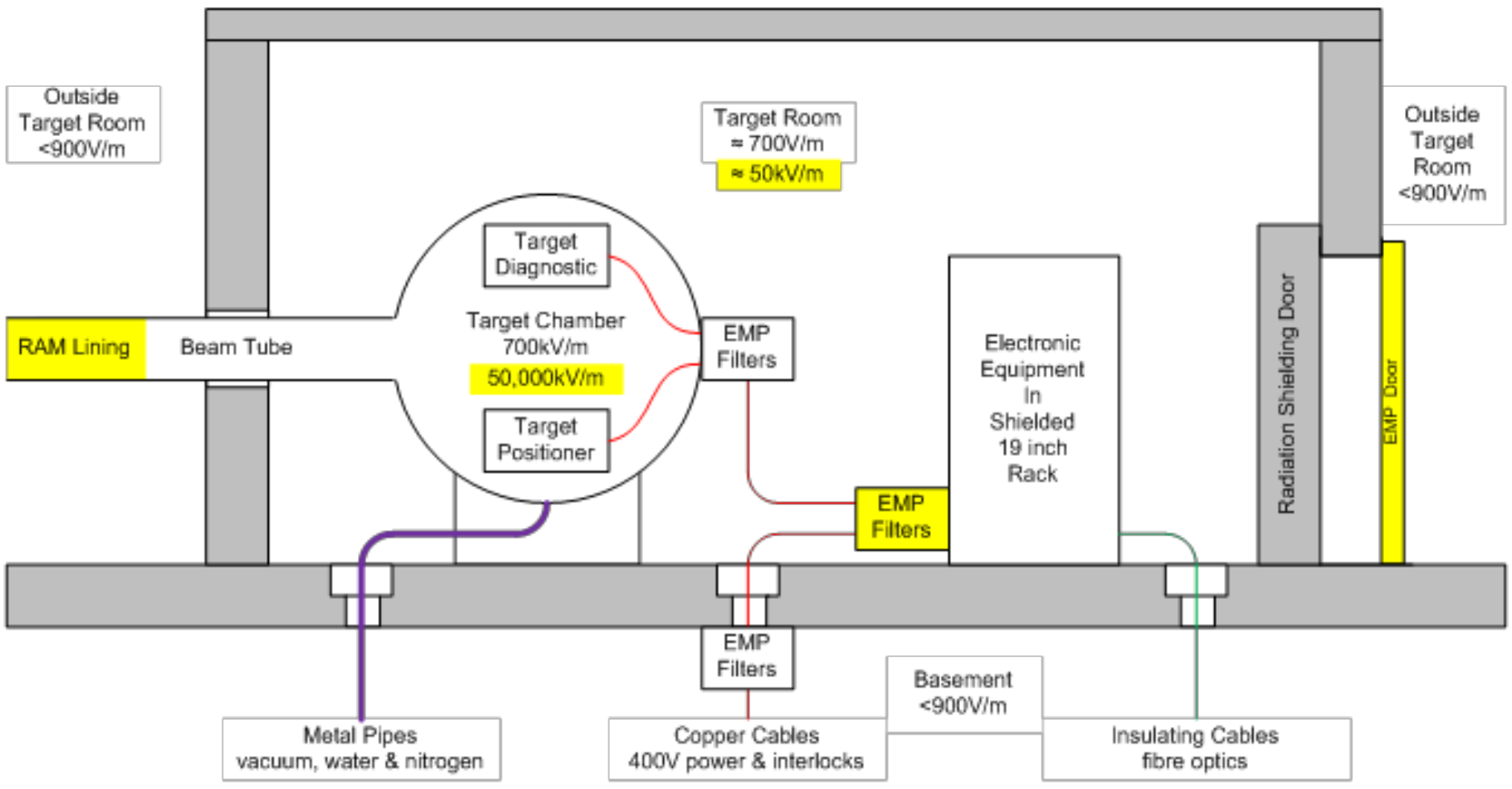

Figure 35 – EMP shielding strategy. Values on yellow background correspond to an EMP estimation based on a scaling with laser power, giving the upper limit.

### 3.4.1 "In-situ" Gamma detectors

The test performed at the ELFIE-100 TW facility with $LaBr_3$ scintillator detectors (see Figure 36), placed at just few centimeters from the target proved that dynode based gated photomultipliers can be used for in-situ gamma spectral measurements down to few milliseconds after the pulse. Better Pb shielding and an adaption of a reference LED signal are promising to provide access to much shorter times. A number of at least 4 detectors of this type are proposed to be developed.

The use of MCP based photomultiplier was less successful due to a strong saturation and long-time recovery of this type of detectors.

Gamma detectors, able to measure the spectrum of the prompt X-ray flash, will be of great importance. Two solutions are under study:

- a foil to produce Compton electrons followed by an electron spectrometer.
- a stack of several high-Z scintillators, possibly interlaced with high-Z materials.

The first option should be considered as a high resolution option, with low efficiency. The last one will have high efficiency but the energy spectrum will be obtained in only few bins.

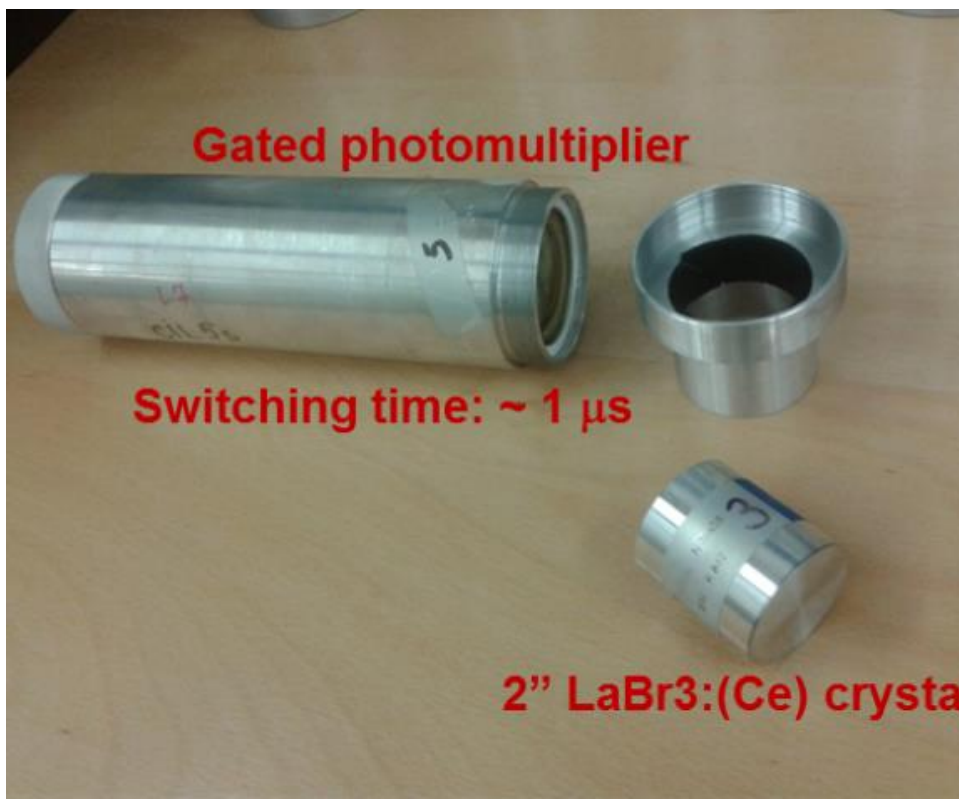

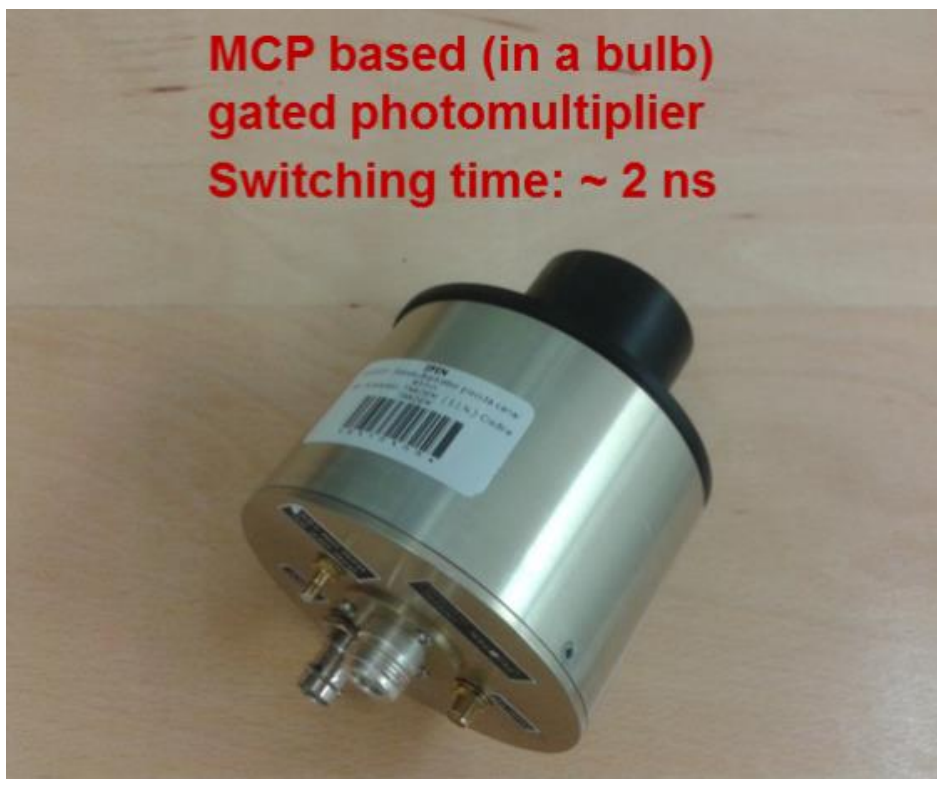

Figure 36 – Left: Dynode based gated-photomultiplier and $LaBr_3$ crystal scintillator. Right: same crystal coupled to an MCP based photomultiplier

### 3.4.2 Neutron Detector and SiC wall

The studies of nuclear reactions in plasma require the construction of a highly segmented detection system for neutrons. The segmentation is needed in order to reconstruct the reaction's kinematic, thus getting information on the center of mass energy distribution of the nuclear cross-sections. The "ideal" detection module must have: high efficiency, good discrimination between gammas and neutrons, good timing performance for TOF neutron energy reconstruction. In addition, it must be able to work in harsh environmental conditions, like the ones established in the laser-matter interaction area.

Very recently, the possibility of manufacturing plastic scintillators with efficient neutron/gamma pulse shape discrimination (PSD) was demonstrated at the LLNL laboratory [162] by using a system of a polyvinyltoluene (PVT) polymer matrix loaded with a scintillating dye, 2,5-diphenyloxazole (PPO) and at INFN-PD by using polysiloxane [163]. First characterization results show that PSD in plastic scintillators can be of similar magnitude or even higher than in standard commercial liquid scintillators. The result is a consequence of the large amount of scintillation dying material used in the polymer, a possibility never tested in the past.

Another recent result obtained by our collaboration is the implementation of new photo-detectors based on silicon technology (Silicon PhotoMultipliers SiPM) [164]. These are now commercial devices, characterized by a high photo-detection efficiency, high gain, single photon sensibility, excellent timing performance, low operative voltage and insensitivity to large electric and magnetic fields. These make such devices particularly suitable for applications in severe environmental conditions, such as those ones foreseen around the laser-matter interaction area at the future ELI-NP facility.

A third important aspect is the signal processing, for which a relevant expertise has been developed within our collaboration [165]. We propose to implement a totally digital acquisition of the multi-hit signals foreseen in the proposed physics case. This is based on the use of commercial digitizers, developed in collaboration with the CAEN (Italy) company, and ad-hoc read-out software.

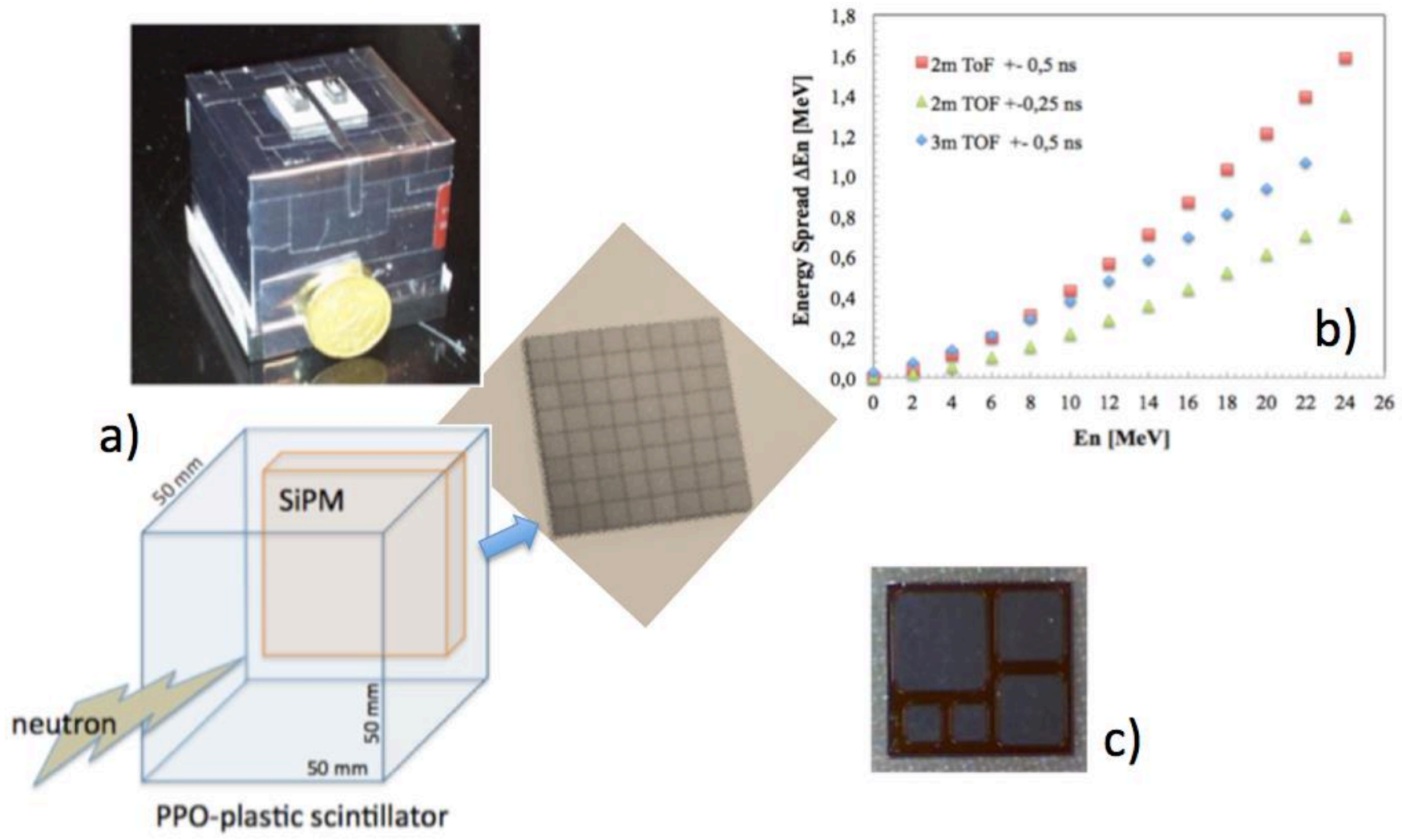

Figure 37 – Left: Neutron detection system. Single module configuration based on 50×50×50 mm PPO-Plastic scintillator and SiPM read-out. Right: Neutron energy spread as a function of neutron energies and for 2 m and 3 m TOF base-line and for 0.5 and 0.25 ns of time resolution.

These basic technologies will be the pillars of our proposed detection system (see Figure 37). It consists on an array of about 400 modules, each including one 25 $cm^2$ area scintillator, 5 cm thick, one SiPM and a digital read-out channel. This allows for a modular structure with easily adapted configurations around the interaction area. As an example, at 2 m distance the total efficiency is estimated as large as about 18 % for 2 MeV neutrons and 6% for 13 MeV neutrons [166]. In such conditions, the array can detect up to about $10^5$ neutrons per shot, which represents a challenging demand for measurement at ELI-NP. Efficient digital shape analyses can handle a multi-component folded signal, still preserving the timing of each neutron detection and the n-γ discrimination. In Figure 37 it is also shown the neutron energy spread as a function of neutron energies, for 2 m and 3 m TOF base-line and for 0.5 (standard) and 0.25 (expected of our module) ns of resolution. We are confident to get in the worst case a neutron energy resolution of the order of 3% at the higher energies.

Concerning the charged particle detectors, INFN is funding the R&D activities on SiC detectors in collaboration with CNR-IMM Catania. The SiC detectors have been proven recently to have excellent properties [101] in terms of high energy and time resolution, resistance to radiation, insensibility to visible light etc. Then, they can be considered as ideal candidates in order to realize a "wall device" to detect charged particles, in some case to be used also in coincidence with neutron detectors, for the ELI-NP experiments. This is what we propose for the study of the $^{11}B(^{3}He, d)^{12}C^{*}$, where the only position and energy measurement of light charged particle can give access to the desired information.

### 3.4.3 Diagnostics for electromagnetic fields

The interaction of ultra-intense laser pulses with the targets will generate electromagnetic fields across the entire electromagnetic spectrum, from the radiofrequency and THz range to the X-ray and gamma ranges, through a large number of mechanisms.

Aside detection tools that would determine the laser accelerated beam properties (kinetic energy distribution, beam divergence etc.), this sections refers to the tools that would probe the laser-target interaction locally. In principle for every laser shot we can record simultaneously a shadowgram, interferogram, Faraday rotation, self-generated harmonic polarization, and secondary radiation (X-ray, THz etc.) as all provide complementary information.

In particular, for the nuclear reaction in plasma studies, a variety of diagnostics equipment is considered to be installed in order to characterize the plasma in terms of energy content, temperature, density, expansion velocity, directivity etc. X-ray and UV-Visible cameras, ions collectors/spectrometers etc., which are the most common diagnostics used in the laser-matter interaction experiments in order to characterize the plasma (temperatures and densities). In addition, in order to monitor the plasma dynamics, we will require the use of two auxiliary probe laser beams, transversal to the main: a) a short, UV probe pulse, (e.g. sub ps, 100 mJ, 200nm) is mandatory to investigate the plasma recovering of the density via the Abel inversion of the plasma induced phase shift; b) a second auxiliary pulse can be used as X-ray "back-lighter", via conversion into K-alpha X-rays, Betatron radiation, Thomson scattering or high order harmonics.

#### *3.4.3.1 Polarimetry*

Self-generated magnetic fields in the laser driven plasma [167, 168] can reveal the dynamics of electrons. Thus, for example magnetic fields due to density and temperature gradients (finite spot effects, thermal instability or filaments) can be important to monitor, as the generation of hot electrons can inhibit the RPA acceleration mechanism. On the other hand, at >1PW laser magnetic fields generated in the plasma have an interest in themselves, as they can be large enough to perturb the quantum motion of electrons in the atom. The method to be used to image the magnetic fields in the plasma is the polarization analysis of the collected harmonics due to laser plasma-interaction. Thus the method is based on the measurement of the depolarization of the generated harmonics propagating through the plasma [169].

Setup items: quartz chamber window (cut off 190 nm), narrow band interferometric filters, beam splitters (neutral density filters), linear polarizers, CCD camera, quarter wave plates, ellipsoidal mirrors

*3.4.3.2 Optical transition radiation (OTR)*

Optical transition radiation as a diagnostic tool will be used to show the presence of currents of modulated electrons, the occurrence of electron beam filamentations or other electron beams in the target. It also gives a micron size view of the laser intensity profile of the laser.

Setup items: CCD camera, bandwidth filters at second harmonic

*3.4.3.3 Terahertz emission detection*

The dynamic process of space charge formation (high density sheet of relativistic electron and respectively ions) is expected to be followed by transient terahertz emission. This was shown in the case when the electron transient sheet forms at the target plasma surface [170, 171]. The strongest THz pulses ever obtained, were reported in these studies. As these rays are emitted noncollinearly, they can be collected separately without affecting the generated ion beam path. Also, they can be used to image in real-time the radiation effects on certain materials and biological tissues.

We propose developing single-shot terahertz transient instrumentation with femtosecond resolution as online monitor of particle density for the ELI-NP laser experiments. The process of particle acceleration from the high power laser matter interaction (in the TNSA mechanism) is followed by coherent transient emission of electromagnetic radiation in the terahertz frequency range. The terahertz field emitted at the rear side of the target, non-collinear with the particle beam, scales with the ion beam particle number. Consequently, optimization of the terahertz emission or of the relativistic charged particle beams will be equivalent. Terahertz radiation can be collected using an elliptical mirror, collimated and taken out of the interaction chamber through a silicon window. Subsequently, the power can be monitored using a pyroelectric detector, and the pulse measured using a single-shot setup that takes use of the synchronized probe beam. The single shot measurements are based on space-to-time conversion: an echelon mirror with 2 μm steps will convert the probe beam into a series of pulses separated in time by ~10fs. These pulses focused on the same spot on an electro-optic crystal and will sample the THz pulse over the entire ps duration (see Figure 38 below).

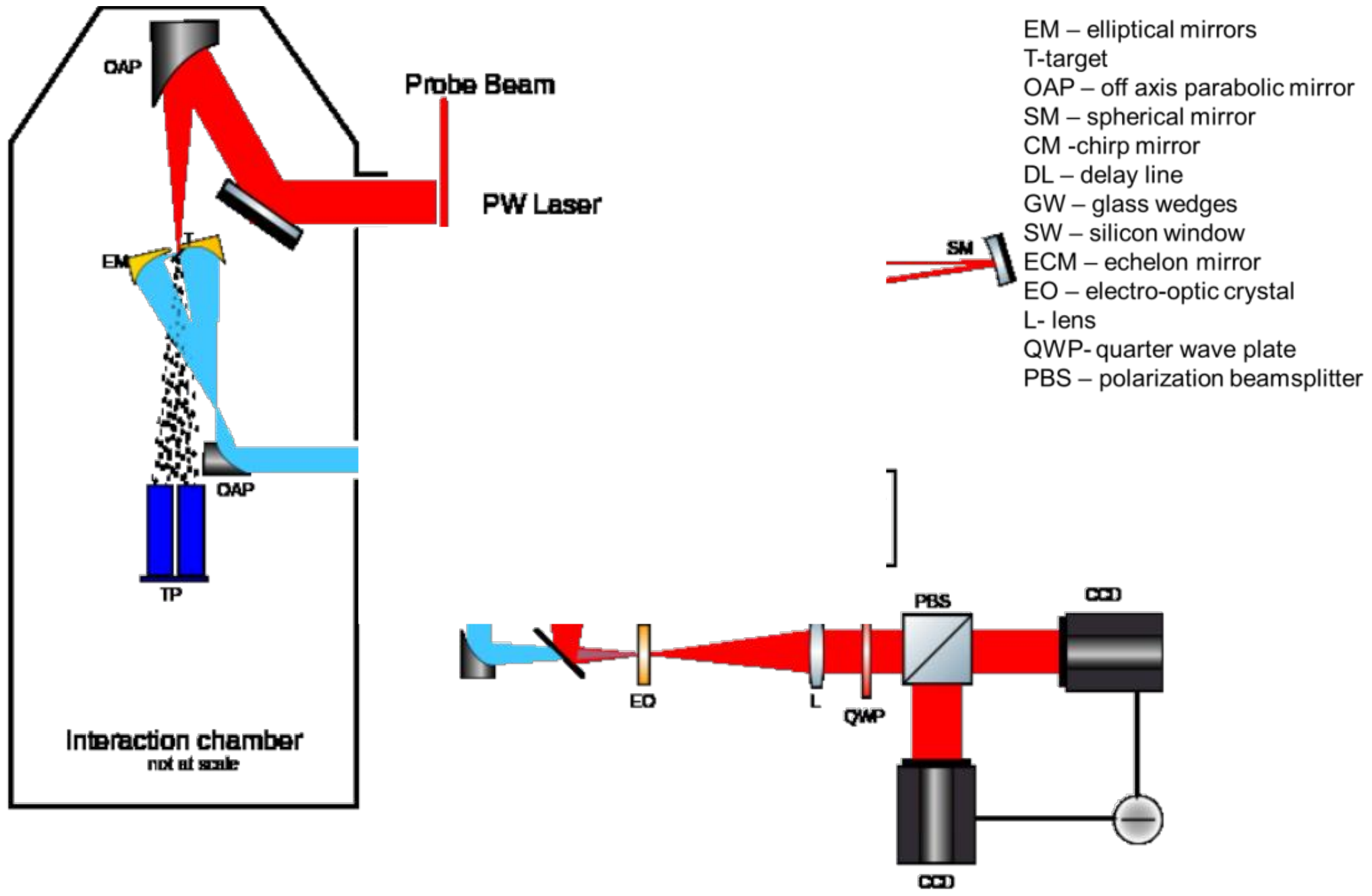

Figure 38 – The sketch of the setup for high temporal resolution THz radiation analysis.

### *3.4.3.4 Optical probe beams for imaging of plasma evolution*

Optical diagnosis tools at ELI-NP are necessary in order to monitor and explore the charged particle acceleration processes in-situ. Together with the ion/electron beam spectrometer, optical probe pulses will explore the laser-target parameter space. These tools are the only tools with access to the time (fs) and the spatial (µm) scale of the laser plasma interaction and provide real-time information on the relevant mechanisms responsible for the generation of the ion/electron beams. This will help in the optimization of the laser and target parameters for an efficient and controllable ion/electron acceleration. Since they are drawn from the main laser (as close to the interaction as possible, to mitigate jitter between shots), it will be easier to synchronize the optical probe beams with the driver pulse. Transverse probing methods, such as interferometry or polarimetry, will produce an image at certain time delays to the driver shot, while longitudinal probing methods, such as frequency domain holography [172] and frequency domain shadowgraphy [173], will create an average over the hole probing time. Transverse probing will benefit from improved temporal sectioning by converting the probe to a few cycle pulse [174].

Interferometric methods are generally based on Mach-Zehnder [175] or Nomarski configurations. Wavefront sensor based measurements [176] can be used for an improved spatial and phase stability. These methods will be used for gas targets (electron densities two order of magnitude lower than the critical density < $10^{19}$ cm$^{-3}$) or to probe the plasma electron expansion in the initial stages of the laser-solid interaction [175]. Electron densities down to $5\times10^{18}$ cm$^{-3}$ and resolutions of ~ 4 ps and ~ 6 μm can be achieved.

The typical setup, for plasma expansion analysis in case of thin film targets, is shown in Figure 39 below.

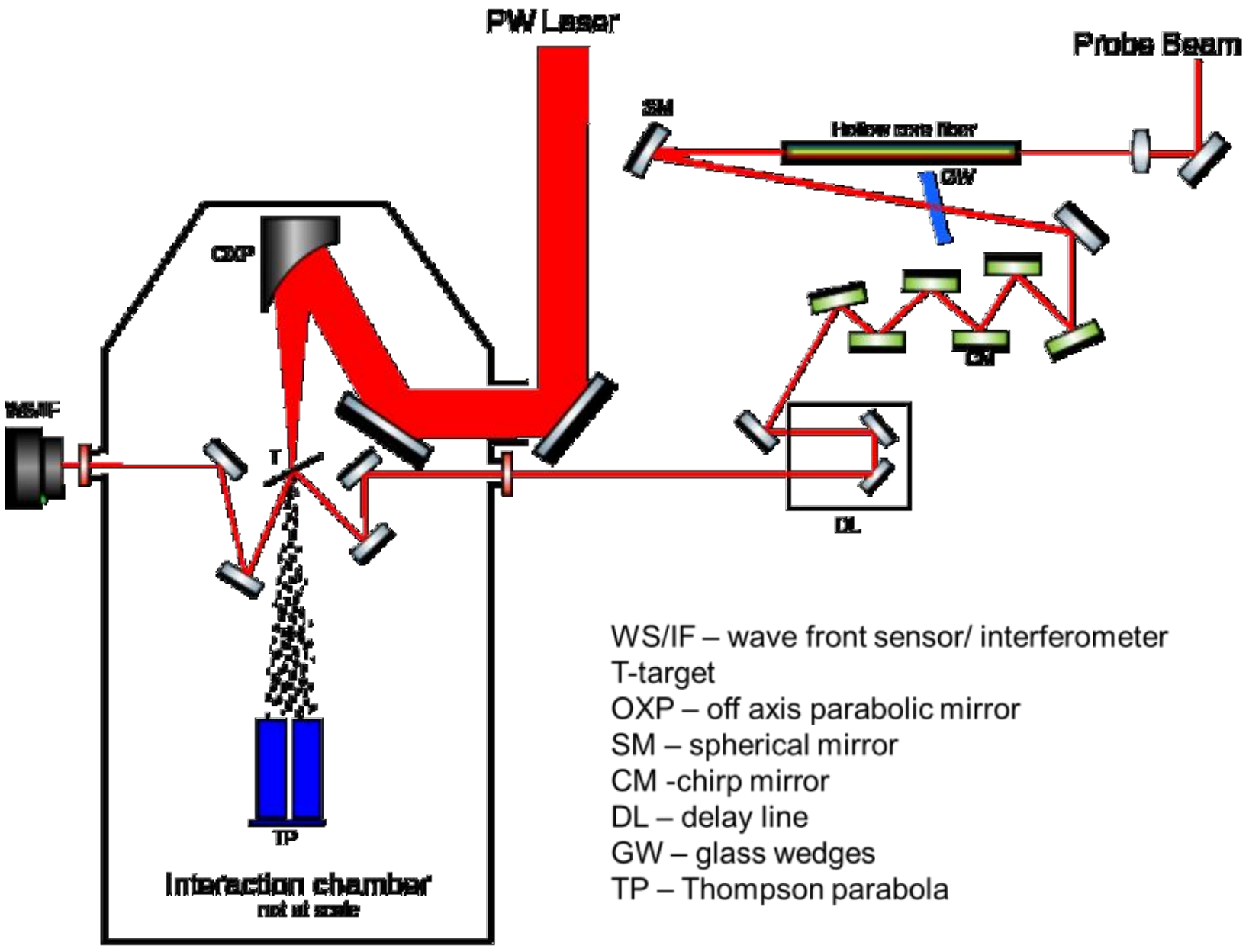

Figure 39 – Typical experimental setup using an optical beam probe for plasma evolution analysis.

In this setup, as well in the previous one, the probe beam is low fraction, of only few mJ, split from the main beam upstream the interaction chamber. Careful design of optical path is required to achieve proper timing at interaction point.

### *3.4.3.5 Diagnostics for XUV spectral range*

One important spectral area of interest covers the wavelength range from 100 nm down to the water window around 4 nm also known as UV to the soft X-ray spectral window (XUV).

Specific sources in the XUV spectral range are high order harmonics generated in the interaction of ultraintense laser fields with gaseous targets or with solid targets. They are used in the generation of attosecond pulses needed as probe beams (e.g. for diagnosing higher density plasma regions then permitted by visible light) or they are analyzed as source of information on the interaction of the laser pulses with the target (for example they can be used to assess the temporal contrast of the pulses). Further sources in XUV range are plasma X-ray lasers; some proposals for ELI-NP include such laser developments for Compton scattering studies. Finally, incoherent emission of the plasma provides valuable information on the state of matter in various experiments. Hence imaging and spectroscopy tools in the XUV range have to be available at ELI-NP.

The XUV spectral window is particularly difficult from the optics and detector point of view. The materials have strong absorption and reduced reflectivity. Specific spectroscopy and imaging devices are available on the market. Some of these are considered for the pool of detectors at ELI-NP. For imaging and spectroscopy, back illuminated cooled CCDs down to -70 °C degree are available from providers such as Andor, Roper Scientific, Photonic Science etc.

EUV and VUV cameras provide resolution in the range of 10 – 24 microns, high dynamic range up to 16 bit when cooled, as well as good temporal resolution of microseconds for small region of interest (ROI). They are available flanged and/or in vacuum versions, depending on the flexibility required for characterization and inspection. The vacuum version requires liquid cooling systems, while the flanged ones can work with air cooler only. Sensitivity response can be 100 nm up to 0.0135 nm. Deep cooling allows for very low dark currents, down to less than 0.005 electron/pixel/sec for extended exposure with low background.

Due to their broad spectral range, sometimes such XUV CCDs are used as large detectors for characterization of keV photons emitted from the plasmas. For imaging of the plasma, the same XUV CCDs are used in conjunction with pinholes having thin film XUV spectral filters and they are essential when XUV imaging of the plasma is required.

XUV cameras are used often in conjunction with XUV diffractive elements such as gratings, in order to provide spectral analysis of the XUV source. The grating developments reached a very high technology level, due to the strong requests of the synchrotron community. Grazing flat field gratings (offered by companies such as Horiba or Hitachi) can cover more than two spectral octaves, while providing a spectral resolution of the order of $10^{-2}$ and better.

Turnkey XUV spectrometers also exist on the market, such as the ones from McPherson. Additional chambers, cooling systems motion and positioning and filters are needed.

*3.4.3.6 X-ray deflectometry diagnostic using betatron emission backlighting*

We propose to implement at ELI-NP an X-ray radiographic system for deflectometric diagnostic of the plasma electron density, at densities above those accessible with laser interferometry up to electron densities corresponding to solid targets ($10^{23}$ $cm^{-3}$). Most of the theoretical predictions are based on particle-in-cell calculations (PIC) which have been only indirectly validated, through measurements of the accelerated particles. For the understanding and optimization of the ion acceleration experiments it will be thus important to measure also the plasma parameters and dynamics, in addition to the particle output. In particular, it will be very useful to measure the initial electron density profile of the target plasma, which is a controlling parameter in the PIC simulations [177]. In addition, comparing the measured and predicted density profiles during the acceleration will help understand the acceleration dynamics and validate the simulations. Lastly, by mapping the total (i.e., hot and cold) electron density, it is possible also to infer the electron energy distribution in the plasma [175].

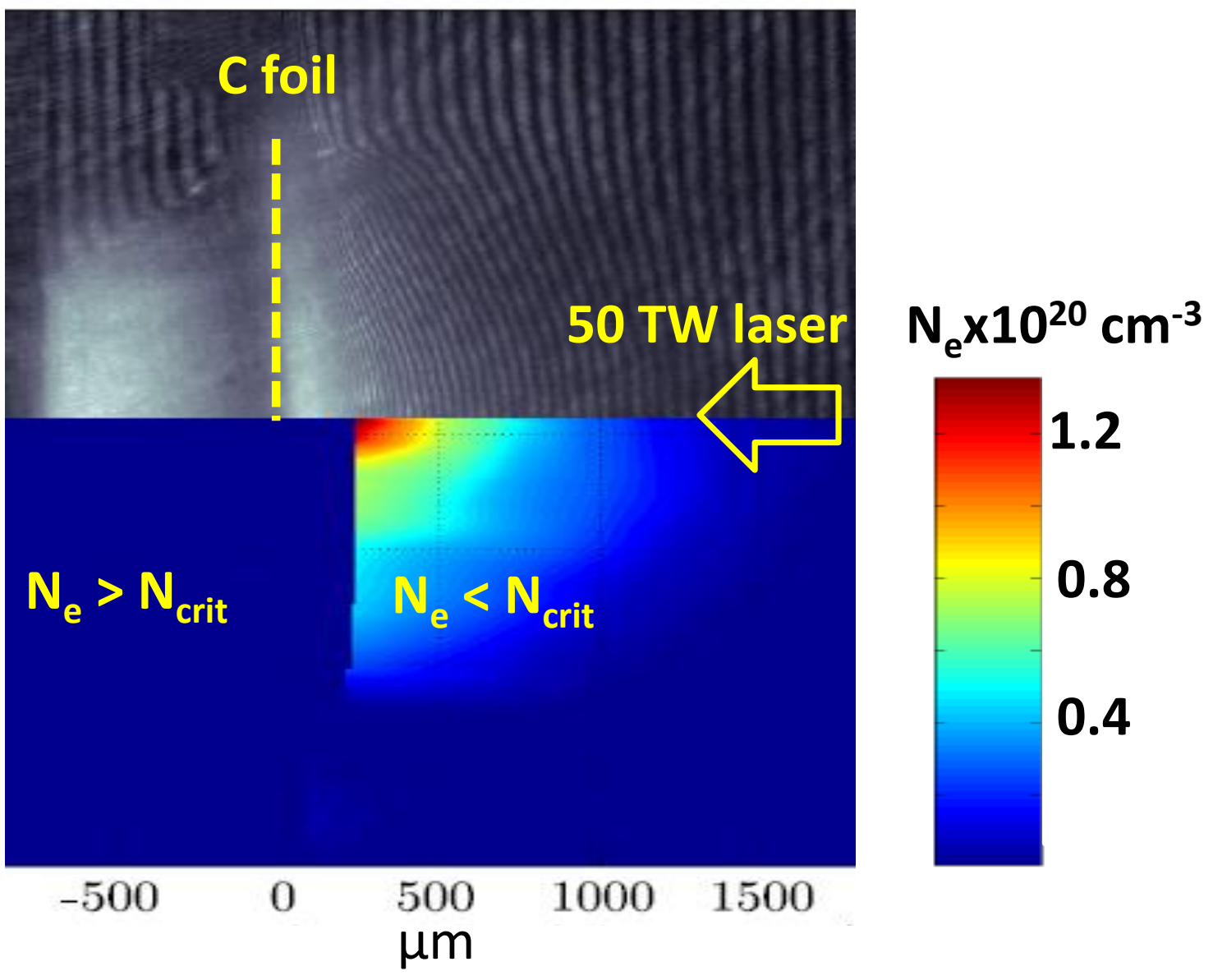

Figure 40 – Interferogram and measured electron density distribution of laser irradiated C foil, obtained using a femtosecond UV probe beam (adapted from Ref. [178]). The density diagnostic is possible only in regions with $N_e << N_{crit} = 8x10^{21}$ $cm^{-3}$.

The most direct plasma density diagnostic is interferometry or deflectometry with a probe or 'backlighter' laser beam [178] as proposed for ELI-NP in section 3.4.3.4 above. In this approach the phase of the probe beam is changed by the areal electron density (interferometry) or by the density gradient (deflectometry) of the plasma. The laser interferometric diagnostic is however limited to electron densities below the critical density ($8x10^{21}$ $cm^{-3}$ for 355 nm UV light). This is illustrated in Fig. 40 with a result adapted from Ref. [178], showing the interferometric diagnostic of a carbon micro-foil irradiated by a 50 TW laser. The backlighter beam is a femtosecond UV laser. The top of Fig. 40 shows the raw interferometric image and the bottom shows the measured electron density. As seen, the highest density that can be measured is $1.2x10^{20}$ $cm^{-3}$, i.e. nearly two orders of magnitude below the critical density. Since most of the ELI-NP experiments will use solid density targets, it is important to extend this diagnostic to higher densities using X-rays for backlighting.

An X-ray interferometric diagnostic for dense plasmas has been recently developed at Johns Hopkins, named *Talbot grating X-ray Deflectometry, or TXD* in short. The technique consists in measuring with micro-periodic gratings the deflection of an X-ray probe beam, caused by refraction on electron density gradients. An example of TXD electron density diagnostic of a solid object with 8 keV X-rays is illustrated in Fig. 41. Recently the technique was demonstrated with a TW laser driven X-ray backlighter. Both 1D and 2D gratings can be used in this method [179-181].

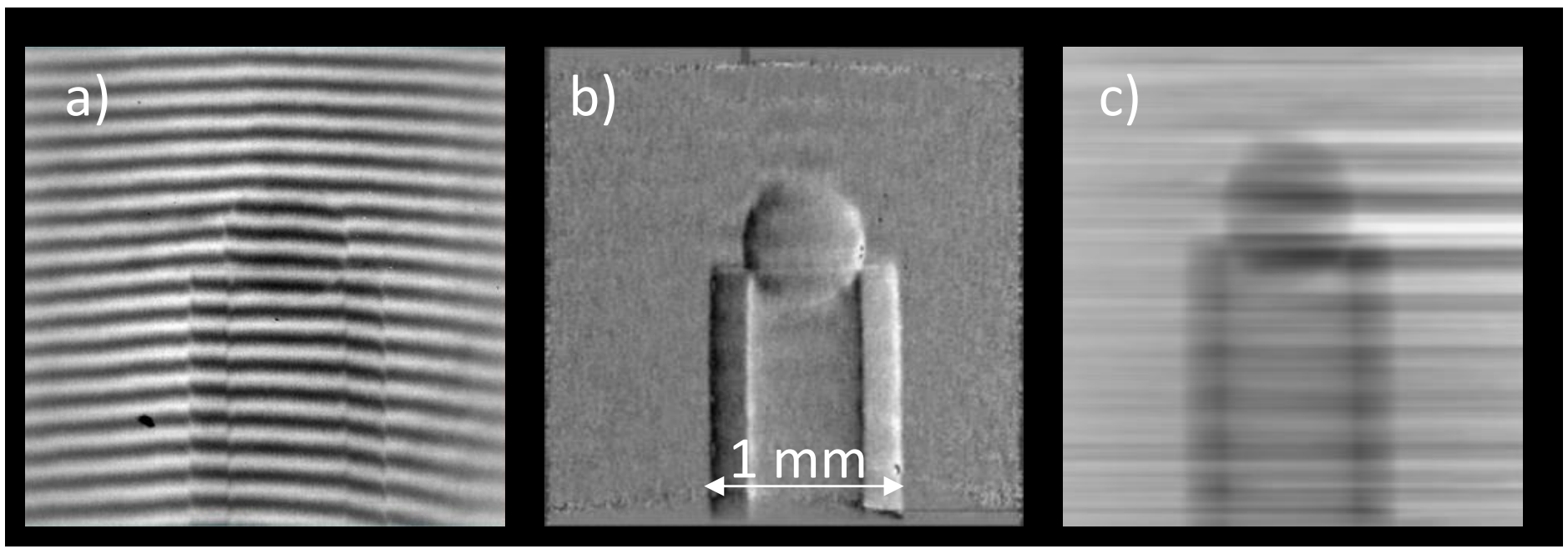

Figure 41 – Talbot X-ray deflectometry of plastic tube and sphere at 17 keV (Ref. [182]): a) Moire fringe image. b) Areal electron density gradient image. c) Areal electron density obtained by integration of the gradient. The artifacts in the density image are due to the use of 1-D Talbot gratings.

K-α backlighters are generally used for X-ray radiography of dense plasmas [183]. These consist of a thin high-Z foil or wire, irradiated by a short pulse, high power laser. K-α backlighters are however not well suited for the μm-size plasmas expected at ELI-NP. First, the large spatial extent of the K-α emission (>10-15 μm) would lead to insufficient spatial resolution. In addition, these backlighters require large laser energy (>100 J) to produce sufficient X-rays at the detector. The K-α emission has also relatively long duration.

To enable X-ray radiography and deflectometry with fs time resolution and μm space resolution, we propose using instead a betatron emission backlighter. Betatron emission occurs during LWFA electron acceleration in gas, when the electrons are transversally wiggled by the plasma potentials. The betatron source typically consists of a 5-10 mm He or H gas cell at ~$10^{19}$ $cm^{-3}$ density, irradiated by a 50-100 TW (e.g. 2J/30fs) laser pulse (see [184] for a recent review).

The betatron emission has ideal characteristics for X-ray interferometric diagnostic of dense micro-plasmas. First, recent studies show that the betatron source extent is only around 1 μm [184-186]. This makes the source highly coherent, which strongly enhances the phase effects, and allows obtaining very high spatial resolution.

Further on, the duration of the betatron emission is extremely short, of only a few fs [184]. This will first enable snapshot measurements of the initial density distribution in acceleration experiments. In addition, by scanning the delay between the backlighter and the main pulse it will possible to follow the dynamics of the acceleration process. (Note that a 'snapshot' diagnostic is possible as long as the target has sub-relativistic velocity. For targets moving with $v \approx c$, the image will be 'streaked' in the motion direction). Lastly, the betatron emission is bright and directional, and has a spectrum well suited for X-ray deflectometry. In optimal conditions about $10^9$ X-ray photons/shot are emitted in a narrow cone beam of 10-15 mrad FWHM [184-186]. The directional emission is beneficial for backlit radiography because it enables placing the X-ray detector far from the target, so as to reduce the background from plasma emission.

The X-ray spectrum has a synchrotron-like energy distribution with critical or mean energy $E_c$, in the 5-20 keV range approximately [184]. As further illustrated, this range is well suited for phase based diagnostic. Additionally, the mean energy can be varied by changing the gas density in the cell, which will enable optimizing the backlighter spectrum for a broad range of target densities. Good shot-to-shot reproducibility has also been demonstrated using dual-stage gas cells which separate the electron injection and acceleration regions [187]. Last but not least, using a gas target for backlighting enables a high repetition rate and avoids the debris problem of solid target backlighters.

Combining the UV deflectometry with the X-ray one, a complete diagnostic of the plasma density would be possible. The development of betatron backlighters for plasma radiography is being considered in the US and in the UK. It is therefore timely to start the diagnostic development also ELI-NP. The CETAL laser could be used for backlighter development and for plasma diagnostic tests, ahead of the ELI-NP commissioning.

*Proposed diagnostic design*

The proposed system is shown in Fig. 42. The preliminary design assumes backlighter parameters from Ref. [9]: 80 TW/30 fs laser pulse, focused with a 1.5 m / F20 off-axis parabola. The target is a 10 mm diameter He gas cell at around $10^{19}$ $cm^{-3}$ density.

The driving beam is derived using a pick-off mirror of 50-70 mm diameter, placed at the periphery of the 10 PW accelerating beam. Since the power intercepted by the mirror is ≤1% of the total beam power, it can be assumed that the acceleration process will not be perturbed. Alternately, the backlighter driver can be derived from the second 10 PW beam. (Note that a linearly polarized beam is generally used for LWFA).

Since the betatron emission has few femtoseconds duration and is synchronous with the laser pulse, the driving beam can be derived inside the interaction chamber. This simplifies the setup in comparison with up-stream beam needed for UV interferometry. A delay line will enable varying the timing of the backlighter pulse over a 1 ns range, for instance.

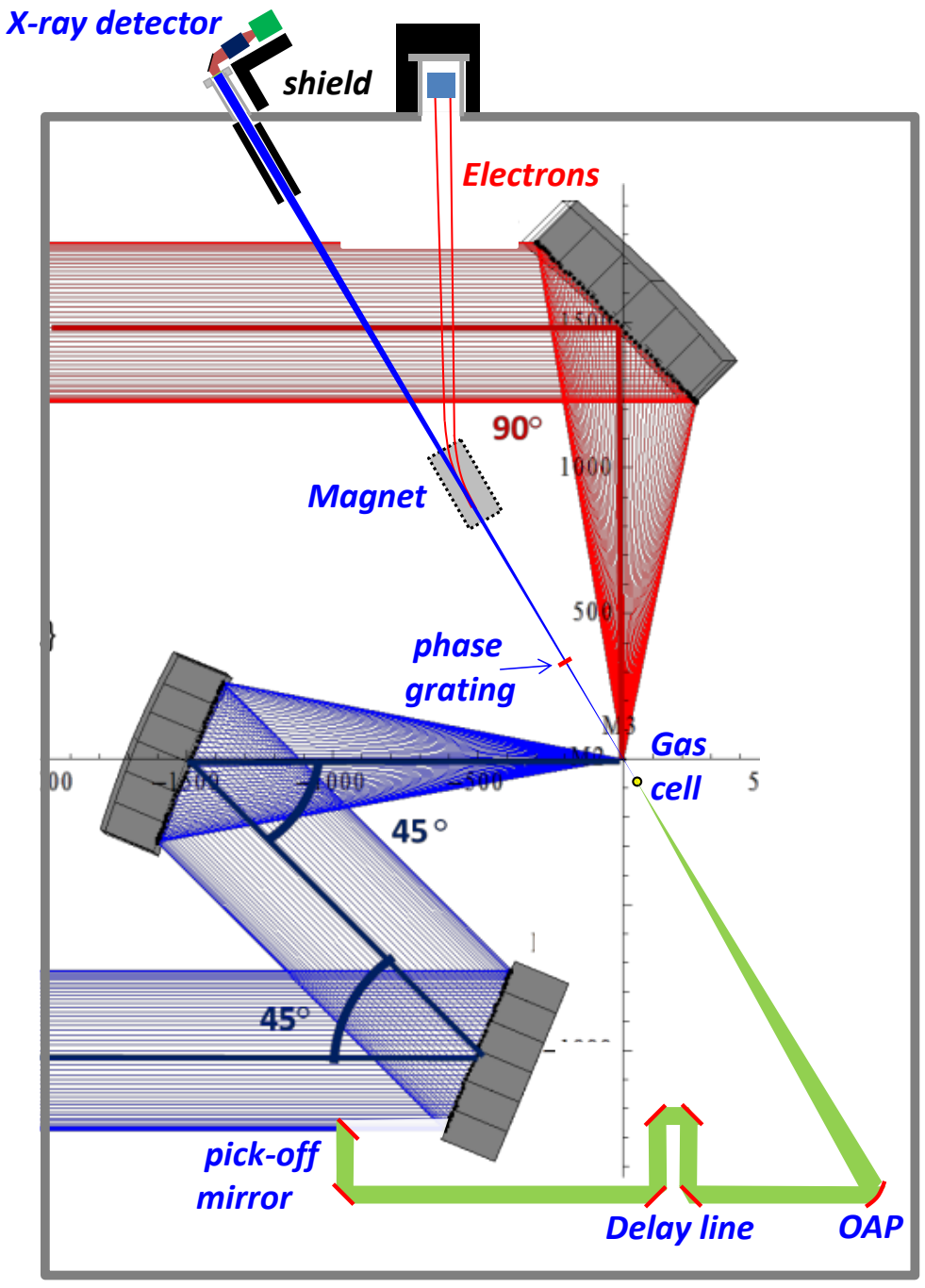

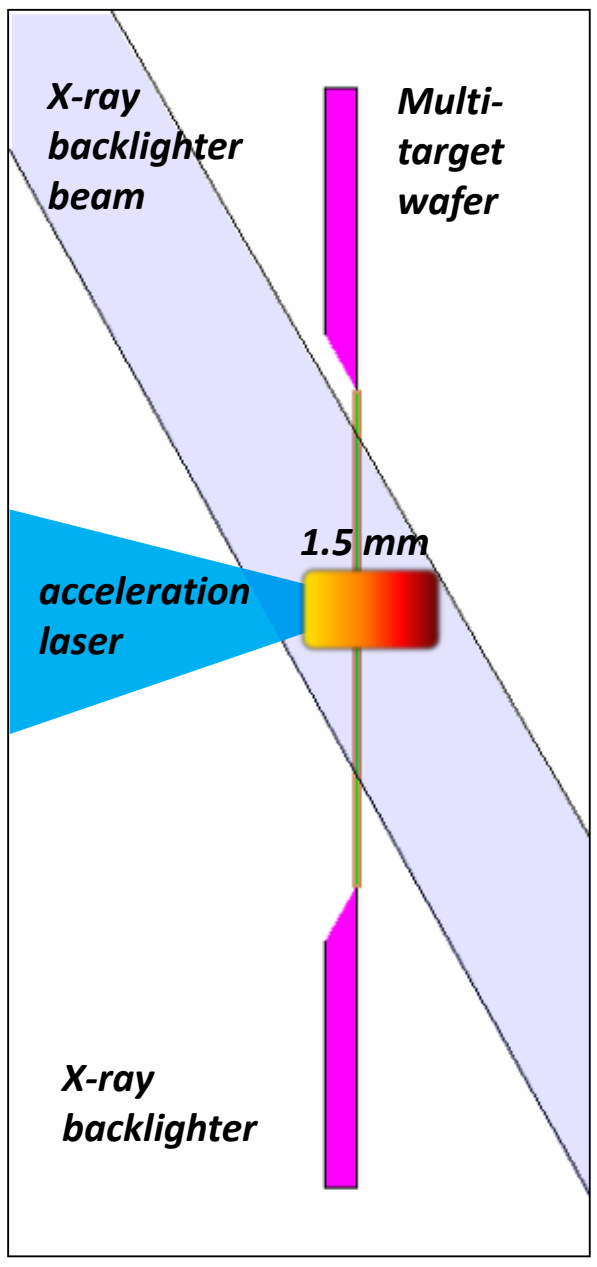

Figure 42 – Right: Layout of the proposed X-ray radiography system for ELI-NP. The pick-off mirror intercepts ≈1% of one the 10 PW beams. Left: Zoom on the intersection of the backlighter, acceleration beam and the plasma (target). The target is mounted on a membrane affixed to the multi-target holder.

The backlighter is inclined with respect to the target axis to allow viewing the plasma through the openings of a multi-target holder (Fig. 42). Assuming a few mm diameter openings, a 15 mrad backlighter beam, and 100 mm backlighter/target distance, a field of view of ~1.5x1 mm can be imaged. The field of view can be increased by changing the magnification and using a lower F# parabola, which increases the X-ray cone beam opening.

The detector is placed at about 2.5 m distance from the target and has active area of about 30x30 mm. The placement far from the target enables good shielding and decreases the X-ray background from plasma self-emission. An X-ray detector well suited for high power laser experiments is a thin (several μm) phosphor, lens coupled to a visible CCD camera. High spatial resolution, detective quantum efficiency, and low sensitivity to gamma-rays can be simultaneously achieved in this configuration. The object magnification is ≥25, which will enable spatial resolution at the target close to 1 μm. The relativistic electron beam produced by LWFA is deflected by a dipole magnet towards an electron imager or spectrometer.

Lastly, the electron beam produced by LWFA could be used for electron radiography, simultaneously with the X-ray radiography. Recently it was shown that electron radiography with a LWFA backlighter can serve for diagnostic of the internal magnetic fields in laser plasmas, similar to the proton radiography [187].

*Diagnostic capability*

To estimate the diagnostic potential of the proposed system, simulations were performed with an X-ray wave propagation code. The target was a C (diamond) or Th foil, having 15 μm diameter and 2 μm thickness. Few μm thick layers of exponentially decreasing density were added at the ends of the foil to simulate plasma sheaths. The target density was varied between 5% of the solid density and the solid density. A typical density profile is shown in Fig. 43(f) for the solid density C case. The target was placed side-on at 100 mm distance from the backlighter, and the detector at 2500 mm. The source size was 1 μm and the detector resolution 15 μm, leading to spatial resolution at the target of 1.6 μm. The backlighter energy was 12 keV.

Fig. 43(d) shows the pure attenuation image of the C target, obtained by simulating contact radiography (no wave propagation). As can be expected, the thin low-Z foil is completely invisible in X-ray attenuation. However, when phase effects are included the foil becomes highly visible through 'propagation' phase-contrast [188]. This is shown in Fig. 43(a)-43(c), which plot the phase-contrast enhanced images for C foils of 100%, 20%, and 5% of the solid density ($N_e$ from 500 x $N_{crit}$ to 25 x $N_{crit}$, approximately). Notably, the phase effects enable imaging the low-Z target with hard X-rays even at low density. In addition, the intensity profiles reflect well the changes in density.

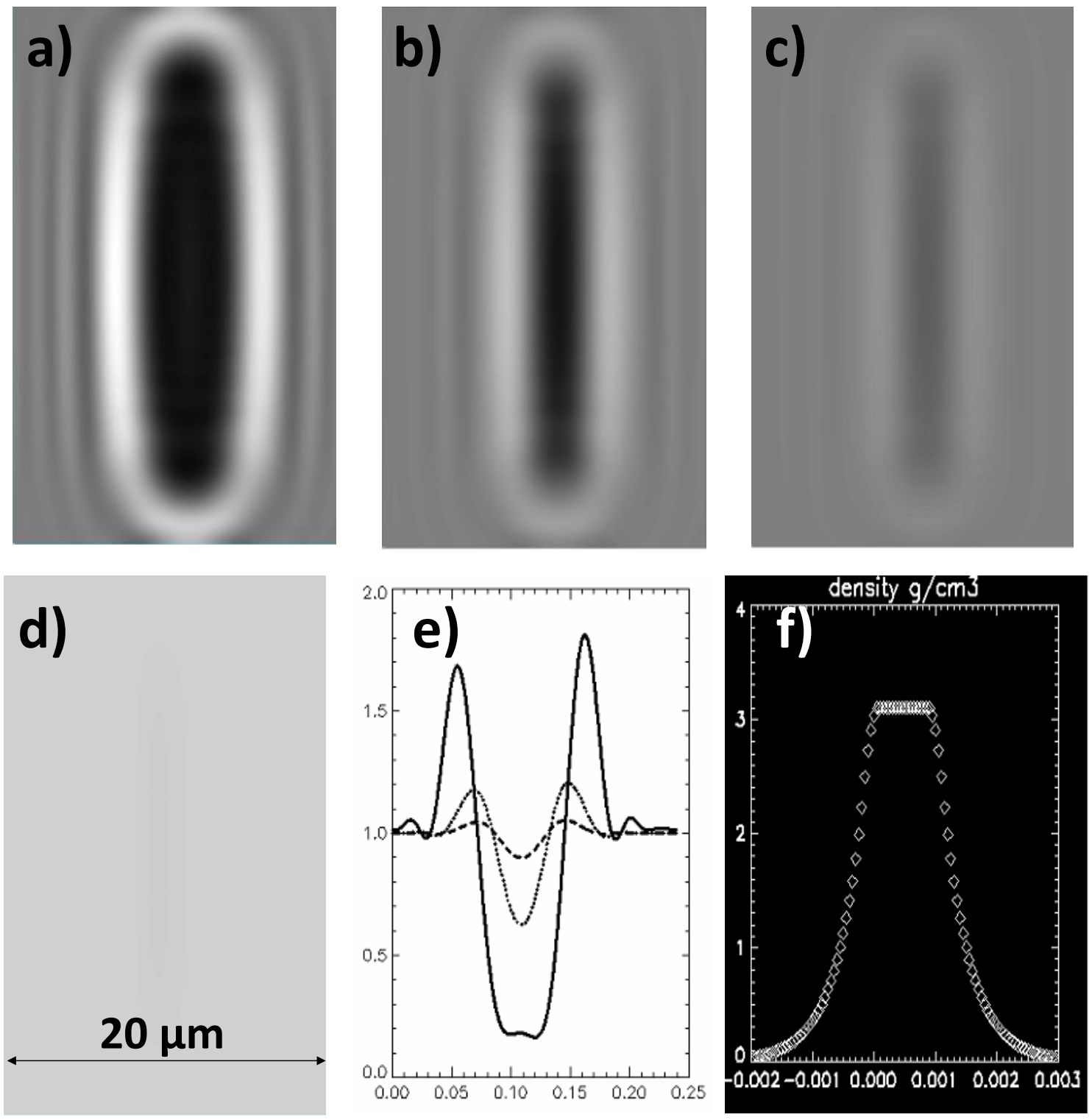

Figure 43 – Simulated side-on X-ray radiography of 15 µm diameter, 2 µm thick diamond-like C target. a)-c) Propagation phase-contrast images for target densities of 100%, 20%, and 5% of the solid density. d) Pure absorption image (contact radiography). e) Lineouts through images a)-c), showing the strong intensity modulation due to phase effects. f) Density profile used in the simulations (100% solid density).

Further on, Fig. 44(a) and 44(b) show phase-contrast images of the Th foil for two different density profiles, linear and exponential (Fig 45(a)). The peak density is 20% of the solid density. The change in gradients is evident in the side lobes of the images (Fig. 44(d)). Note that even for the high-Z target the phase-effects dominate over attenuation, due to the sharp density gradients.

The simulations show that due the high spatial coherence of the betatron source, X-ray deflectometry could be performed at ELI-NP using a single Talbot phase grating at high magnification (Fig. 42). This greatly simplifies the technique and avoids any photon loss in the grating. Fig. 45(a) shows that high fringe contrast can be obtained in this setup using only a few µm thick Si grating.

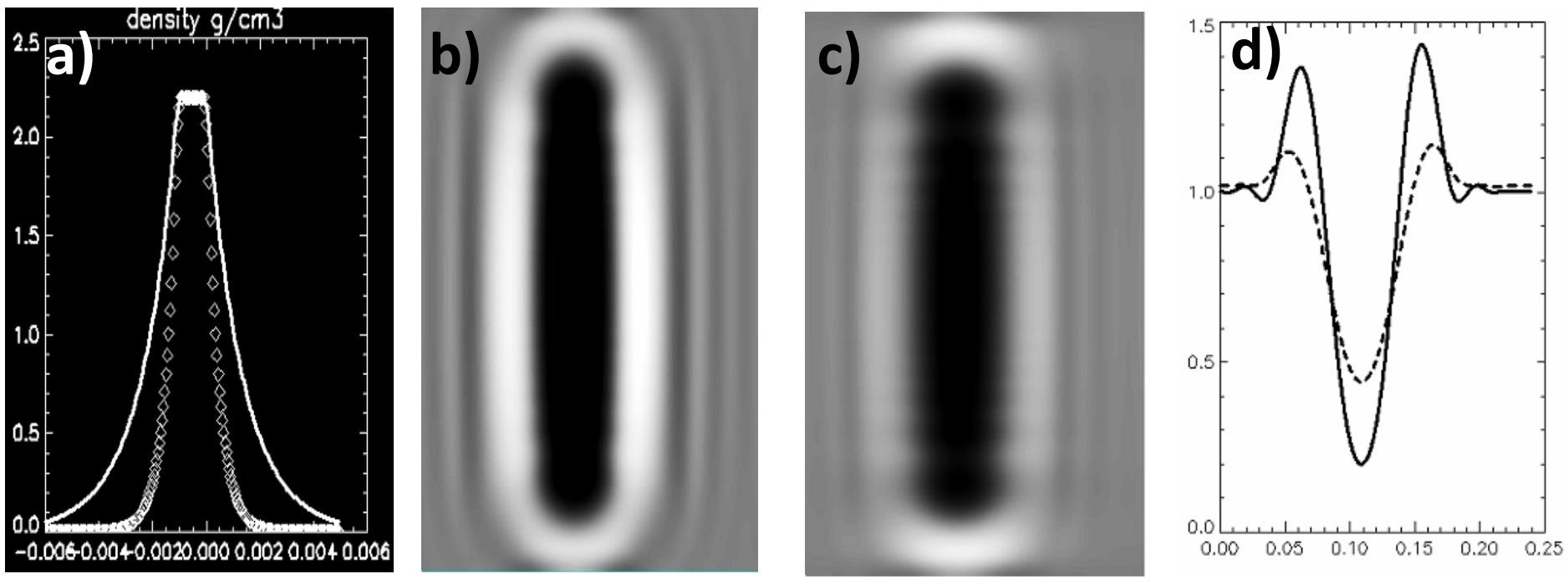

Figure 44 – Side-on X-ray radiography of Th target of 20% the solid density. a) Density profiles. b) Phase-contrast image for the steep density profile. c) Image for the broader profile. The different density profiles show as changes in the intensity peaks at the ends of the foil, in d).

When the target is present the backlighter beam will be deflected and produce a fringe shift or perturbation with respect to the unperturbed fringe pattern, as illustrated in Fig. 45(b). The fringe shift is then analyzed by Fourier methods to obtain the phase and amplitude changes caused by the target. Figures 45(d) and 45(e) show examples of such an analysis for a C target having 20% of the solid density, and sheaths with exponential and linear density profiles (Fig. 45(c)). As seen, the fringe shift images clearly reflect the different density gradients. Assuming axial symmetry, the fringe shift can be Abel inverted to directly obtain the electron density profile. A more detailed analysis would include also the wave propagation effects.

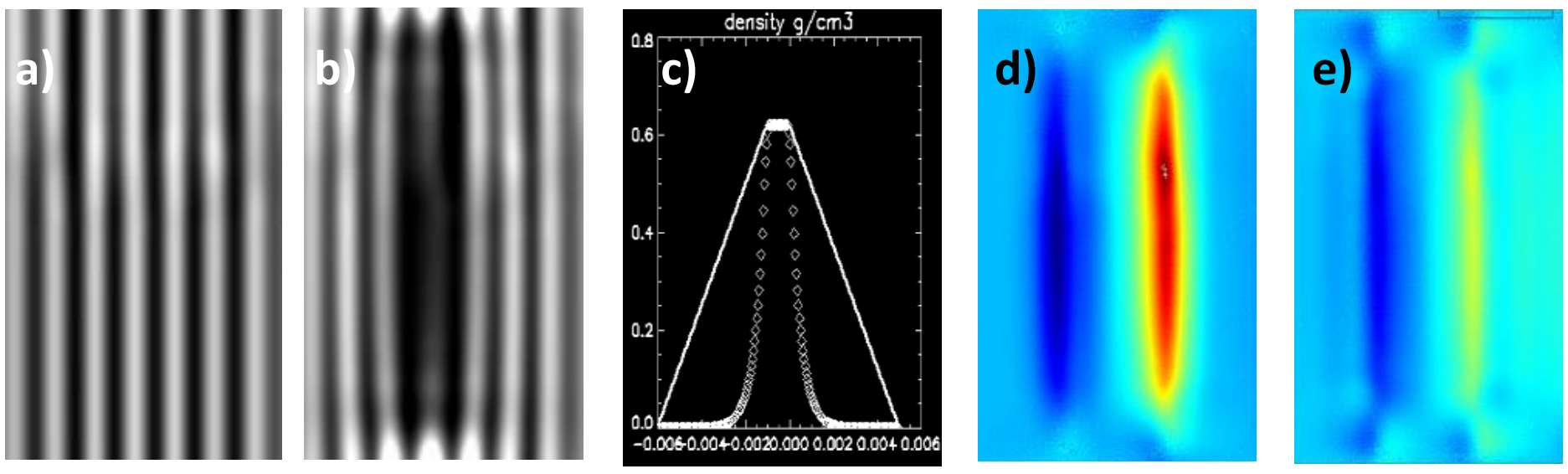

Figure 45 – a) Fringe image obtained by adding a Talbot phase grating with few μm period, to the radiographic system. b) Fringe image including a C foil of 20% of the solid density. c) Exponential and linear sheath density profiles assumed. d), e) Fringe shift images for the two density profiles.

*Photon budget*

Assuming $10^9$ X-ray photons are emitted in a 10 mrad angle [185], in the geometry in Fig. 42 the expected photon count is ~$10^3$ X-rays per pixel in a single shot. This is more than adequate, since the TXD method needs few photons per pixel to produce useful information. This is evident in Fig. 41, which was obtained with only several tens of X-rays per pixel. Further on, it is expected that the betatron source will be further improved.

The main question is the proportion between the useful light and the background light. Two main contributions to the background can be expected. The first is the X-ray K or L-shell emission from the target, falling in the same energy range as the backlighter emission, and the second is the high energy gamma background. To estimate the first contribution we use the $\approx 10^{-4}$ conversion efficiency of laser light into K-α X-rays and take into account that K-α light is emitted in 4π, whereas the betatron emission in a narrow cone beam. Using the conversion efficiency from laser energy to betatron X-rays from Ref. [185], one obtains then that the K-α background will be at most about 25% of the useful signal.

More difficult to assess is the gamma background. At ultra-high laser intensities, the conversion efficiency of laser energy into gamma rays is predicted to be very high (up to 30%). Nevertheless, the gamma emission will be directional, with essentially all of the radiation emitted in a forward $\approx 90^{o}$ cone beam [189]. The background will arise in this case from the scattered gamma rays and from the target positron annihilation radiation. Assuming that this background is 1% of the maximum gamma output, and that the X-ray detector is a few μm thick phosphor layer, only about one in 100 pixels will be recording a high energy photon. Further on, since RPA is based on circularly polarized laser light which reduces electron heating, it can be expected that the gamma background will also be reduced in solid foil acceleration experiments.

### 3.4.4 Focal intensity diagnostics

We propose also to implement a direct diagnostic of the focal intensity, based on a simple physical process which can easily scaled to intensities of $>10^{22}$ W/cm$^2$. The motivation for such a diagnostic is that the traditional, indirect intensity diagnostic based on measuring the pulse energy, duration and focal spot size, will become uncertain at intensities $> 10^{22}$ W/cm$^2$ [190-193]. The focal intensity diagnostic will confirm the achievement of exceptional intensities at ELI-NP independent of any acceleration mechanism and will enable progressing with confidence to the Day One proton acceleration experiments.

There are several physical processes on which a direct focal intensity diagnostic for ultrahigh intensity lasers can be based [190-193]. Of these, field ionization seems the simplest. The ionization method relies on the fact that the electric field in high charge states ions (e.g. H-like) of mid to high-Z atoms, is comparable to that of lasers with intensity in the $10^{22}$ to $10^{25}$ W/cm$^2$ [193].

The ion charge state distribution can be measured simply using an X-ray spectrometer and will serve as a direct measure of the laser focal intensity [193].

To spatially localize the interaction region at the laser focus, thin foils or low density foam layers of high-Z materials can be used. Further on, a mixture of several materials with increasing Z can be used for a wide dynamic range of the intensity measurement.

This direct intensity diagnostic will be complementary to the traditional focal intensity diagnostics. Implementing such a "machine diagnostic" will benefit also the operation of ELI-NP laser as a user facility, by confirming prior to experiments that the desired focal intensity is achieved. Further on, a "focal spot spatial and temporal contrast" diagnostic could also be implemented, based for instance on electron density profile measurements in thin foils with short wavelength interferometry. Such machine diagnostics are of interest for all HPLS TDRs as well as for the other ELI pillars.

## 3.5 PASSIVE OR OFF-LINE DIAGNOSTICS

Passive detectors based on dosimeter film, track detectors, imaging plate etc. are less interesting for high repetition rate systems. However, they have some advantages. For example, spatial resolution with large acceptance combined with energy resolution offered by stacks of radiochromic films is not possible otherwise. Sometime they may be used (in single-shot operation mode) to cross check the results obtained using the active detectors.

In this category of detectors, we include also off-line activation measurement setup. As reference the NATALIE detection system [194] developed at CENBG has ~40 $LaBr_3$/NaI/Ge detectors in order to measure activation stacks with up to 20 foils.

A low background gamma spectroscopy facility is also high desirable for long-lived isotopes measurements. Such facilities exist in IFIN-HH in vicinity of the ELI-NP site. For short lived isotopes (minutes) the setup may be installed in the ELI-NP building. It needs some 10 $m^2$.

The space needed for processing radiochromic films (RCF), CR39, image plates (IP) or bubble detectors is about 3x3 $m^2$ and will be available in the Dosimetry Laboratory.

### 3.5.1 Radiochromic (Gafchromic) Films

A radiochromic film changes its optical density when irradiated by ionization radiation (electrons, X-ray, ions etc.). The optical density will depend linearly on the dose delivered up to high values (1000 Gy). Radiochromic films(RCF), placed in a stack configuration and optionally sandwiched between copper or aluminum absorbing layer, are going to be used to obtain 3D image of the spatial and energy distribution of the laser accelerated ions (radio-chromic imaging spectroscopy). They will be also used in conjunction with a Thomson parabola or an ion wide angle spectrometer to obtain the ion species composition (for light ions) and charge states. As a passive, plastic detector, it is immune to electromagnetic pulse and it can be placed close (few centimeters) behind the target.

RCF films batches need to be calibrated with a known ion beam flux and are post-processed at latest 24h after exposure. For post processing a CCD based document scanner can be used. Examples of suitable scanners are Microtek ArtixScan 1800f and Shimadzu UV-1800. A gray scale wedge (e.g. #T4110cc Step Transmission Calibrated & Certified Stouffer Industries Inc.) can be used to convert the scanned data to optical density values.

Gafchromic films are commercialized by Ashland Inc, Covington USA. The following products are going to be used

- HD-V2 (8 μm active layer – 97 μm polyester substrate) 10 Gy to 1000 Gy dose range, resolution 5μm
- MD-V3 (120 μm polyester – 15 μm active layer – 120 μm polyester) 1 Gy to 1000 Gy dose range, resolution 5 mm

MD-V3 films will be used for lower fluxes, or for the end layers of the stack. The errors in inferring the laser beam spectral energy is due to the uncertainty in the dose, variation in the film properties, calibration errors etc. and can reach up to 20%. Disadvantages: consumable, large thickness and therefore suited for light ions only.

### 3.5.2 CR39

CR39 plates are plastic polymer track detectors that can be used in a stack configuration for beam profiling as well as detectors for Thomson parabolas or ion wide angle spectral detectors. The advantage of using them is their insensitivity to photons, electrons or above 10 MeV protons. Thus they are good options when different species of ions must be identified (see Jung et al. (2013)). An example of commercial CR39 are TASTRAK films, produced by Track Analysis Systems Ltd.UK. Ion tracks are revealed by slow etching and read by an automated microscopy system. TASTRAK comes in 0.1, 0.2, 0.5 and 1 mm thicknesses and costs 1 euro per sheet (10x10 mm$^2$). The CR39 plates are not reusable.

### 3.5.3 Image Plates (IP)

Image plates are 2D imaging sensors used in conjunction with Thomson parabolas and ion spectrometers. Similar to RCF, image plates are sensitive to a broader range of radiation kinds. The IP films contain layers of phosphorus elements that are driven in metastable excited states when excited by radiation and post processed through luminescence by a laser scanner. The spatial resolution is in the range of 25-50 μm. Also IP have a large dose dynamic range sensitivity and linearity. One advantage of IPs is that they are reusable. Examples of commercially available IP films used in laser driven experiments are FujiFilm BAS-SR (high resolution) and BAS-MS (high sensitivity). Examples of laser scanners are FujiFilm FLA 7000 and VMI 5100MS scanner from VMI (http:/www.starview.com).

### 3.5.4. Bubble detectors

Bubble detectors are one of the most convenient neutron dosimeters. They consist of tiny superheated droplets in a clear polymer. A neutron hit of a droplet determines the vaporization and formation of visible bubble. A set of bubble detector with different thresholds (6 values) on neutron energies for bubble formation represents a low-cost, low-size neutron spectrometer with no gamma sensitivity. Bubble detectors can be reset (re-zeroed) by compression for about 30 minutes. The automatic reader, the recompression chamber and unfolding software are commercialized by BTI (http://www.bubbletech.ca).

*Table 1. Diagnostics (and detectors) for HPLS-TDR1 – Laser Driven NP*

<table>
<tr><td>1</td><td colspan="3">Plasma diagnostics setups</td><td colspan="4"></td></tr>
<tr><td></td><td></td><td>Spatial Resolution</td><td>Time Resolution</td><td>Diag-nostics</td><td colspan="3">Features/references</td></tr>
<tr><td></td><td>Interferometry/ Shadowgraphy</td><td>several $10\mu m$</td><td>$fs \div ps$</td><td>N</td><td colspan="3">For gas targets, and plasma expansion in thin films</td></tr>
<tr><td></td><td>Optical Transition Radiation</td><td>several $100\mu m$</td><td>No</td><td>$I(n_{ions})$</td><td colspan="3">Spot profile</td></tr>
<tr><td></td><td>Polarimetry</td><td>No</td><td>No</td><td>B</td><td colspan="3">Self-generated magnetic fields</td></tr>
<tr><td>2</td><td colspan="7">Ion detectors</td></tr>
<tr><td></td><td>Active (on-line)</td><td>Spatial resolution</td><td>Time resolution</td><td>Processing time</td><td>Dynamic range</td><td>Single Particle</td><td>Features</td></tr>
<tr><td></td><td>Micro-channel plate (MCP) + phosphor screen + CCD</td><td>several 10 $\mu m$</td><td>few 100’s ps</td><td>1 s (CCD readout)</td><td>$10^3$</td><td>Yes</td><td>$e^-$, X-ray sensitive require vacuum $10^{-6}$ mbar</td></tr>
<tr><td></td><td>Scintillators + ICCD or EM-CCD</td><td>several 100 $\mu m$</td><td>few 100’s ps</td><td>1 s (CCD readout)</td><td>$10^3$</td><td>yes</td><td>Stackable depth profile</td></tr>
<tr><td></td><td>Photodiode array</td><td>50 $\mu m$</td><td>few ms</td><td>0.5 s</td><td>$10^3$</td><td>yes</td><td>e.g. Radeye1 does not require vacuum</td></tr>
<tr><td></td><td>Passive (off-line)</td><td colspan="6"></td></tr>
</table>

| | | | | | | | |
|---|---|---|---|---|---|---|---|
| | CR-39 tack detectors | several$\mu m$ | No | Few hours | $10^6$ | yes | Sensitive to ions only |
| | Image plates (IP) | several$\mu m$ | No | several min | $10^5$ | no | Reusable |
| | Radiochromic films (RCF) | 3-10 µm | No | several min | $10^4$ | no | self-developing |

| | | |
|---|---|---|
| **3** | **Ion (including electron) diagnostics setups** | |
| | Thomson parabola + active or passive detector | ion species, charge states and energy |
| | Wide angle spectrometer + active or passive detector | ion species, energy and spatial profile |
| | Radiochromic imaging spectroscopy | spatial profile, energy distribution with stacks |
| | Terahertz spectroscopy | ion flux, electron bunch temporal profile |
| | Time of flight | Energy |
| **4** | **Electromagnetic fields** | |
| | Polarimetry, optical & THz detectors | |
| | Optical probe beam | Pick-up from main beam |
| | XUV diagnostics | |
| | Betatron emission backlighter | Pick-up from main beam |
| 5 | **Gamma** | |
| | $LaBr_3$+gated PMT+LED +DAQ | Online |
| | NATALIE-like (activation) system | Offline |
| | Compton/scintillator spectrometer | Prompt |
| 6 | **Neutron** | |
| | Time-of-Flight (plastic scintillator with lead shielding) | |
| | ERBSS (Extended Range Bonner Spheres System) | |
| | Bubble detectors | passive detector |
| | Neutron imaging detector | |

### 3.6 DECAY STATION AND TRANSPORT SYSTEM

The decay station will offer the possibility to perform high resolution spectroscopy using Ge detectors at shortest possible time scales after the laser shot. Ge detectors could not survive the strong X-ray flash and EMP in the vicinity of the target, such that they have to be well shielded and a fast transport system has to be employed to move the produced nuclei from the collection point to the Ge detectors measuring point. The proposed position of the decay station is behind the beam-dump, in order to benefit also from its shielding, but not in the central position, where the recoil separator is proposed to be installed (see below). Thus the distance from collection to measuring points is:

- minimum 15 m for collection inside the interaction chamber
- minimum 6 m for collection in the final focal plane of the recoil separator.

Fast tape transportation systems, as employed currently in various nuclear physics facilities [195], are optimized for transportation over short distances (approximately 1 m). In aerosol gas-jet systems, instead of being implanted in the tape, the radioactive nuclei are stopped in the gas and attached to aerosol molecules for transportation over long distances (~100 m) before being captured by filters [196] in front of detectors or on tapes that move over a short distance [197] to reach the final measuring position. In all cases a total transportation time of about 1 second can be achieved. The solution of implantation directly in the tape has a higher collection efficiency, but mechanical reliability is the main concern, especially that the path of the tape can be straight only if the implantation is done in the focal plane of the recoil separator. The gas-jet can easily be re-routed, but the collection efficiency (reported up to 70%) and its stability has to be studied. In conclusion, implementation of the combined solution is proposed offering highest flexibility.

### 3.7 RECOIL SEPARATOR

A recoil separator consisting of a large dipole (warm electromagnetic) magnet and focusing elements (electromagnetic quadrupoles), both before and after the dipole, has been proposed (see Figure 46).

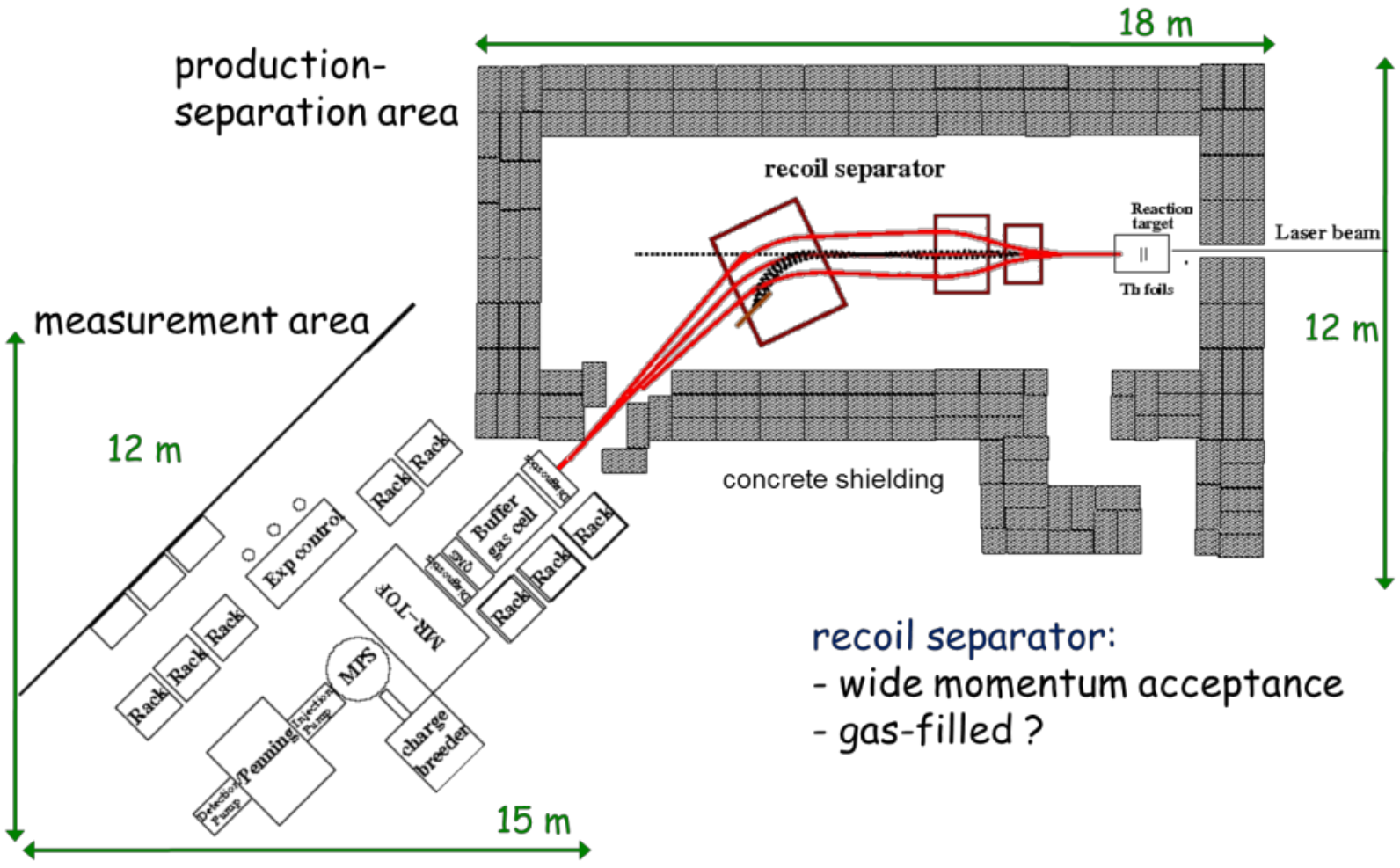

Figure 46 – Experimental setup including mass measurement devices fed by a recoil separator.

Being intended for separation of high Z elements with a large charge state distribution at the input, gas filling could be employed to reduce the dispersion and increase efficiency for selection of desired isotopes. Separation of the gas section from vacuum could be done with very thin foils imposing adequate pumping/venting procedures. Diagnostics such as Faraday cups and beam position monitors have to equip the separator, provided that they can work properly in the EMP environment that is expected to be rather high inside the spectrometer, because it is coupled directly through a large aperture to the target chamber.

A concept based on a double stage Wien (velocity) Filter was proposed by Univ. Giessen (see Figure 47), comparable to the SHIP separator installed at GSI. The proposal takes advantage of the fact that nuclei of interest from the fission-fusion mechanism are produced with quite large differences in velocity compared to both in-flight and at-rest fission products. This solution is therefore preferred and will be further studied in collaboration with Univ. Giessen and GSI.

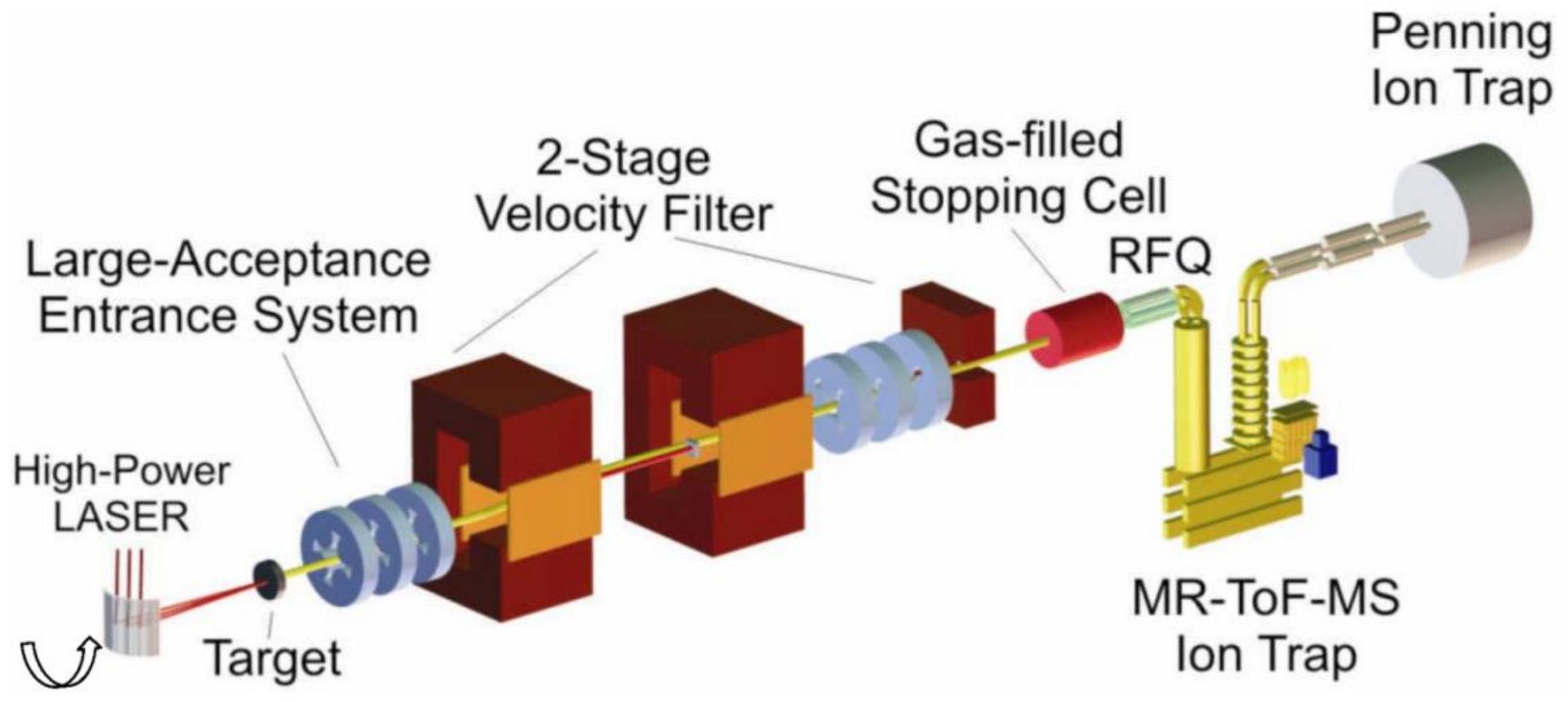

Figure 47 – Concept of the recoil separator for fission-fusion products proposed by Univ. Giessen

### 3.8 GAS CELL AND MASS ANALYSER (MR-TOF AND/OR PENNING TRAP)

Installation of a gas cell to thermalize the nuclei selected by the recoil separator, followed by a mass measurement setup represents an option for extending the study of N~126 nuclei, after their production with sufficiently high rates has been confirmed by decay spectroscopy. A possible experimental arrangement is shown in Figure 47. The gas cell, MR-TOF spectrometer and Penning trap employ high voltages, RF power, ultra-high vacuum and sophisticated control systems that have to be developed with the additional condition to withstand the EMP generated by the laser-target interaction. A research contract with GSI has been initiated, aiming at the definition of these devices that are similar to those foreseen by the ELI-NP photo-fission TDR for stopping, separation and study of fission fragments.

### 3.9 LB-CMPF DEVICE

The Laser Based Compact Magnetic Photo-Fusion devices described in section 2.3.3 is of high relevance for the proposed nuclear studies and application in this TDR. It will be developed in close collaboration between ELI-NP and Technical Univ. of Crete. The capacitors back and associated power supplies are to be placed in the vicinity on the interaction chamber, possible on the platform at the height of the interaction chamber lid. The integration of high current lines and feedthroughs needs a careful design but the feasibility is not a concern. Measures for protection of expensive 10 PW optics in case of the coil break have to be implemented.

### 3.10 CONTROL SYSTEM

The ELI-NP E1 area dedicated to the Laser Driven Nuclear Physics experiments will have the experiment monitoring and control systems architecture similar to the HPLS and Laser Beam Transport System (LBTS) control systems. The architecture is based on TANGO which will permit local distributed control of the equipment and additional clients to remotely supervise/control the experiment.

This solution will allow a standardization of the control systems inside ELI-NP, while providing easy maintenance, better security, better logging and interfacing methods between the experimental areas, HPLS and LBTS.

A dedicated User Room will be used to remotely control from outside the E1 area the equipment as the experiment is running. A TANGO framework will be developed to link the experimental area to the User Room using a dedicated client – server architecture that will allow maintenance and upgrades to be performed without interacting with other systems.

A data storage server will be available for short term experimental data saving and this shall benefit in general from dedicated data busses, separated from the client – server TANGO architecture that controls and monitors the equipment itself, in order to achieve the highest data throughput.

Dedicated TANGO servers are envisaged to interface the equipment necessary in the experiment such as:

- Focal spot monitoring
- Solid target alignment
- Solid target manipulation
- Target insertion system
- Delay generators
- Monitoring CCDs

- DAQ system
- Vacuum system for the E1 interaction chamber

Each of the above equipment will have a Human Machine Interface able to locally (from inside the interaction area) or remotely (from the User Room) monitor and control the parameters needed to run the equipment and the experiment.

For the equipment where the API is not provided, or the required development time does not fit into the general schedule, the link between the User Room and the experimental area will be made using remote desktop (or similar) and by using their proprietary software.

The User Room will also provide through the TANGO architecture information to the user regarding the HPLS parameters and LBTS configuration. The HPLS and LBTS parameters will be controlled from the HPLS control room by the operators from personnel and machine safety reasons.

### 3.11 RADIATION SOURCE TERM AND BEAM DUMP

The beam dump that has been designed for the E1 area in the preliminary radioprotection study uses a proton source term produced by one or two 10 PW lasers with $1.4\times10^{12}$ protons/pulse, a quasi-monoenergetic energy distribution around 3 GeV and an angular distribution of ±20°. The resulting beam-dump turned out to be very large and heavy (about 2000 tons). This beam bump was used for specifying the building size and in order to obtain building permit license. However, the installation of this beam dump is not mandatory from the start. It can be sized and installed gradually, depending on the results obtained during the first period of operation. Following the discussions within the working group, the limit of 500 MeV for protons produced through the RPA mechanism has to be considered to be more realistic, considering also the 250 J per pulse available at ELI-NP, comparable with few running 1 PW laser systems. Moreover, most of the proposed experiments are requesting heavy ions at moderate energies, such that protons will appear only as a contaminant in the beam, originating from depositions on the target surface in the vacuum chamber. This deposition can be reduced quantitatively by heating the target before firing the laser. The RPA mechanism is expected to produce a low beam divergence of 5°. However, in the two 10 PW beam configurations shown in Figure 32(b), the angle between the central axis and the focal direction is about 10°, which may produce (e.g. due to beam pointing fluctuations) two particle beams. Larger angles will result from shorter focal lengths. Therefore, the divergence for the RPA source term (labelled with A in the table below) was kept 40°, similar to the case of the TNSA-like proton source term (labelled with B).

*Table 2. Radiation source terms for E1 experimental area radioprotection calculations*

| Source Label | Particle | Mean Energy (GeV) | Particles / Pulse | Freq. (Hz) | Div[2] | Spectrum | FWHM (GeV) |
|---|---|---|---|---|---|---|---|
| A | *Proton* | 0.5 | 1.40E+12 | 1/60 | $40^0$ | Rectangular | 0.05 |
| | *Electron* | 1.5 | 8.00E+10 | | | Gaussian | 0.15 |
| B | *Proton* | 0.04 | 1.00E+14 | 1/60 | $40^0$ | Boltzmann[1] | n/a |

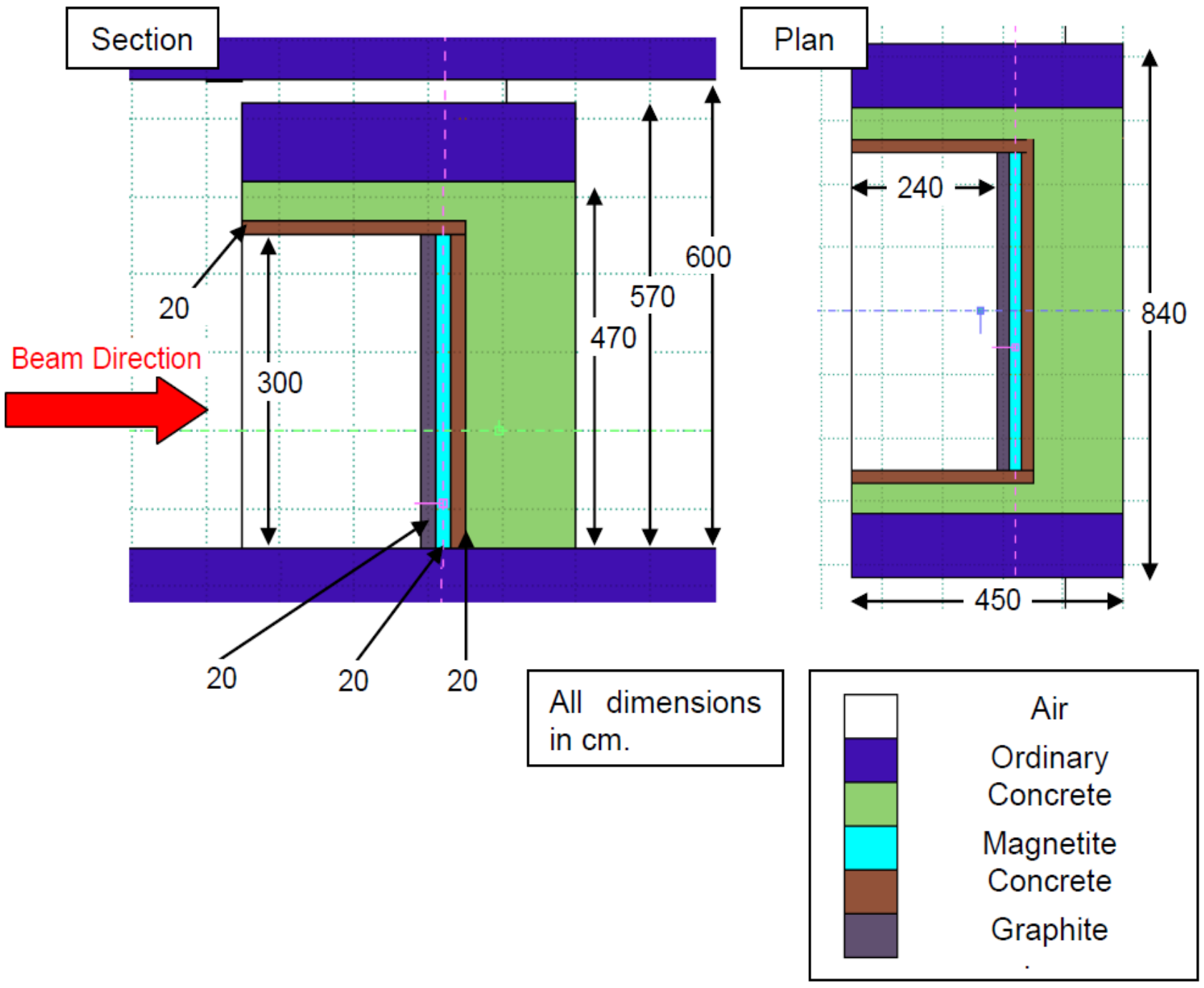

Figure 48 – The beam-dump dimensions and composition assumed in the radioprotection calculations.

The proposed beam-dump design and the dose calculations performed by Nuclear Technologies are shown in the Figures 48-49. For source term A, the doses on the corridors reach ~8 μSv/h, thus being higher than the targeted[1] 1 μSv/h and suggesting that additional shielding might be needed or the repetition rate may have to be reduced. However, we note that these calculations were done for an extreme case, which will be encountered in very few experiments. For the TNSA-based proton source term, the shielding considered in the calculations is just enough to obtain 1 μSv/h on the hot-spot of the cold-side (corridor). As expected, the calculations have shown that the main contribution to the dose levels inside and outside the E1 area are due to high energy neutrons generated by high energy protons interacting with the forward chamber wall, consisting of 10 cm Al. The use of a thin window moves most of the proton interactions into the beam dump and dose levels are significantly reduced. However, this solution was considered not truly applicable in practice, because usually at forward angles many diagnostics devices will be positioned, having the same effect as the thick chamber walls. Nevertheless, in some cases this method of reducing the dose rates can be applied. The heavy ion source terms are not considered, because they are producing much less radiation than the proposed proton terms.

Dose rates calculated for the cold side of the roof impose that it will be inaccessible when the beam is present in E1 (a condition that will be imposed actually for all laser experiments).

Monitoring the dose rates produced on-line by the high-power lasers with active detectors requires special pulsed field detectors. Such detectors for both gamma and neutrons are developed for example by ELSE Nuclear (see http://www.elsenuclear.com) being able to measure up to several μSv/pulse at repetition rates well above 10 Hz. Standard commercial detectors will saturate in the pulsed radiation field specific to ELI-NP. If such detectors are installed inside the experimental area, they have to be switched off during laser shot sequences (runs), while those installed outside will just record one more pulse for each laser shot.

An analysis of the operation modes of the ELI-NP facility concluded that the experiments in E1 experimental area will run, in average, for 90 days/year at 300 pulses/day (excluding the days for setting-up/dismounting the experiments). We distributed therefore 15 days for source term A and 75 days for source term B. With this assumption and a supposed (typical) geometry of the experimental equipment inside and in vicinity of the interaction, calculations of activation have been performed for each term source for a long operation time (20 years) and various cooling time after the last run stop. The results for source term A are shown in Figure 50.

---

[1] We note that the 20 mSv limit for the total dose per year allowed by the national Regulation for workers in radiation fields corresponds to 10 μSv/h for 2000 working hours. The design criteria at ELI-NP have been chosen a factor of 10 below allowed limits.

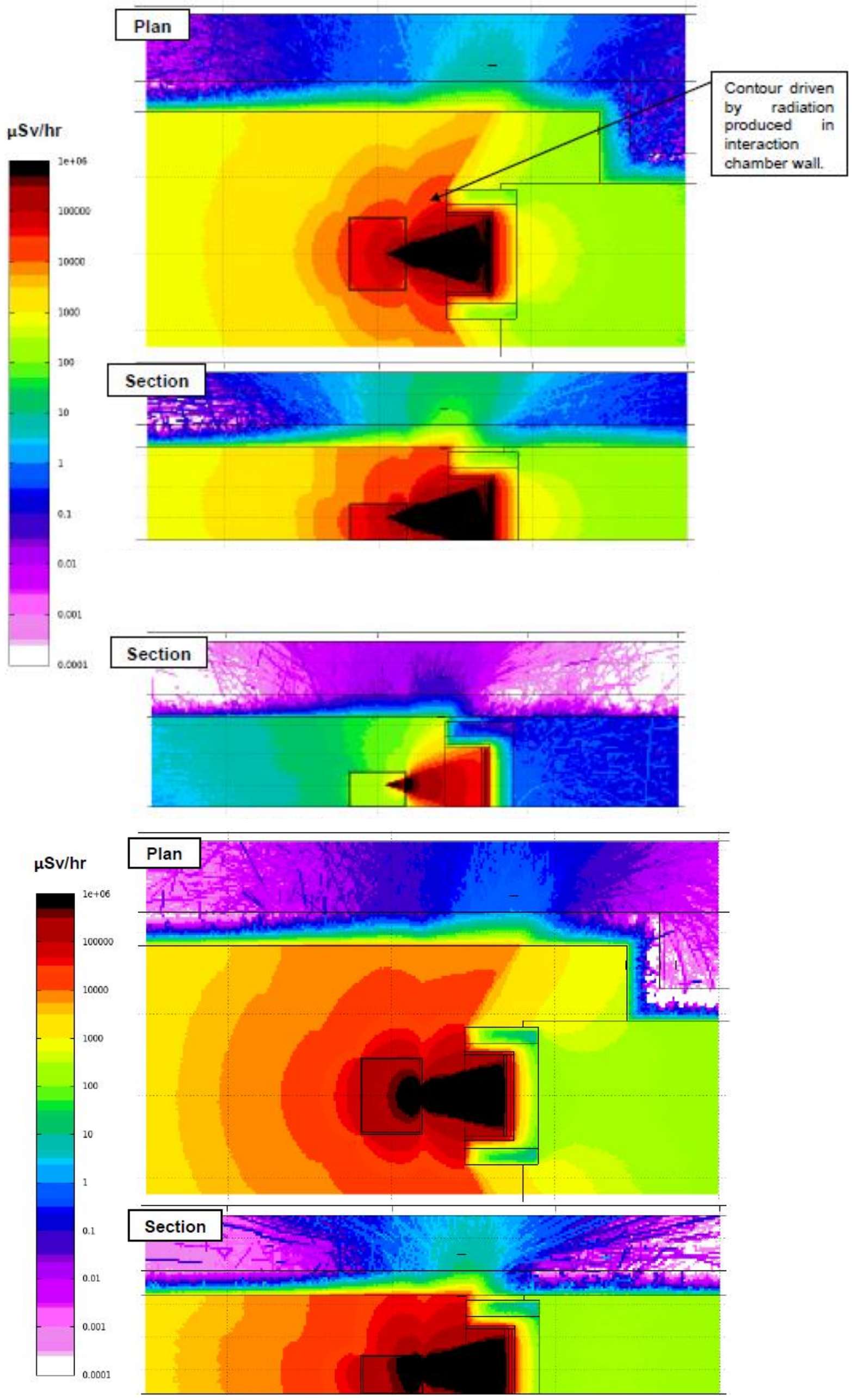

Figure 49 – Calculated dose rate contours for the E1 area.

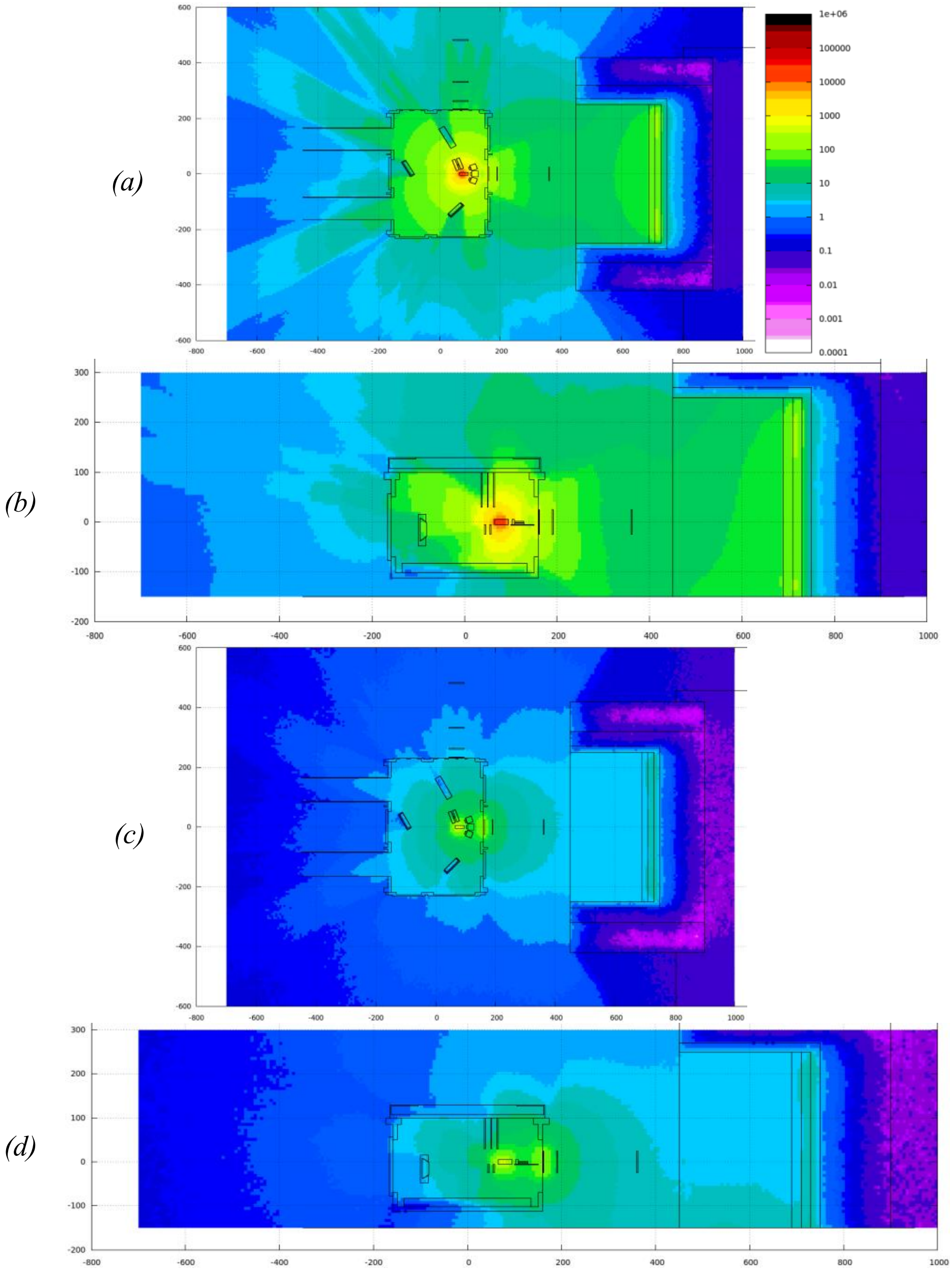

Figure 50 – Activation contour doses rates calculated for source term A for 300 shots per day and 15 days irradiation followed cooling times of 10 minutes (a and b) and 2 hours (c and d). The subfigures (a) and (c) present a top view, while (b) and (d) a side view. Right-up color scale is in μSv/h.

As expected, all the materials placed at forward angles are strongly activated and require very long cooling time (days) before being accessible. However, the source term A is an extreme case, most of experiments will produce significantly less activation. We remind also that the 300 shots per day represent only 5 hours of continuous run, letting long time for target and diagnostics exchange using the target insertion system installed on top of the E1 chamber.

### 3.12 ACCOMMODATING THE EQUIPMENT REQUESTED BY HIGH FIELDS AND PHOTO-FISSION TDRS

Several requests for space in the E1 area have been formulated by other TDRs:

- A long focal mirror (and associated laser beam lines) is requested by the HFQED TDR. The mirror should point towards the electron beam dump foreseen in the E6 area.
- A mass separator and low energy ions beam lines are required to bring photo-fission IGISOL beams to the decay station
- A collinear laser spectroscopy setup is requested also by the photo-fission TDR and, if installed, may be used for recoil ions produced by laser interaction in E1, behind the gas cooler.

All these requirements are taken into account in the layout presented in the next section.

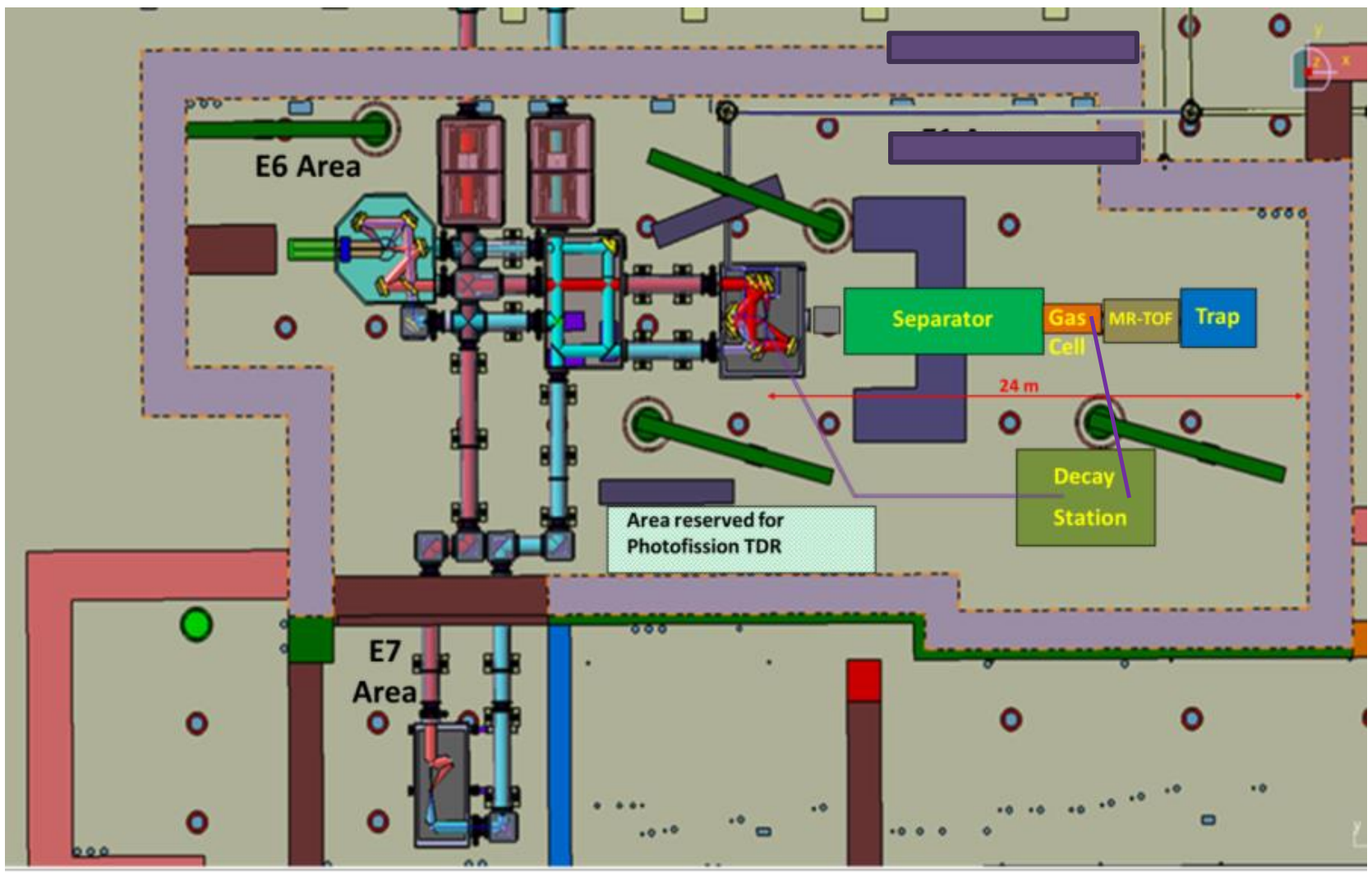

Figure 51 – Layout of experimental areas where the two 10 PW beams are distributed. Main equipment in the E1 area is drawn schematically.

### 3.13 PROPOSED LAYOUT OF THE E1 AREA

The layout shown in Figure 51 takes into account the most demanding devices (in terms of footprint) presented in sections 3.1 – 3.12. Additional shielding (purple rectangles) are foreseen in order to decrease the dose in adjacent areas, to be confirmed by on-going radioprotection calculations. The access in the interaction chamber will be done through a softwall clean room of a surface of about 3x4 $m^2$.

## 4. ESTIMATE OF COUNT RATES/FEASIBILITY OF PROPOSED DEVICES

### 4.1 NUCLEAR FISSION-FUSION REACTIONS

Table 3 summarizes the main quantities mentioned in section 2.1.3 concerning the expected yield for the production of neutron-rich nuclei around N~126 through the fission-fusion mechanism (on the basis of 300 J available laser pulse energy and assuming a factor of 100 stopping range reduction by collective effects).

*Table 3. Compilation of relevant parameters determining the expected yield (per laser pulse) of the fission-fusion reaction process.*

| | Normal stopping | Reduced stopping |
|---|---|---|
| Production target: | | |
| $^{232}$Th | 560 nm | 560 nm |
| $CD_2$ | 520 nm | 520 nm |
| Accelerated Th ions | $1.2 \times 10^{11}$ | $1.2 \times 10^{11}$ |
| Accelerated deuterons | $2.8 \times 10^{11}$ | $2.8 \times 10^{11}$ |
| Accelerated C ions | $1.4 \times 10^{11}$ | $1.4 \times 10^{11}$ |
| Reaction target: | | |
| $CH_2$ | 70 μm | – |
| $^{232}$Th | 50 μm | 5 mm |
| Beam-like light fragments | $3.7 \times 10^{8}$ | $1.2 \times 10^{11}$ |
| Target-like fission probability | $2.3 \times 10^{-5}$ | $2.3 \times 10^{-3}$ |
| Target-like light fragments | $3.2 \times 10^{6}$ | $1.2 \times 10^{11}$ |
| Fusion probability | $1.8 \times 10^{-4}$ | $1.8 \times 10^{-4}$ |
| Fusion products | 1.5 | $4 \times 10^{4}$ |

The first and most important milestone on the way to prove this laser-driven nuclear reaction mechanism is to develop the acceleration of high density Th ion bunches with the desired energy. We note that the energy of 7 MeV/u means a conversion efficiency of about 13% for the 250 J pulses, which is indeed predicted for the RPA process.

### 4.2 NUCLEAR (DE-)EXCITATION INDUCED BY LASERS

In this section we will refer mainly to the case of $^{84m}Rb$ described in section 2.2. The laser intensity needed to produce plasma in the favorable conditions for NEET process (density of $10^{-3}$ – $10^{-2}$ g/cm$^2$ and temperature around 400 eV) is around $10^{15}$ W/cm$^2$. For the uncompressed ELI-NP pulses of 250 J and ~1 ns duration, this laser intensity is obtained for a focal spot of about 150 μm diameter. These conditions are rather similar to those achieved [55] during the experiment at the PHELIX laser at GSI. The thickness of the ablated layer is about 30 μm. If we denote with $N_{84m}$ the number of nuclear isomeric states in the ablated volume, the number of NEET transitions per laser shot is:

$$N_{NEET} = N_{84m}\, \tau\, \lambda_{NEET}$$

where $\tau$ is the time spent by the isomers in the plasma volume having favorable parameters, and $\lambda_{NEET}$ is the NEET process rate as defined in section 2.2. Due to the fast expansion of plasma, the typically values of $\tau$ are 30 ps ($=3\times10^{-11}$ s).

The values of NEET rates are obtained based on refined atomic transition calculations (MCDF) for the different charge states present in the plasma and the energy of the nuclear transition which has been measured with an uncertainty of +/- 6 eV (1σ) around 3,498 keV [55] in two experiments performed at the Tandem accelerator of IPN-Orsay and the ELSA accelerator of CEA-Bruyères le Châtel. In the following estimations we will consider the range of possible $\lambda_{NEET}$ values [55]:

$$4.8\times10^{1}\ s^{-1} \le \lambda_{NEET} \le 6.68\times10^{4}\ s^{-1}.$$

In the case of $^{84m}Rb$, the NEET process corresponds to the excitation of a nuclear state with 9 ns half-life, emitting 218.3 keV and 248 keV gamma rays. To avoid pile-up in a $LaBr_3$ gamma detector, not more than one count per detector per laser shot is imposed. Supposing an efficiency of the gamma detector of $\varepsilon_{eff}=10^{-3}$, we conclude that for $N_{NEET}=1/\varepsilon_{eff}=10^{3}$ we need to have:

$$5\times10^{8}\ \text{isomers} \le N_{84m} \le 7\times10^{11}\ \text{isomers}.$$

These isomers can be obtained accelerating carbon ions with the ELI-NP 100TW@10 Hz outputs. On a similar laser, at Astra-Gemini, they succeeded to accelerate carbon ions in a huge quantity with the TNSA process [94]. The energy distribution of the laser-accelerated carbon ions is shown on Figure 15(a). Such ion energy distribution can create $2.3\times10^{6}$ $^{84m}Rb$ isomers per shot in a 30 μm thick germanium target via the $^{76}Ge(^{12}C, p+3n)^{84m}Rb$ reaction at incident energies around 60 MeV. If the germanium target is irradiated with the 100TW@10Hz laser during one hour, we obtain 3.5 $\times 10^{10}$ $^{84m}$Rb nuclei available to be excited in a hot and dense plasma. This number of isomers is already two orders of magnitude higher than the needed one taking into account the most optimistic NEET rate of $6.68\times10^4$ s$^{-1}$. In the worst case, where we consider the lowest NEET rate, to produce $10^{12}$ isomers, one needs two orders of magnitude more carbon ions per laser shot compared to the Astra-Gemini experiment. Such numbers of accelerated particles are compatible with the yield estimations for the RPA regime.

However, the techniques of plasma trapping using pulsed magnetic fields described in section 2.3.3 allow for significant increase of plasma lifetimes in the favorable conditions of density and temperatures. If $\tau$ is increased from 30 ps to the nanosecond–microsecond range, the above required number of isomeric states is reduced by the same factor. Under these conditions, the observation of the NEET effect becomes indeed feasible. In the promising situation of plasma trapping for a duration of ~1 μs, the NEET/NEEC/BIC effects become observable even for short-lived isomers, which may be produced by one 10 PW pulse and followed shortly by plasma creation with the second high-energy (250 J) uncompressed pulse. For example, the 18.8 s – 124.7 keV isomer in $^{90}$Nb (see Figure 10) is measurable in these conditions, with further advantage that the lower excitation energy isomer has 63 μs half-life, which allows to detect hundreds of counts per detector in a shot without pile-up. The disadvantage is that the isomer decay should be observed immediately after the 10 PW pulse. Reduction of prompt gamma ray interactions in the detector can be achieved placing the secondary target far from the acceleration target, installing adequate shielding, and combining the acceleration methods with refocusing techniques mentioned in section 2.4.2, in order to keep the ion beam size on the second target at the level of 150 μm, i.e. equal to the laser spot for plasma generation.

### 4.3 NUCLEAR REACTIONS IN A LASER PLASMA

In this section the technical details of the proposed activity in section 2.3 are presented. The simulated ion energy distributions [91] for carbon and lithium ions, generated by the interaction of a fs laser pulse impinging onto a thin target at two different laser intensities, are shown in Figure 52. The predictions were validated also by comparison with experimental data reported in literature [93, 94]. It is evident that, through the laser focal spot tuning on target, we are able to change both the beam intensity and the ion energy distribution, which can be properly tailored according to the energy intervals one would explore.

When working in the range of few $10^{18}$ W/cm$^2$, it turns out that the energies of the produced ions are suitable to generate reactions of astrophysical interest. Due to hydrogenated contaminants on the target surface, we expect a fraction of protons, as shown in Figure 52, during the production of carbon, lithium or boron. Such protons could be a source of secondary neutrons when colliding with the aluminum chamber walls, generating an unwanted background in the neutrons' detectors. This possibility is of course strongly dependent on the initial proton energies. Working at an intensity of the order of $10^{18}$ W/cm$^2$, we expect maximum proton energies of about 3 MeV, which is below the neutron production energy threshold on nuclei (including the weakly bound deuterons of the D gas-jet target), thus producing a negligible neutron background contribution as originating from target protons.

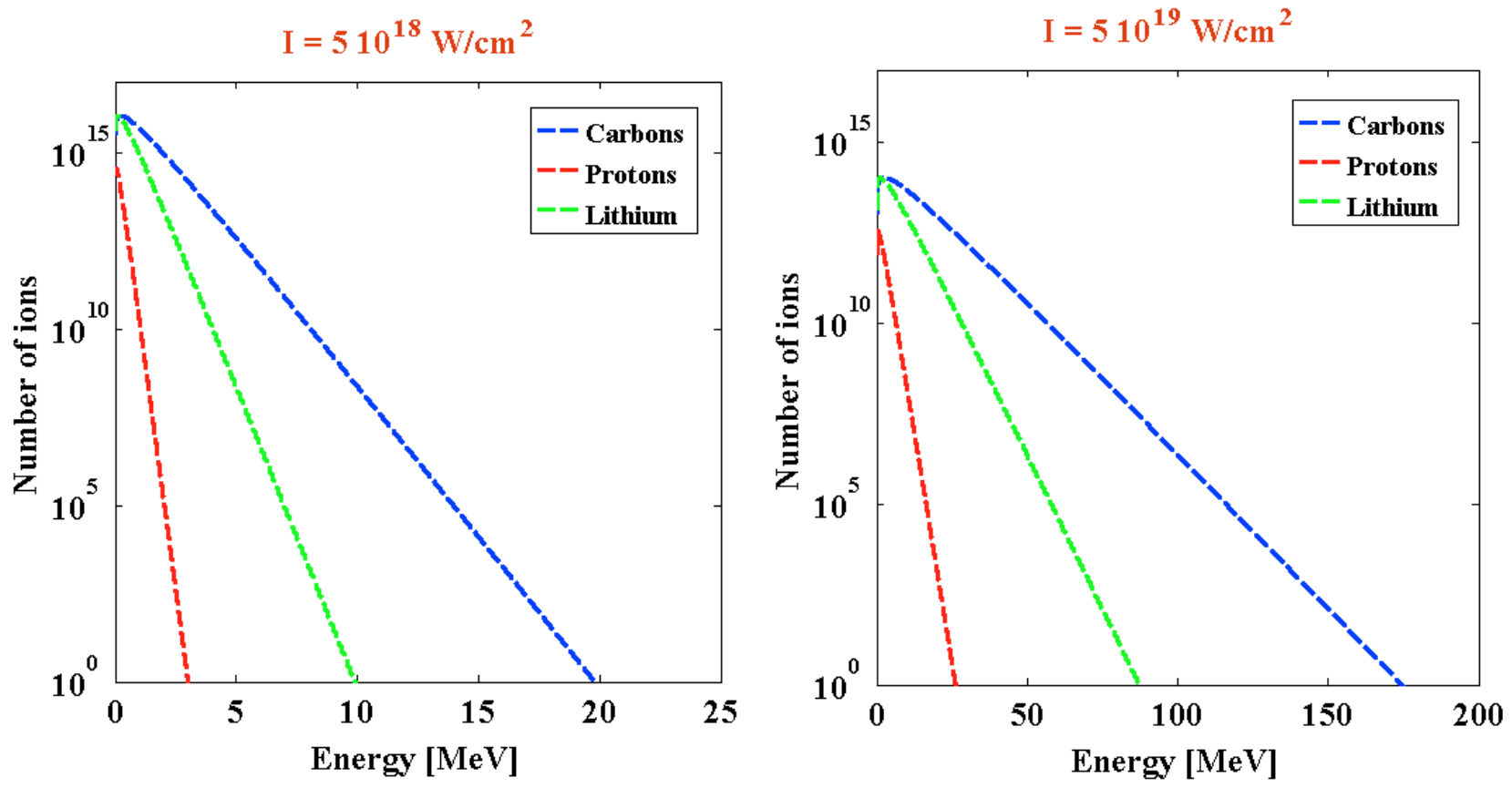

Figure 52 – Energy distributions of carbon and lithium ions as well as of protons, simulated assuming the interaction of one petawatt laser on a 5 μm thick carbon or lithium foil. The total yield of protons was evaluated assuming a 10 nm layer of hydrogen contaminant on the surface [89]. The distributions were obtained for two intensity values, 5 x$10^{18}$ and 5 x $10^{19}$ W/cm$^2$, obtainable by working with two focal spot radii, 160 and 50 μm, respectively.

In order to maximize the ion yields, and to better and better tailor the ion energy distributions, several techniques in the optimization of the laser-matter interaction can be attempted. We will particularly focus our efforts in R&D activities about nano-structured target surfaces as described in Section 2.3.2.1, which can remarkably modify the efficiency of laser radiation absorption, as well as the properties of the nascent plasma.

### 4.3.1 Gas-Jet Target: Thin Mode

In this configuration, we intend to minimize the "plasma-plasma friction", i.e. the energy dissipation of the fast flowing plasma colliding with the gas-jet plasma, in order to work in a more "classical" nuclear physics experimental scheme (i.e. projectiles on a fixed target).

Specific simulations have been done in order to describe the experimental conditions under these assumptions. In particular, complete neglecting the plasma effects and setting the helium or deuterium gas density at about $10^{18}$ atoms/cm$^3$ and using a gas jet thickness of few mm (e.g. $d$ = 5 mm) we estimate an energy threshold of few keV above which the stopping power of the incoming carbon ions can be neglected (i.e. no thermalization takes place in the gas-jet target); for boron and lithium ions these energies are even lower. Practically, in this way all carbon or lithium ions pass through ionized gas (a plasma target), except the ones producing the nuclear reactions we want to study. In such conditions, the reaction rates can be easily estimated, once the corresponding astrophysical factor can be assumed to be known. Of course, the simulations do not take into account the effects of the plasma environment on the reactions (i.e. how does the plasma affect the reaction rate), but this is actually the main aim of the proposal.

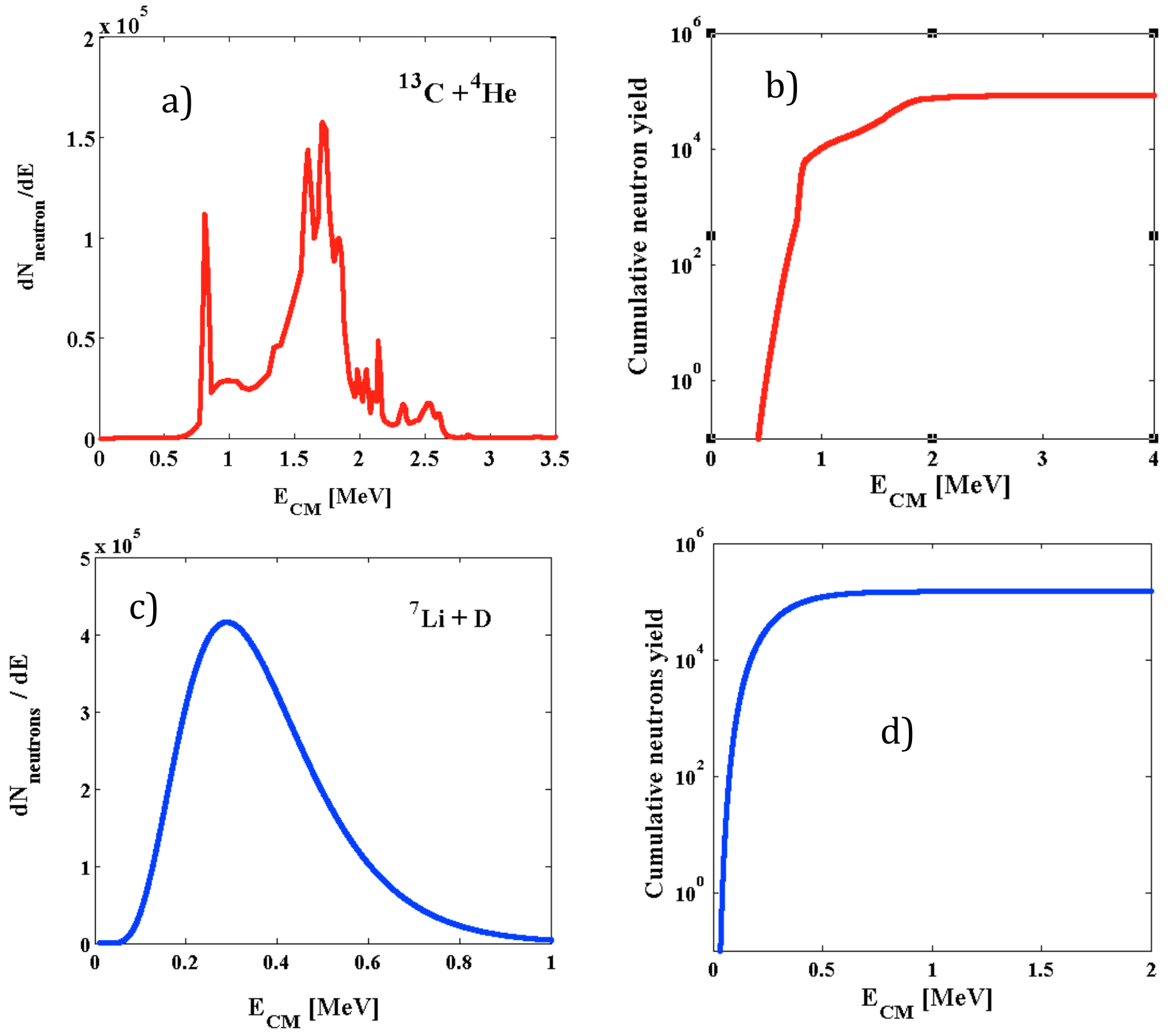

Figure 53 – Expected number of neutrons as a function of the center of mass energy for the: $^{13}C$ + $^{4}He$ a) and b), $^{7}Li$ + d reactions c) and d). The yields are corresponding to one single laser shot of 0.1 petawatts on, 1 μm thick, carbon or lithium target, assuming an intensity of $3\times10^{18}$ W/cm$^2$ for the $^{13}C$ + $^{4}He$ and $2\times10^{18}$ W/cm$^2$ for the $^{7}Li$ + d.

In Figure 53 the expected number of neutrons as a function of the center of mass energy for the $^{13}C$ + $^{4}He$ and $^{7}Li$ + D reactions are plotted, along with the corresponding cumulative yields. These calculations are related to "single shot" laser pulses of 0.1 petawatts on carbon or lithium target, 1 μm thick, assuming an intensity of $3\times10^{18}$ W/cm$^2$ for the $^{13}C$ + $^{4}He$ and $2\times10^{18}$ W/cm$^2$ for the $^{7}Li$ + D reaction. Such parameters were chosen to ensure the detection of about one neutron for each detection module (see section 3.4.1) per shot; this also takes in to account that the overall efficiency for a single neutron detector placed at two meters from the target is about $10^{-5}$ [166].

Under these conditions, we can reach a good sensitivity in the determination of center of mass energies in the range between 0.8 and 2.6 MeV for the $^{13}C$ + $^{4}He$ reaction, and between 0.1 and 0.8 MeV for the $^{7}Li$ + D, while the statistical significance of the collected data is guaranteed by the performance (repetition rates up of 10 Hz) which will be achievable at the ELI-NP facility.

In such operative conditions we are also able to estimate the astrophysical factors by reconstructing the neutron energies using time of flight (TOF) measurements. In Figure 54 we plot the center of mass energy uncertainty as a function of the neutron-detected energy for both reactions. The calculation was done assuming a standard 1 ns overall time resolution, but we are confident to improve this value up to 0.5 ns or better, which is particularly useful in the case of the $^{7}Li$ + d reaction.

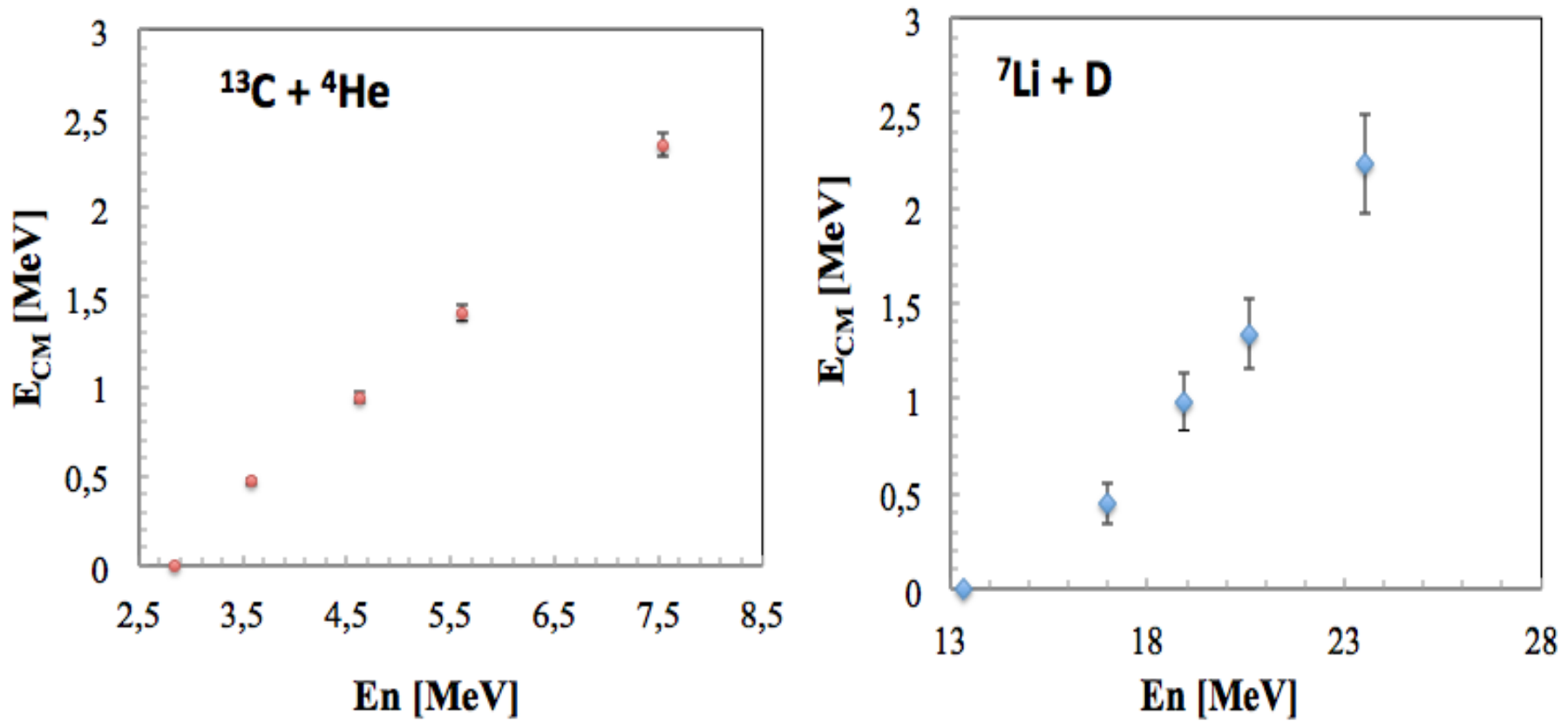

Figure 54 – Center of mass energy as a function of the detected neutron energy for $^{13}C$ + $^{4}He$ and $^{7}Li$ + D reactions. The error bars are obtained assuming an overall time resolution of 1 ns.

In order to change the working energy range for both reactions at higher values, it is possible to increase the laser intensity (up to $5\times10^{18}$ W/cm$^2$) and use a thin degrader (few μm of Mylar), introduced between the carbon or lithium target foil and the gas-jet. Such operative conditions allow the tuning of the carbon or lithium energy distributions in order to significantly reduce the contribution of the low energy part of the spectra. Increasing the laser intensity, the laser spot size considerably reduces thus producing a smaller amount of C/Li ions available for the reactions; this problem can be partially compensated thanks to the unique ELI-NP laser, which allows to operate at the full laser power (10 PW), but still in high repetition rate mode.

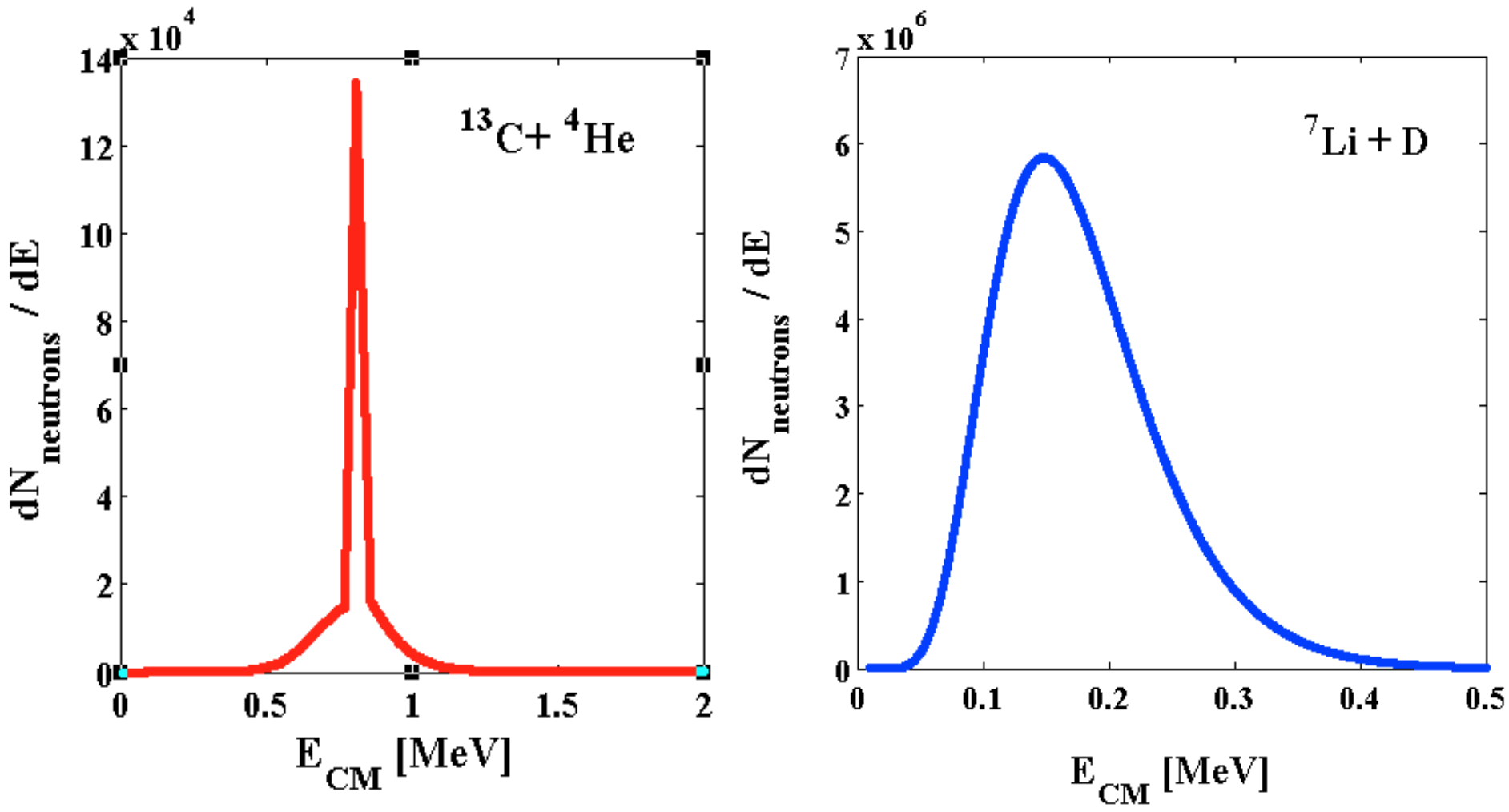

Figure 55 – Neutron energy distributions for the $^{13}C$ + $^{4}He$ and $^{7}Li$ + D reactions at 10 petawatts laser power at $0.7\times10^{18}$ w/cm$^2$ of intensity, obtained by using 10 μm of carbon foil and 1 μm lithium foil. The gas jet density is $10^{18}$ Atoms/cm$^3$

For the low center-of-mass energy domain (i.e. <0.8 MeV), the cross-section is drastically reduced, but also in this condition the formidable ELI-NP laser performances should allow to obtain – with sufficient accumulated statistics – meaningful physical results. By decreasing the laser intensity (to $0.7\times10^{18}$ W/cm$^2$ for example), the number of ions produced at each laser shot considerably increases, obtaining a significant increase of the rates in the following energy ranges: 0.5 – 1 MeV for $^{13}C$ + $^{4}He$ and 50 - 400 keV for $^{7}Li$ + d. When working at the full power pf 10 PW and maximum repetition rate, without changes in the experimental setup, we are also able to extract information on $^{7}Li$ + D. In Figure 55 the neutron energy distributions for both reactions are plotted, again we still expect to detect – for the $^{7}Li$ + d – a single neutron per each detection module per shot. For the $^{13}C$ + $^{4}He$ reaction (as can be noticed by the expected yields) it will be necessary to change the TOF path, reducing it at 1 m, in order to increase the overall efficiency of a single neutron detection module up to $4\times10^{-4}$, thus also increasing the detection efficiency to about 0.5 neutrons per shot.

As an alternative to the aforementioned technique, and for further investigation at lower energies, we also plan to work in a thick gas-jet target configuration, which will be discussed in the next section.

### 4.3.2 Gas-Jet Target: Thick Mode

Operations in the "thick" gas-jet target mode (that means most of the streaming primary plasma is stopped/thermalized into the secondary one) are possible, when the gas density (up to $10^{20}$ atoms/cm$^3$) is increased in addition to its physical thickness (Figure 14(a)); the changes should contribute to enhance the total reactions rates at lower energy (or lower temperature). In particular, the fast streaming carbon or lithium plasma will be stopped at the plasma jet target, forming a new plasma made by a mixture of carbon and helium, or lithium and deuterium. After the thermalization phase (tens of pico-second), such plasma has a final temperature that is determined by both the initial temperatures of the two starting plasmas. This temperature, as well as the final plasma volume, can be properly measured.

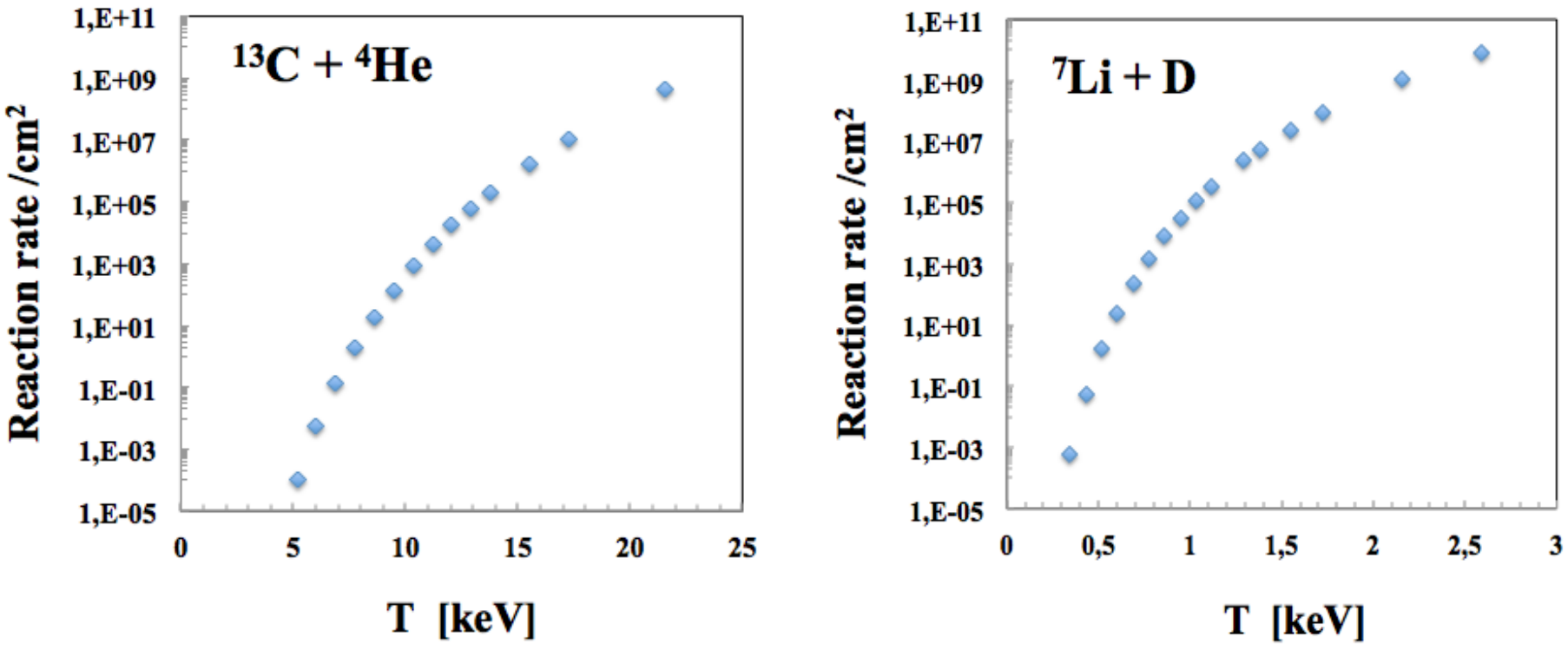

Figure 56 – Neutron yields/cm$^2$ as a function of the final plasma temperature for the $^{13}C$ + $^{4}He$ and $^{7}Li$ + D reactions at 10 PW. The estimates have been obtained assuming to use a carbon foil of 10 µm thickness and $10^{20}$ He/cm$^3$ as gas jet density for the generation of $^{13}C$ + $^{4}He$ plasma and a 1 µm thick lithium foil and $10^{20}$ D/cm$^3$ for the $^{7}Li$ + D plasma.

Also in this case, the plasma temperature tuning is enabled by changing the laser parameters, the solid target thickness, the gas-jet target density and the secondary plasma temperature etc. The expected neutron yields for both reactions are shown in Figure 56. As can be noticed, there is a direct access to the reaction rates as a function of the temperature, which is the main goal of the astrophysicist, moreover, also in this case it is possible to extract in an indirect way the astrophysical S factor by de-convolving the thermal distributions.

In conclusion, through the proposed setup and thanks to the unique characteristics of the ELI-NP laser facility it will be possible to explore nuclear reaction rates in energy regions otherwise inaccessible and, more importantly, inside a plasma. These studies can be, obviously, extended to many others astrophysical relevant reactions.

### 4.3.3. Multifluid code development for investigations on dense plasmas of different compositions in high magnetic field for fusion and astrophysics studies

The numerical investigations describing the experiments making use of the proposed LB-CMPF described in section 2.3.3 are based on the development of two separated codes. The first concerns a 2-D single fluid resistive MHD code in cylindrical geometry, describing the spatio-temporal evolution of the high density and high temperature plasma in a high external applied mirror-like magnetic field. The second is a 2-D multi-fluid code in cylindrical geometry. The development of a multi-fluid code allows describing in more details the spatio-temporal evolution of the state parameters of each of the species of the plasma and of the fusion reaction products, including specific topological profiles for the externally applied magnetic field. The use of the multifluid code allows to study the burning process of high density ($10^{19}$ cm$^{-3}$ or higher), high temperature (from tens to hundreds of keV) plasmas trapped and oscillating in a high (up to 120 T) external applied magnetic field. The computer running time for the 2-D multi-fluid code is extremely long, so for the needs of our proposal we split the numerical simulation in two 1-D multi-fluid codes, enabling to describe the spatio-temporal evolution of the state parameters of the plasma and for different external magnetic field topologies. We present the numerical results for two different fusion fuels. The first concerns the D-D nuclear fusion reaction and the second the p–$^{11}$B reaction. Both the configuration and the initial plasma production in the LB-CMPF device are described in section 2.3.3. For both fuels, the initial plasma density (about $10^{18}$ e/cm$^3$) is the same and we applied the same external magnetic topology with initial magnetic field up to 110 T-120 T.

*(a) 2-D single fluid MHD simulation in cylindrical geometry*

Experimental studies [103-106] confirm that high intensity ultrashort laser beam interaction with deuterated clusters could produce a plasma with a density up to $10^{19}$ cm$^{-3}$ and D ions with high kinetic energy from 10 keV to 50 keV, capable to initiate D-D (or D-T) ion fusion nuclear reactions. The plasma expands very fast in vacuum, decreasing both the local density and the number of D-D ion nuclear fusion reactions. The application of an external high magnetic field in mirror-like topology enables to decrease the plasma expansion velocity, increases the trapping time of the plasma and improves the neutron production. The number of the produced neutrons depends strongly on the initial spatial plasma distribution (‘plasma volume’) which is limited in the case of the ‘single-focal-spot’ of the optical focusing system. A new idea to improve the initial ‘plasma volume’ is based on the effect of the self-guiding propagation of an ultrashort laser beam in the plasma. Under these conditions of propagation, the interaction 'plasma volume' increases considerably, preserving all the plasma parameters as in the case of a single focal spot. The laser propagation produces a relatively long filament [111] with an important number of multi-focal spots (or hot spots), due to the non-linear self-guiding propagation of the laser in the plasma. The proposed new scheme of interaction allows to improve both the number of neutrons and the coupling efficiency of the energy transfer from the laser pulse to the plasma. Figure 18 shows experimental results of the ultrashort laser beam propagation in a neutral deuterated cluster jet and the production of a relatively long filament with an important number of multi-focal spot (or hot spots) in the plasma volume. This experimental result confirms the increases of the interaction volume by an important factor (see the number of hot spots in Figure 18) due to the self-guiding propagation of the ultra-short laser beam in the clusters.

*A 2-D single fluid resistive MHD code*

A 2-D resistive MHD code [108, 109] in axisymmetric cylindrical geometry, coupled with the magnetic field equations, has been developed in order to study the global trapping time of the plasma in the LB-CMPF device, the radial and axial plasma losses from the device, the burning efficiency (neutron production and alpha production) and the effect of the initial plasma conditions in the case of the non-linear laser propagation in the clusters. The code can handle very large magnetic fields and very steep gradients of the plasma parameters; with mirror-like magnetic field. A double-plasma spot was considered [112], describing the initial plasma distribution with a plasma density up to $10^{18}$ cm$^{-3}$ and temperature of 10-50 keV for the self-guided (non-linear laser beam propagation) region and a plasma density of $10^{16}$ cm$^{-3}$ for the surrounding the multi-plasma spot region. We investigate the trapping time and the plasma spatio-temporal evolution of the plasma density for two cases. The first corresponds to the double-plasma spot and the second to the single focus case. For both cases, an external mirror-like magnetic field up to 120 Tesla was considered with a ratio of $B_{max}/B_{min} \approx 2$. Figures 57(a) and 57(b) present the comparison of the spatio-temporal evolution of the plasma density for the double-plasma spot and the single-focal spot respectively [112]. The time scale corresponds to an interval of 10 ns. The horizontal axis in Figure 57 represents the plasma dimensions in cylindrical (r, z) coordinates and the vertical axis represents the value of the plasma density. For the double-plasma spot case, we observe an oscillation of the plasma density producing 2.5 times longer trapping times with high density compared to the single focus case. These numerical calculations show that in such a configuration the neutron production is up to $5\times10^{9}$ neutrons per laser shot, when the external applied magnetic field is up to 120 Tesla and the time interval 10 ns. For longer trapping time up to 1 μs the proposed configuration allows to improve the neutron flux (see next section on multi-fluid code).

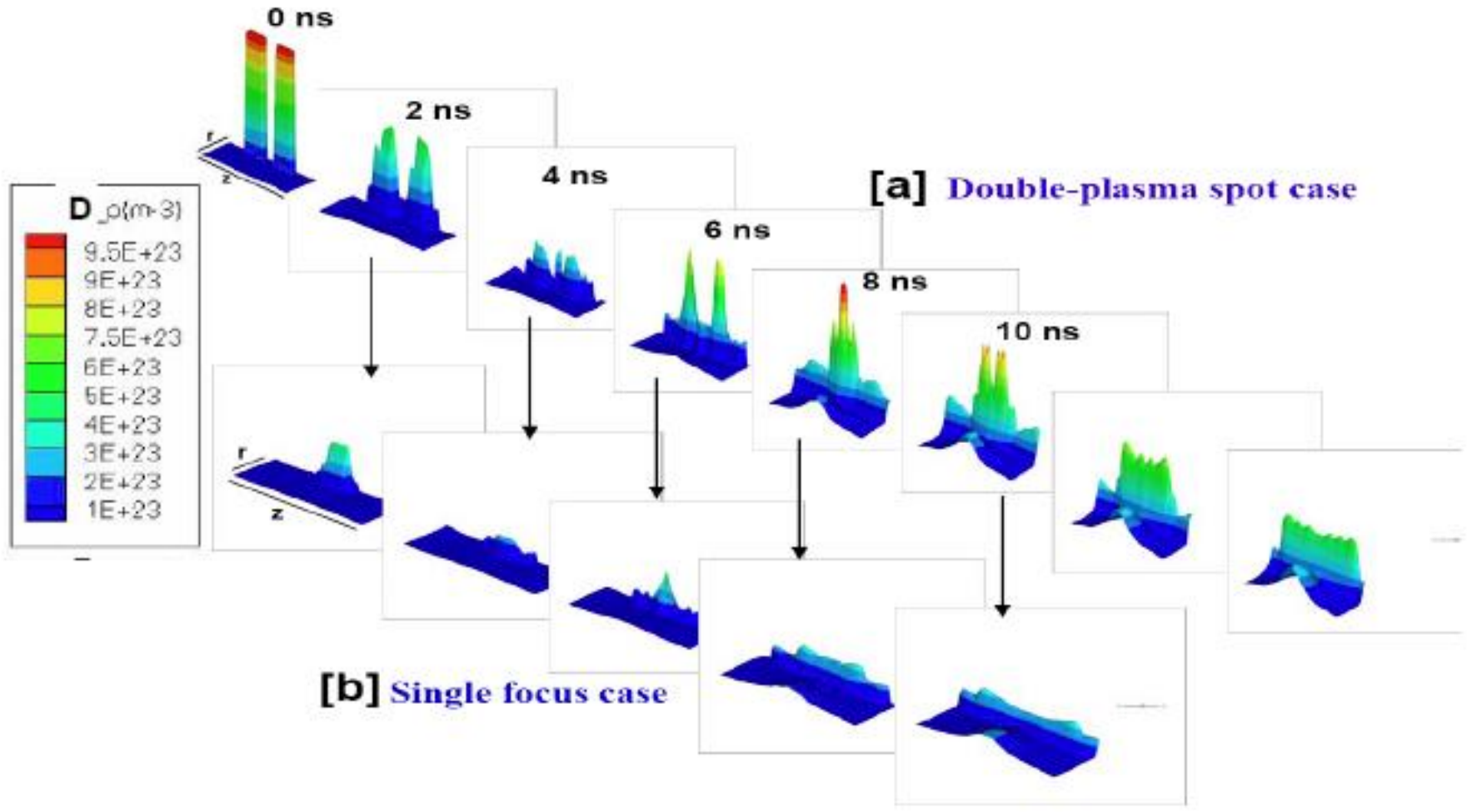

Figure 57 – The spatio-temporal evolution of the 2-D (r: radial and z: axial) plasma density for the [a] double-plasma spot and [b] single-focal spot. Both correspond to a time interval of 10 ns.

As we mentioned previously, in the context of the present proposal the laser beam intensity necessary to produce a high density ($10^{18}$ cm$^{-3}$) and high temperature (20 – 70 keV) plasma from laser beam interaction with clusters or thin solid target is up to $10^{17}$ W/cm$^2$. However, the volume of the produced plasma is small due to the dimension (section) of the focal spot of the laser beam, as is consequently the number of nuclear reactions (small reaction rate). The use of PW laser beams enables to increase both the energy in the laser beam and the dimension corresponding to the laser-plasma interaction, conserving the intensity up to $10^{17}$ W/cm$^2$. The non-linear propagation of the laser beam and the produced filaments increases the length of the produced plasma in the proposed LB-CMPF device. Both considerations, concerning the PW laser beam and the filaments, allow for increasing the interaction volume and improving the reaction rates for the different fuels. For the numerical simulation concerning the spatiotemporal evolution of the plasma in the proposed device, we consider a high plasma density up to $10^{18}$ cm$^{-3}$ for both fuels, DD (or DT) and the $p^{11}B$, but different plasma temperatures, because the cross section of the fusion nuclear reaction, for each fuel depend strongly on the temperature. The initial temperature, for the following numerical simulations, is 30 keV for the DD plasma and 120 keV for the $p^{11}B$ plasma. For both cases (fuels), the external applied magnetic field is up to 110 T.

*The multi-fluid code*

The multi-fluid code describes the spatiotemporal evolution of the reacting plasma species (electrons, protons, deuterons, boron and alphas where are treated as fluids) plus the electric and the magnetic fields, in different geometries (spherical, cylindrical or Cartesians in the case of the laser beam interaction with thin discs). The evolution of the electric and externally applied magnetic fields (which are described by Maxwell's equations) has important effects on the expansion of the plasma species (species separation and trapping of the plasma species). The magnetic trapping of the plasma is very important, because the plasma density (ion species density) remains high for a relatively long time, improving the fusion nuclear reaction rates by a substantial factor. The pulse duration of the high magnetic field driver described in the section 2.4 is up to a few microseconds, enabling to investigate simulations using the multifluid code for different plasma fuels in the proposed LB-CMPF device for at least 1μs. This time interval is relatively long and very favorable for the fusion nuclear reaction rate if the external applied magnetic field is sufficient for trapping of the initial high plasma density for the same temporal interval.

*1-D radial multi-fluid simulations for $p^{11}B$ fuel*

In the case of the $p^{11}B$ fuel, the cross section for efficient reaction rates corresponds to a temperature from 120 keV to 600 keV. To achieve this relatively high plasma temperature in the proposed LB-CMPF device with a moderate laser beam intensity of the order of $10^{17}$ W/cm$^2$, a new scheme on laser plasma production is proposed. This scheme is based on the plasma block acceleration, due to the interaction of high contrast, short laser pulses with thin solid targets. Figure 58 presents two different schemes, capable to initiate high density and high temperature of a $p^{11}B$ plasma, in the proposed device. In the first scheme two thin solid discs are placed near the magnetic mirrors of the LB-CMPF device. Two high contrast PW laser beams interact with the targets and produce high temperature protons and boron ion plasmas from the two thin solid discs, respectively (see also the description in the previous section). The plasmas are propagated inwards the magnetic configuration, following the magnetic flux tubes (see Figure 58(a) and Figure 58(b)).

The expansion of both plasmas (fluids) in the magnetic device enables to form a high density ($10^{18}$ cm$^{-3}$) and high temperature (> 120 keV) plasma composed of protons, boron ions, electrons and alphas. The spatiotemporal evolution of the plasma species described by the multi-fluid code present similar effect (see oscillations in Figure 59) as these presented for the DD fuel (Figure 58(a)). Figure 60(a) presents the density (in m$^{-3}$) of p and B ions as a function of the radial dimension of the plasma and Figure 60(b) presents the alpha density (in m$^{-3}$) distribution as a function of the radial dimension of the plasma. Both figures correspond to a magnetic trapping time of 0.4 μs. Numerical simulations for longer trapping times up to 1μsec show that approximately $8\times10^{10}$ alphas (see Figure 59) are produced from the fusion nuclear reaction of $p^{11}B$. If the plasma temperature is higher, up to 300 keV the alpha production increases approximately by a factor of 5 due to higher value of the cross section of the fusion nuclear reaction.

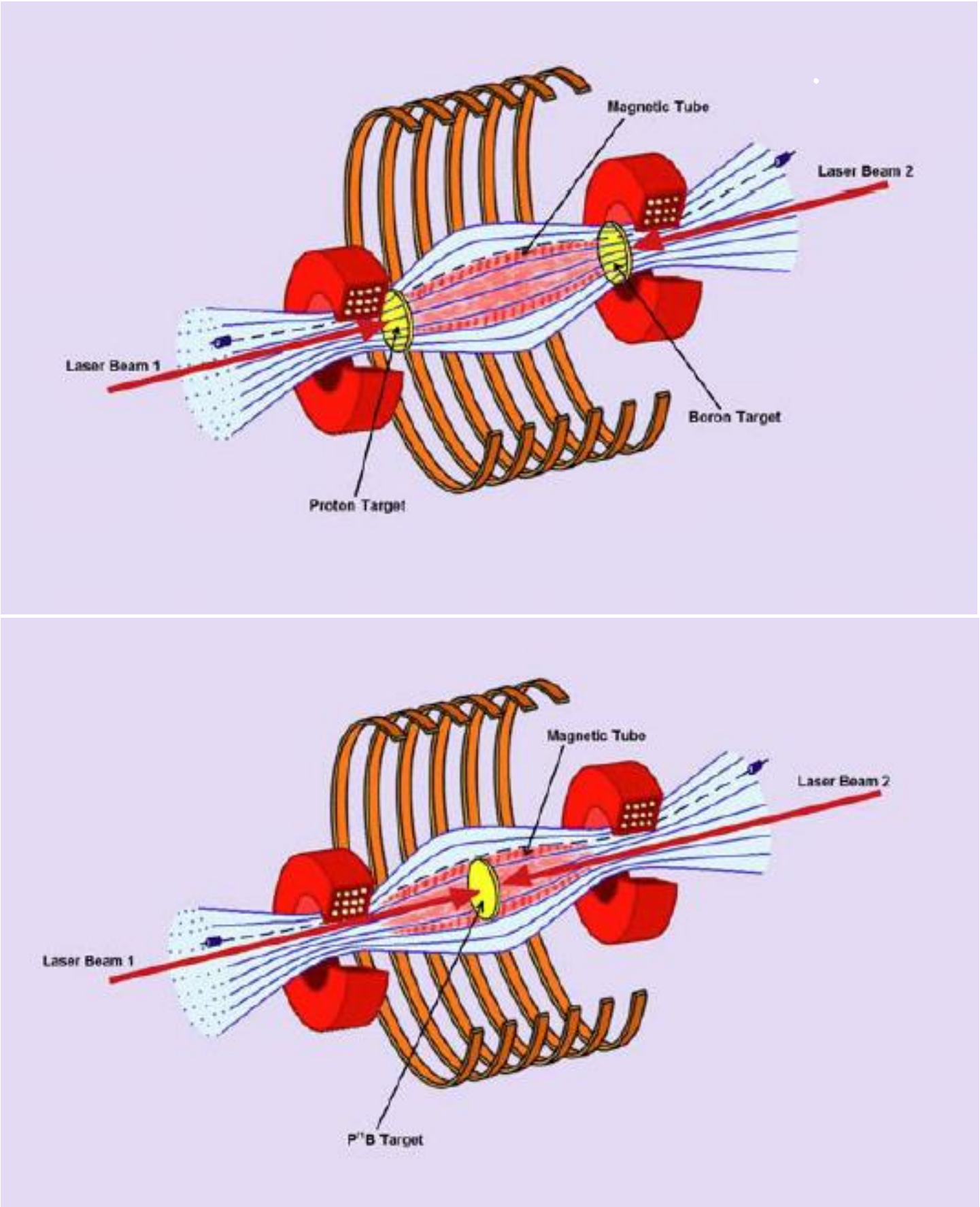

(a)

(b)

Figure 58 – (a) and (b) The proposed new schemes for numerical and experimental investigations on the $p^{11}B$ fusion nuclear reaction and alpha production. The external applied magnetic field is 110 T.

The second scheme concerns to place a thin disc containing both the protons and the boron, in the center of the LB-CMPF device and irradiate each surface of the disc with a PW laser beam in order to produce two high-temperature ion beams. The trapping of both beams in the mirror-like magnetic topology for 1 μs, forms a high temperature and high density plasma (electrons, p, B, alphas). The plasma density remains relatively high during the trapping time with high reaction rates for the produced alphas. The numerical simulation gives similar results as for the previous configuration (Figure 58(a)) concerning the alpha production.

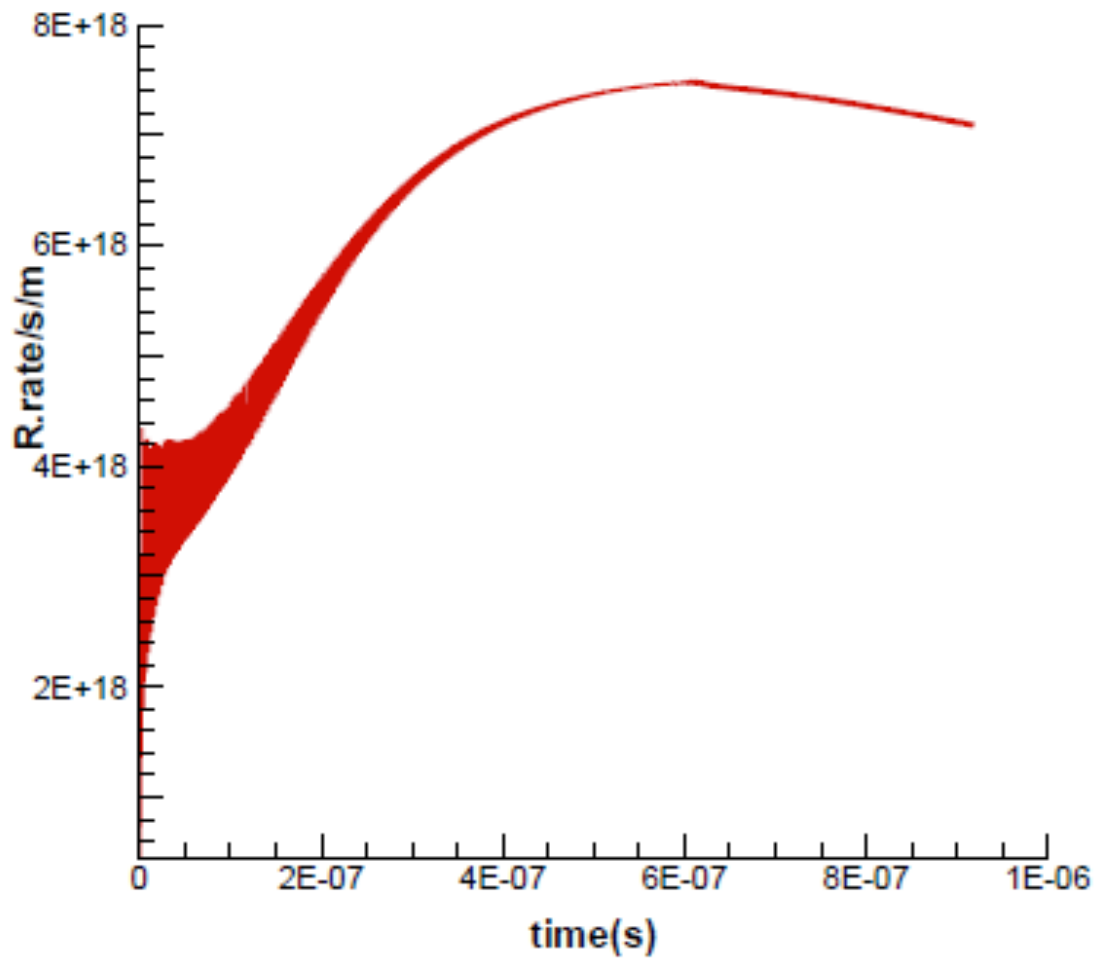

Figure 59 – The $p^{11}B$ fusion reaction rate as a function of the trapping time of the plasma in the LB-(CMPF) device. The plasma temperature is 120 keV, the external applied magnetic field is 110 T and the alpha production is approximately up to $8x10^{10}$.

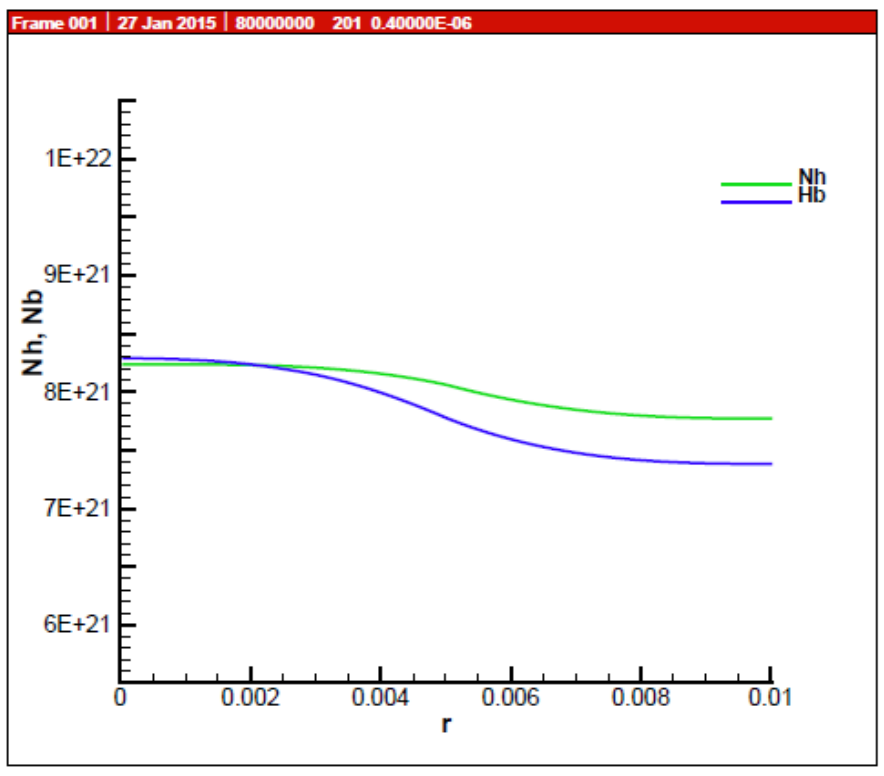

(a)

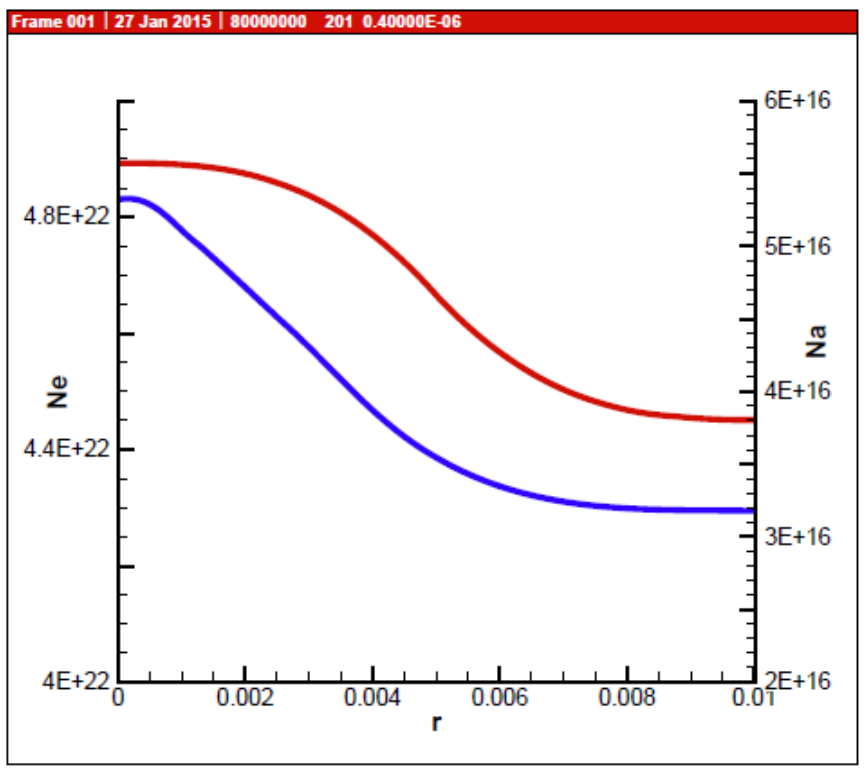

(b)

Figure 60 – (a) Spatial density profile of the protons (blue curve) and the boron ions (green line) for 0.4 µs trapping time. (b) Spatial distribution of the produced alphas (blue line) after 0.4 µs from the laser beam impact on the target.

This result of alpha particles generation may be supported by the measurements of approximately $10^8$ alphas in the experiment reported by G. Korn et al. [114]. This high value above the earlier experimental result [113] permits the conclusion that there is an important experimental proof of the secondary reactions with an avalanche increase of the gain of $p^{11}B$ fusion used for the computations of plasma block ignition of solid fuel within a cylindrical trapping by the measured ultrahigh magnetic field of 10 kT [121, 127, 198]. The nonlinear – force-driver block acceleration by picosecond pulses in the PW range was summarized before [199] and the combination with the magnetic trapping for both the $p^{11}B$ fusion [200] and the DD (DT) fusion [201] is one of the mechanisms studied in the present numerical investigation.

The multifluid code (see section 2.3.3) enables to study the production of high energy ions (hydrogen or deuterium) up to hundreds of MeV from laser beam interaction with thin solid targets. The use of the multifluid code for the ion acceleration allow to describe the spatiotemporal evolution of the initial produced plasma and the ions during the acceleration process The code enables to describe the ion beam parameters (energy and density) as a function of the initial thickness of the solid target. The interaction of high contrast short laser pulses with thin solid targets allows plasma block acceleration and production of ion beams with densities very close to the solid density.

## 4.4 NEUTRON GENERATION AND OTHER APPLICATION

The neutron yields for the laser-ion driven method described in section 2.4.1 are evaluated by scaling the yields measured at the 200 TW Trident laser facility [131]. These experiments measurements yielded $10^{10}$ n/sr per shot. Assuming scaling of the ion energy with the square root of the laser peak power [202], will give a scaling factor of:

$$\sqrt{P_{ELI-NP}/P_{Trident}} = \sqrt{10\ \text{PW}/0.2\ \text{PW}} = 7.1.$$

In Ref. [203] the total neutron yields for beams of various ion species impinged on neutron-converters of different materials are shown. For ion energies of about 100 MeV, the increased ion energy corresponds to about an order of magnitude increase of the neutron yields. Thus, we can project $10^{11}$ n/sr per laser shot at ELI-NP.

To evaluate the expected neutron yields for the laser-electron driven method described in section 2.4.2 we scale the measured yields from the Texas Petawatt laser [142]. The electron temperature in this study was measured to be $T_{e-Texas} = 10.5$ MeV. Assuming a $T_e \sim \sqrt[3]{P_L}$ of the electron jet temperature with the laser power, we can expect:

$$T_{e-ELI-NP} \sim T_{e-Texas} \times \sqrt[3]{P_{L-ELI-NP}/P_{L-Texas}} = 10.5\ \text{MeV} \times \sqrt[3]{10\ \text{PW}/1\ \text{PW}} = 23.1\ \text{MeV}.$$

The neutron yield as function of electron energy for various material are given in Ref. [204]. The increased electron temperature will correspond to about an order of magnitude increase of the neutron yields. We therefore project a few times $10^{10}$ neutrons per shot emitted isotropically.

## 5. SPECIFIC NEEDS AND UTILITIES, TRANSVERSAL NEEDS

The needs derived from the experiments proposed in this TDR for the vacuum systems and the control system have been forwarded and included in the corresponding TDRs.

For the building construction, besides the distribution of the weights in the E1 area according to the layout in Figure 51 (this information is requested for the system of springs and dampers used for the vibrational floor isolation), several requests have been transmitted:

- availability of local cranes and mobile cranes
- access to the plates closing the penetration through the floor in view of future addition of various feedthroughs (electrical/optical/cooling water) towards equipment installed in the basement
- penetration towards the E5 area

No liquid nitrogen pipe has been requested in the E1 area, taking into account that at maximum few detectors will be used within the decay station for a rather limited period of time, when $LN_2$ could be supplied to the detectors via mobile Dewar vessels of medium capacity.

The requests addressed to the electronics laboratory include a cubic vacuum chamber with a volume of about 0.5 $m^3$, where detectors can be tested under high vacuum.

The Target Laboratory is essential for the success of experiments in E1. The thin and ultrathin foils of various materials are needed for the RPA acceleration of heavy ions. The best identified methods for the production on large quantities at reasonable costs are Si wafers processing and deposition of desired materials. Analyzing tools for produced (or procured) targets are mandatory. Complementing the target laboratory with a well-equipped micromachining is also necessary.

## 6. SAFETY REQUIREMENTS

Radioprotection and laser safety are obviously the main aspects to be considered in the design of the equipment and procedures to be employed during the experiments presented in this TDR. Concerning the radioprotection, in area monitoring it is important to be able to measure the prompt radiation flux, such as to prove the fulfilment of the radioprotection goals (total dose rates less than a given level) for a number of actual pulses, that can be much larger than the number of pulses with maximum radiation emission that will be taken as a basis of the radioprotection calculation. Active detectors used currently in radiation monitoring may be saturated in case of laser pulses, when the dose will be delivered in a very short time interval. Therefore, special detectors for pulsed neutron and gamma fields have to be used.

The experiments foreseen in Laser-TDR1 are probably producing the largest EM pulses, compared to other ELI-NP experiments. Beside the shielding put in place to protect the personnel and the equipment, the installation of an EMP monitoring system with fixed and mobile antennas will be very useful for a real understanding of the origin and the effects of EMP, and a clear separation from radiation damages.

Beside the above mentioned hazards and other types of physical hazards obviously associated with experimental work, in the experiments proposed in this TDR present also additional risks that should be mitigated by adequate engineering and procedures:

- explosion due to the use of high pressure gas bottles/tanks
- explosion due to the use hydrogen gas targets
- electroshocks due to high voltages used by Thomson Parabola Ion Spectrometers or by the charging system of the LB-CMPF device
- noise can be associated to capacitor bank discharge of the LB-CMPF device
- chemical hazards may appear for example in processing in CR39 track detectors.

## 7. COLLABORATIONS

The Laser Driven Nuclear Physics TDR is benefiting from collaborations under running research contracts with:

- Nuclear Technologies (UK) on radioprotection studies and
- ICIT - Ramnicu Valcea (Romania) on the target exchange system.
- INFLPR (Romania) on development and test of new diagnostics and targets at CETAL – 1 PW facility.

A similar R&D contract is under preparation with GSI (Germany) on gas cell, RFQ and MR-TOF devices.

MoU’s have been signed also with other institutions collaborating to this TDR:

- MoU with Univ. Giessen (Germany)
- MoU with TU Darmstadt (Germany)
- MoU with INFN LNS (Catania, Italy)
- MoU with Univ. Strathclyde (UK)
- MoU with STFC (UK)
- MoU with IPPLM Warsaw (Poland)

Strong collaboration will continue with the researchers proposing the experiments included in this TDR, affiliated to many other institutions as mentioned in the Contrıbutor list on the second page. Common developments and experiments at running high power laser facilities are essential for the success of future experimental program at ELI-NP.

Together with French teams from CENBG/IN2P3 (Bordeaux), LULI/Ecole Polytechnique (Palaiseau) and CELIA/Univ. Bordeaux, experiments devoted to the optimization of heavy ion acceleration schemes and new diagnostics for the studies of nuclear excitation induced by lasers (section 2.2) are planned at CETAL, ELFIE, LULI2000 or other facilities. Funding of such activities is possible through project(s) within ANR (France) – ANCS (Romania) joint calls for proposals.

The collaboration with LNS-INFN (Catania), University of Catania and INO, CNR-Pisa during next couple of years will be centered on prototyping and testing few modules for the high granularity detection systems described in section 3.4.2 for neutrons and charged particles, emitted by nuclear reactions in the plasma (sections 2.3). Both the new materials foreseen for the detectors and the proposed studies of direct measurement under plasma conditions, of astrophysical relevant nuclear reaction cross sections are expected to attract a wider nuclear physics community, some groups from GANIL (Caen) and IPN-Orsay already expressed their interest to join.

The collaboration built around the proposed dense and hot plasma trapping in high magnetic field for fusion-astrophysics studies and neutron generation schemes (section 2.3.3 and 4.3.3), involves teams from 8 institutions besides ELI-NP:

1. Technical University of Crete (TUC)
2. Institute of Electronic Structure and Laser (IESL-FORTH), Heraklion, Crete
3. Laboratoire de Physique des Plasmas (LPP-EP), at École Polytechnique, Palaiseau, France
4. Bordeaux University I and CELIA Laser Facility, Bordeaux, France
5. Laser und Plasmaphysik, Technische Universität Darmstadt (LP-TUD), Germany
6. Racah Institute of Physics, the Hebrew University of Jerusalem (RI-HUJ), Israel
7. Department of Theoretical Physics, University of New Wales (DTPUNW) in Sydney, Australia
8. Electronic Structure and Laser (IESL-FORTH) in Heraklion, Crete, Greece

The main objectives of the collaboration in the current preparatory phase concern the improvements of the existing advanced technologies, necessary for the development, the integration and the operation of the proposed LB-CMPF device:

a) Production of high density and high temperature plasmas in a relatively big volume and in the presence of an externally-applied high magnetic field in a mirror-like topology, in order to trap the plasma and to increase the rate of nuclear fusion reactions (neutron production for DD fuel). The initial high density plasma will be produced by the interaction of an ultrashort, ultra-high intensity laser beam with clusters and /or thin solid targets (thin discs).

b) Optimization of the non-linear propagation of the ultrashort laser beam in the clusters, in order to improve the interaction volume.

c) Investigation on the production of high density ion beams with an energy up to 600 keV, from short laser beam interaction with thin solid target (thin discs) in the external applied high magnetic field in the mirror-like topology, especially for the case of the p-$^{11}$B nuclear fusion reaction.

d) Experimental studies on existing configurations to increase the externally-applied magnetic field up to extreme values.

e) Improvements on existing multifluid numerical codes, describing the spatio-temporal evolution of the high-density and high-temperature plasma in the externally applied high magnetic field.

f) coupling of the laser-produced high-density and high-temperature plasma with the pulsed high magnetic field driver. There is a proposition from M. Roth to use their laser facility using a long laser pulse.

g) A new experimental chamber will be installed in the IESL-FORTH laser facility (member of the European LaserLab) in Crete (Greece) in order to perform specific experiments concerning the non-linear propagation of ultra-short laser beam in clusters and initiate experiments on ion beam acceleration produced by high contrast and ultra-short laser beam interaction with thin solid targets (thin discs).

***Acknowledgements.*** F.N., G.A., S.B., M.O.C., I.D., P.G., M.G., C.I., V.L., L.N., I.O.M., C.P., M.R., F.R., D.S., M.T., B.T., I.C.E.T., D.U, S.G. and N.V.Z were supported by the Project Extreme Light Infrastructure - Nuclear Physics (ELI-NP) - Phase I, a project co-financed by the Romanian Government and European Union through the European Regional Development Fund. This work was in part supported by BMBF (project 05P15WMENA) and by MEN, program RO-CERN (ELI-NP; FAIR), contract E08.